\documentclass{article}
\usepackage[francais,english]{babel}
\usepackage[latin1]{inputenc}
\usepackage{geometry}
\usepackage{graphicx}
\usepackage{cite}
\usepackage{amssymb,amsmath}

\usepackage{pstricks}

\usepackage{amsfonts}

\def\BC{\begin{center}}
\def\EC{\end{center}}
\def\BF{\begin{figure}}
\def\EF{\end{figure}}
\def\BM{\begin{displaymath}}
\def\EM{\end{displaymath}}
\def\BE{\begin{equation}}
\def\EE{\end{equation}}
\def\BA{\begin{eqnarray}}
\def\EA{\end{eqnarray}}
\def\BAN{\begin{eqnarray*}}
\def\EAN{\end{eqnarray*}}

\title{Ginzburg--Landau description of laminar-turbulent oblique band formation in transitional plane Couette flow}
\author{Joran Rolland%
\thanks{{\it Corresponding author:}\protect\\
{\tt joran.rolland@ladhyx.polytechnique.fr}}, 
Paul Manneville\footnote{Laboratoire d'hydrodynamique de l'\'Ecole Polytechnique,
91128 Palaiseau, France}}

\date{\today}

\begin{document}

\maketitle

\textbf{abstract :}
Plane Couette flow, the flow between two parallel planes
moving in opposite directions, is an example of wall-bounded flow
experiencing a transition to turbulence with an ordered coexistence of turbulent and laminar domains in some range of Reynolds numbers
$[R_{\rm g}, R_{\rm t}]$. When the aspect-ratio is sufficiently large,
this coexistence occurs in the form of alternately turbulent and laminar oblique bands. As $R$ goes up trough the upper threshold $R_{\rm t}$,
the bands disappear progressively to leave room to a uniform regime of
featureless turbulence. This continuous transition is studied here by
means of under-resolved numerical simulations understood as a modelling approach adapted to the long time, large aspect-ratio limit. The state of the system is
quantitatively characterised using standard observables (turbulent
fraction and turbulence intensity inside the bands). A pair
of complex order parameters is defined for the pattern which is further
analysed within a standard Ginzburg--Landau formalism. Coefficients of
the model turn out to be comparable to those experimentally determined
for cylindrical Couette flow.

\sloppy

\section{Introduction\label{S0}}

In their way to turbulence, wall-bounded shear flows display cohabiting
turbulent and laminar regions. This striking phenomenon can even be 
statistically permanent and spatially organised, as for the flow between
counter-rotating cylinders (cylindrical Couette flow, CCF) or
counter-translating plates (plane Couette flow, PCF, Fig.~\ref{fig1},
top-left). Cohabitation then takes the form of alternately turbulent and
laminar oblique bands. This peculiar pattern was first discovered by Coles
and Van Atta in CCF (barber-pole or spiral turbulence) \cite{CVA66}, the
corresponding domain in the control parameter space being next charted
by Andereck {\it et al.} \cite{Aetal86}. These experiments were restricted
to the observation of a single spiral arm due to limited aspect-ratio
(the ratio of the gap between the cylinders to the perimeter).

Later on,
Prigent {\it et al.} \cite{Petal} performed studies at larger aspect-ratios, which allowed them to observe several intertwined spiral arms
and to show that the oblique bands in plane Couette flow (Fig.~\ref{fig1},
bottom-left) were, qualitatively and quantitatively, the zero curvature
limit of the spirals:
\BF
\BC
\includegraphics[width=0.30\textwidth]{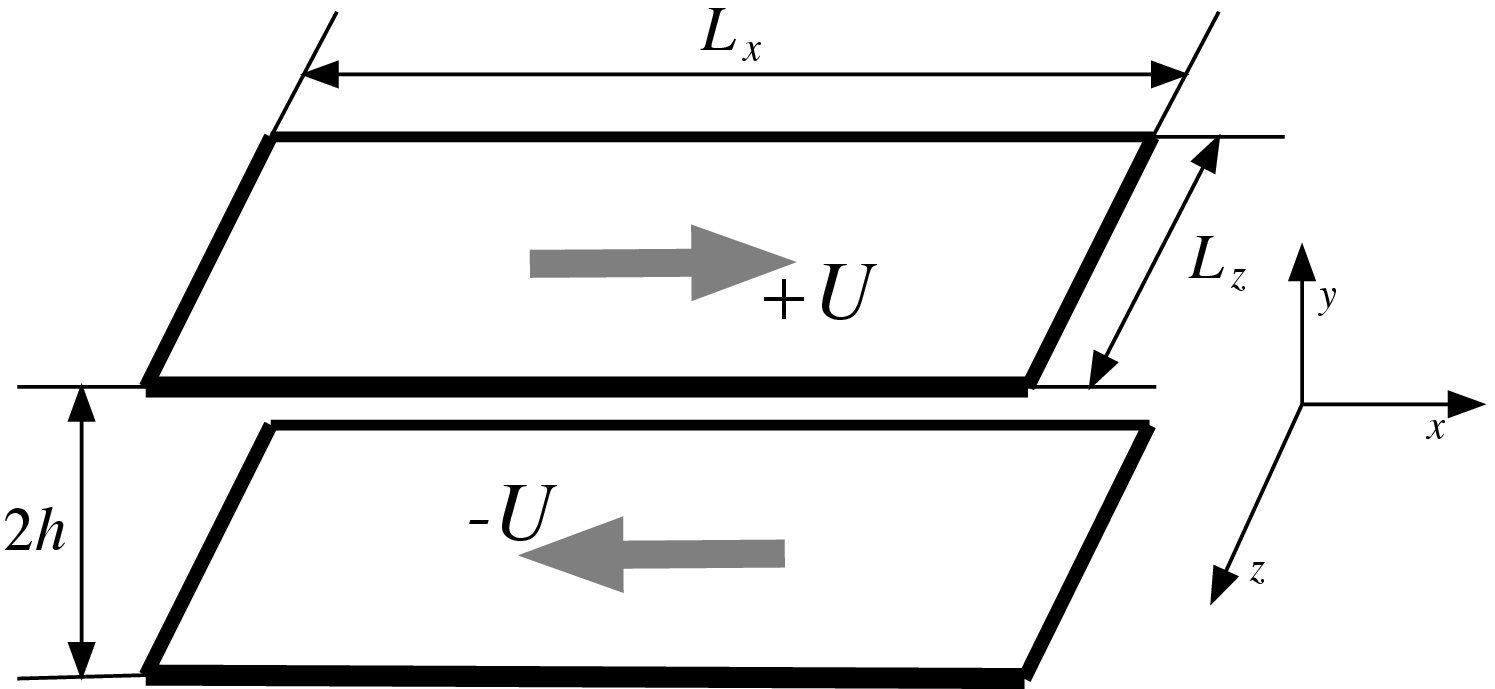}%
\hspace{2em}
\includegraphics[width=0.47\textwidth]{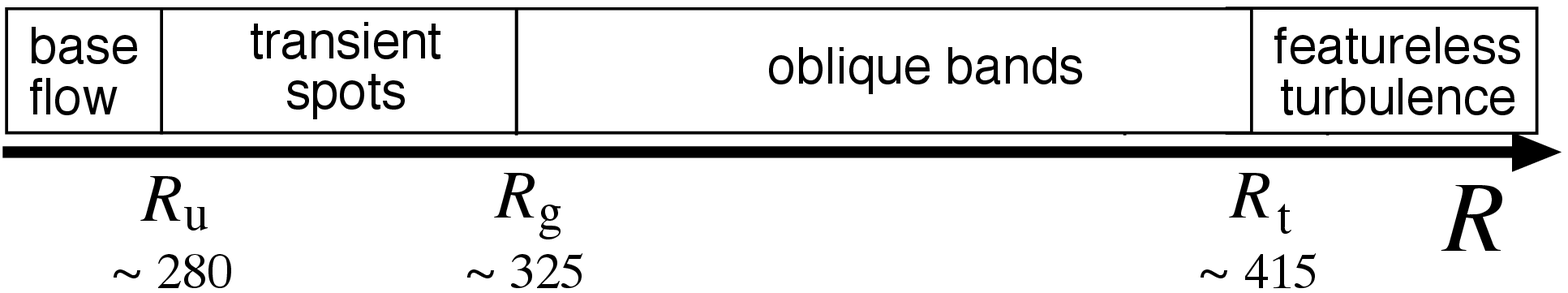}\\[6ex]
\includegraphics[height=0.16\textwidth,clip]{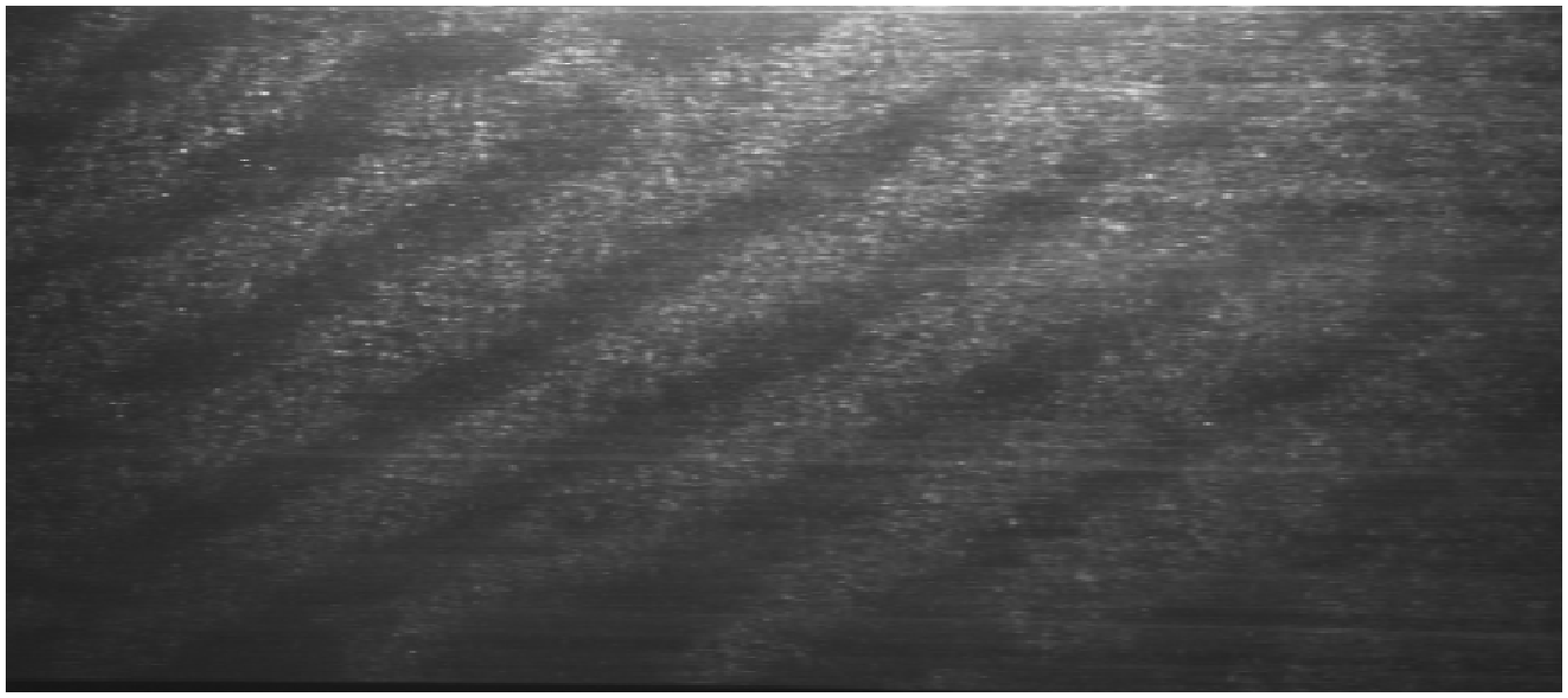}\hspace{2em}
\includegraphics[height=0.12\textwidth,clip]{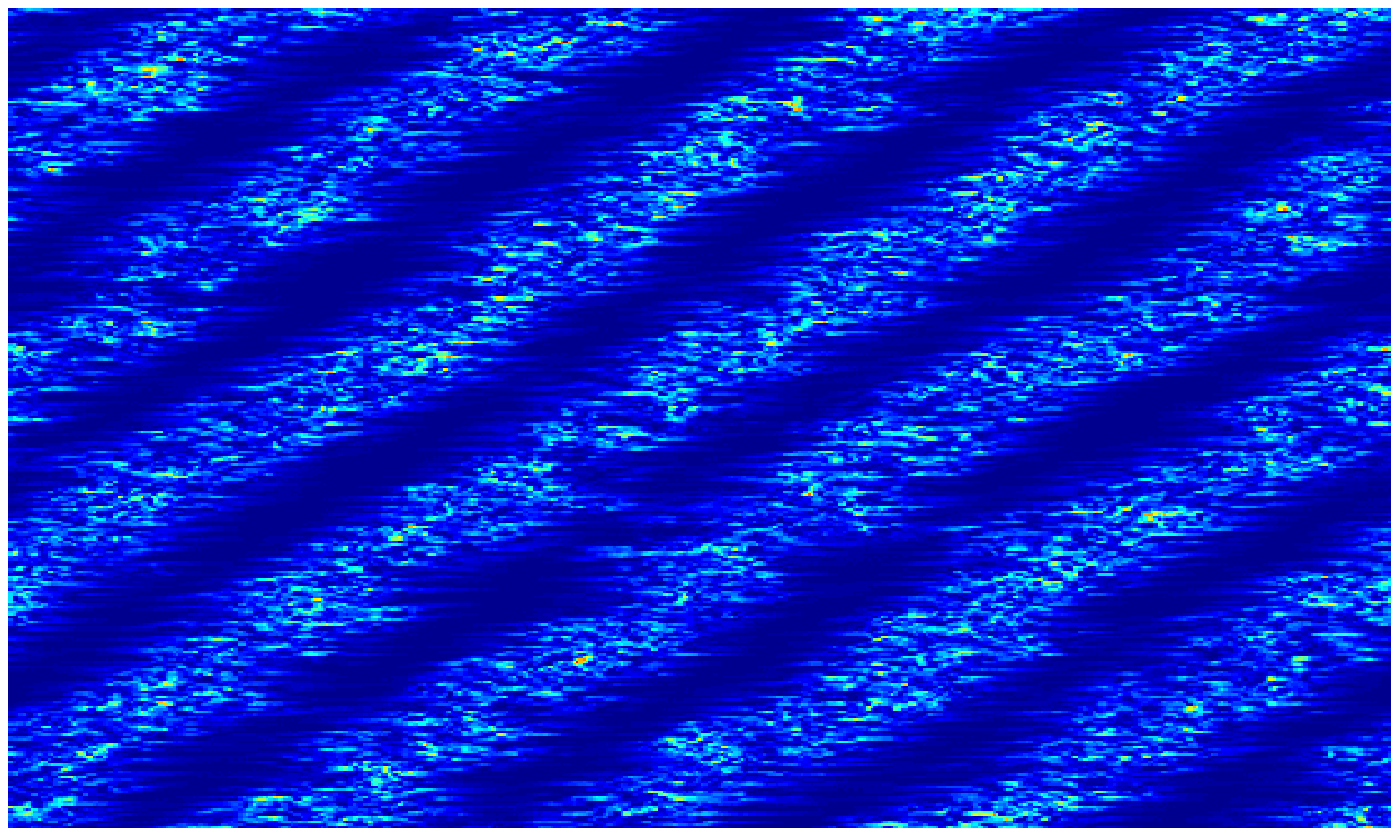}
\EC
\caption{Top-left: Geometry of the plane Couette flow experiment, $2h$ is the gap, $L_x$ and $L_z$) the streamwise and spanwise dimensions, $+U$
and $-U$ the wall speeds. The Reynolds number is defined as $R=Uh/\nu$
where $\nu$ is the kinematic viscosity. Top-right: Experimental
bifurcation diagram after Prigent \cite[(a)]{Petal}. Bottom: Picture of
experimental banded turbulence in plane Couette flow apparatus with
$L_x=770h$, $L_z=340h$ at $R=358$ (courtesy Prigent) and colour level
representation of the perturbation energy averaged over $y$ in our
under-resolved {\sc ChannelFlow} simulations with periodic boundary
conditions at $L_x=432$, $L_z=256$, $R=290$ and $t=18000$.
The two pictures are at roughly the same scale and similar Reynolds
numbers after correction for the transitional range $R$-shift due to
numerical under-resolution\cite{MRxx}.\label{fig1}}
\EF
Upon appropriate definition of a Reynolds number $R$ based on the nominal
shear rate, ({\it i\/}) these patterns bifurcate
continuously at similar values of a well-defined upper threshold
$R_{\rm t}$ above which turbulence is featureless, ({\it ii\/})
the spirals/bands are observed upon decreasing $R$ down to comparable
values of a lower stability threshold $R_{\rm g}$ below
which laminar flow eventually prevails, and ({\it iii\/}) the streamwise
and spanwise wavelengths are similar \cite{Metal01}. Figure~\ref{fig1} (top-right) recapitulates the experimental findings for PCF.

Direct numerical simulations (DNS) of the Navier--Stokes equations for
PCF were performed by Barkley \& Tuckerman \cite{BT05-07} who could
obtain the band patterns in fully resolved, elongated but narrow, tilted
domains. Their choice of boundary conditions however precluded the
occurrence of patterns with defects or orientation changes inside
the flow. This was not the case of the DNS by Duguet {\it et al.}
\cite{Detal10} who recovered the experimental findings of Prigent
{\it et al.} in fully resolved very large aspect ratio domains.
Similarly, the spiral regime was numerically obtained by Meseguer
{\it et al.} \cite{ref10} and Dong \cite{ref10b} in CCF and
the oblique band pattern in plane channel flow by Tsukahara
{\it et al.} \cite{ref9}.

Up to now, there is no clear physical explanation for the formation
of the spirals/bands from the featureless turbulent regime when $R$ is
decreased below $R_{\rm t}$ \cite[b]{BTD}. We however do have a consistent
phenomenological description of the transition in CCF by Prigent
{\it et al.} \cite{Petal} in terms of two coupled Ginzburg--Landau
equations with (strong) external noise added, introducing two complex
amplitudes, one for each possible pattern orientation. Most of the
coefficients introduced in these equations could be fitted against the
experiments. In a similar vein, Barkley {\it et al.} \cite{BTD}
introduced the phase-averaged amplitude of the dominant Fourier mode
of the turbulent mean flow modulation \cite[b]{BT05-07} as an order
parameter for the PCF transition. The emergence of the bands
was then identified from the position of the peak in the probability
distribution function (PDF) of this order parameter, shifting from zero
in the featureless regime to a nonzero value in the banded regime.  

In the present article, we come back to the quantitative characterisation
of the patterns in terms of order parameters. In contrast with
\cite{BT05-07,BTD}
we consider a configuration that does not freeze the orientation and allows 
for defective patterns. We keep the general noisy Ginzburg--Landau
framework introduced in \cite{Petal} for CCF and validate the approach in
terms of amplitude equations at a quantitative level for PCF by means of numerical experiments.
We take advantage of our previous work where the recourse to under-resolved DNS using Gibson's public domain code {\sc ChannelFlow} \cite{Gibson} was introduced \cite{RM09}. In \cite{MRxx} we brought evidence that this procedure could be viewed as a consistent systematic modelling strategy permitting simulations in wide domains during long time lapses at moderate numerical load.
We indeed showed that all qualitative aspects of the transitional range
are preserved  at the recommended resolution (Fig.~\ref{fig1},
bottom-right) and that, in the (slightly better) numerical conditions
chosen here, the resolution lowering amounts to a 15--20\% downward shift
of $[R_{\rm g},R_{\rm t}]$ from the experimental findings. This resolution
reduction will allow us to accumulate statistics on moderate aspect ratio
systems during very long times. We surmise that our results can be carried over to the realistic case of fully resolved simulations or experiments up to an appropriate adaptation of the Reynolds scale. We shall support this point of view briefly in \S\ref{Sb-R}.

We first recall the numerical procedure in \S\ref{Sb-implementation}, next we turn to the extraction of the turbulent fraction and the turbulence intensity (\S\ref{Sb-averaging}) and to the definition of order parameters
able to include information about the spatial organisation,
\S\ref{Sb-orderparameter}. Results are then analysed in the successive
subsections of \S\ref{S2} devoted to the determination of the
phenomenological parameters introduced by the Ginzburg--Landau formalism
and accounting for the spanwise, streamwise and $R$ dependence
of the pattern. Section~\ref{S3} summarises our findings.

\section{Simulations and data processing\label{S1}}

\subsection{Numerical implementation\label{Sb-implementation}}
\BF
\BC
\includegraphics[width=0.64\textwidth,height=0.05\textwidth,clip]%
{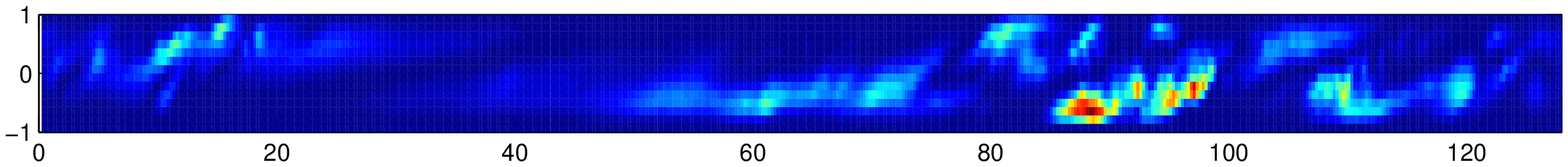}
\vspace{2ex}

\includegraphics[width=0.9\textwidth,height=0.05\textwidth,clip]%
{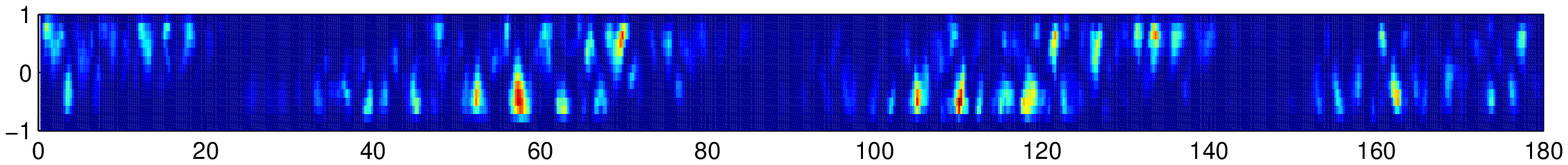}
\vspace{4ex}

\includegraphics[width=0.20\textwidth,height=0.318\textwidth,clip]%
{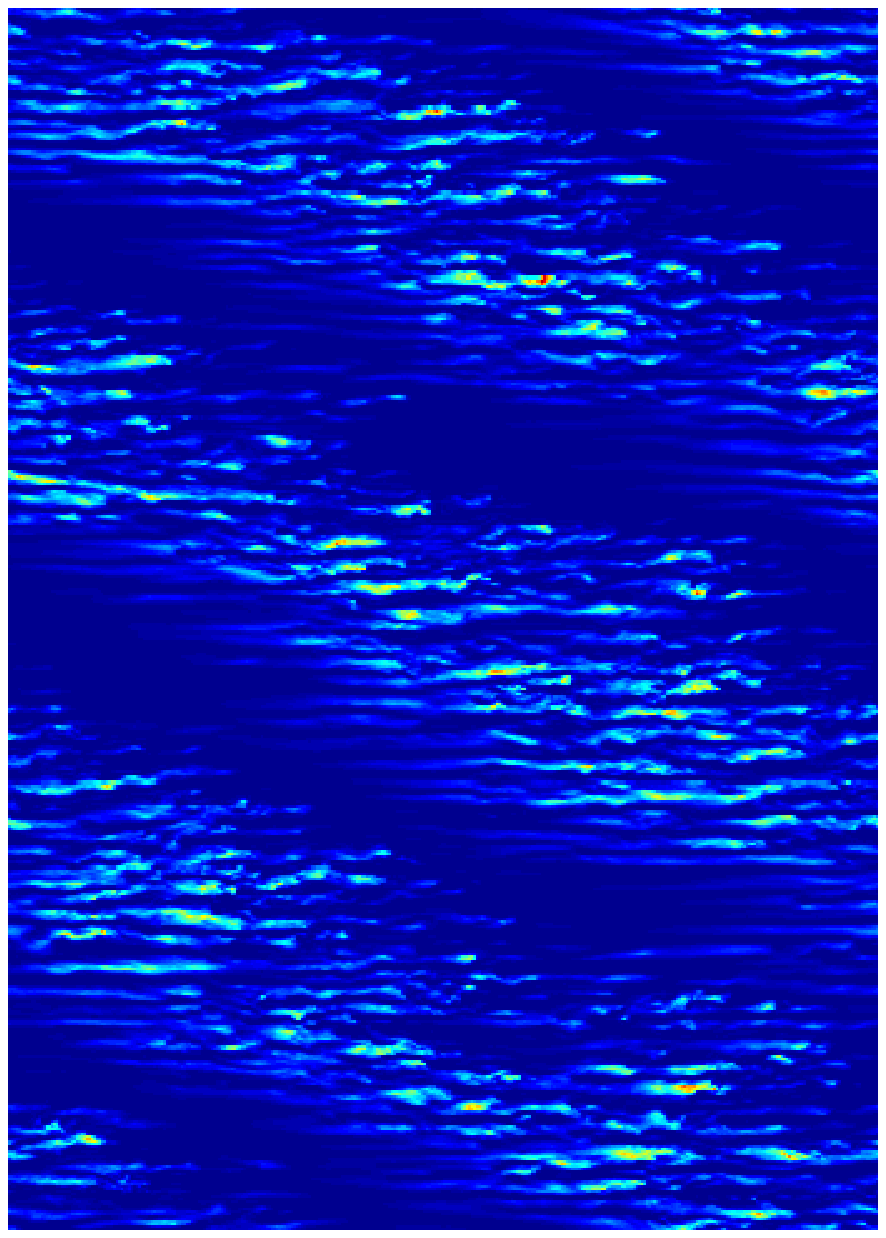}
\hskip2em
\includegraphics[width=0.20\textwidth,height=0.318\textwidth,clip]%
{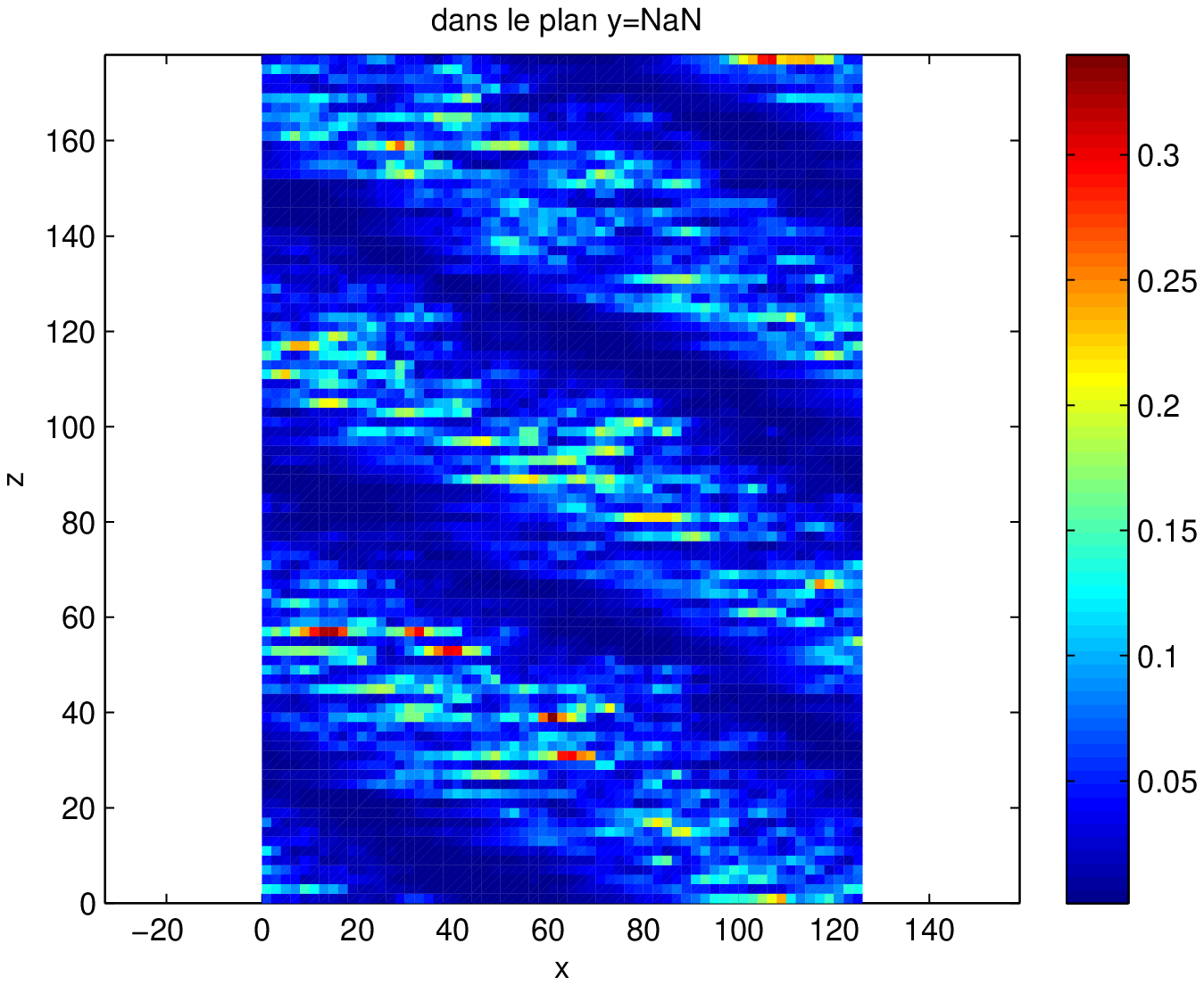}
\hskip2em
\includegraphics[width=0.20\textwidth,height=0.318\textwidth,clip]%
{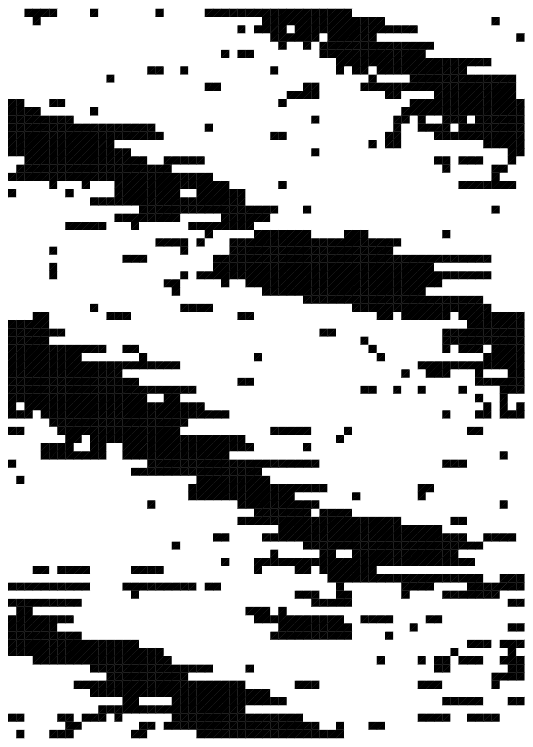}
\hskip2em
\includegraphics[width=0.20\textwidth,height=0.318\textwidth,clip]%
{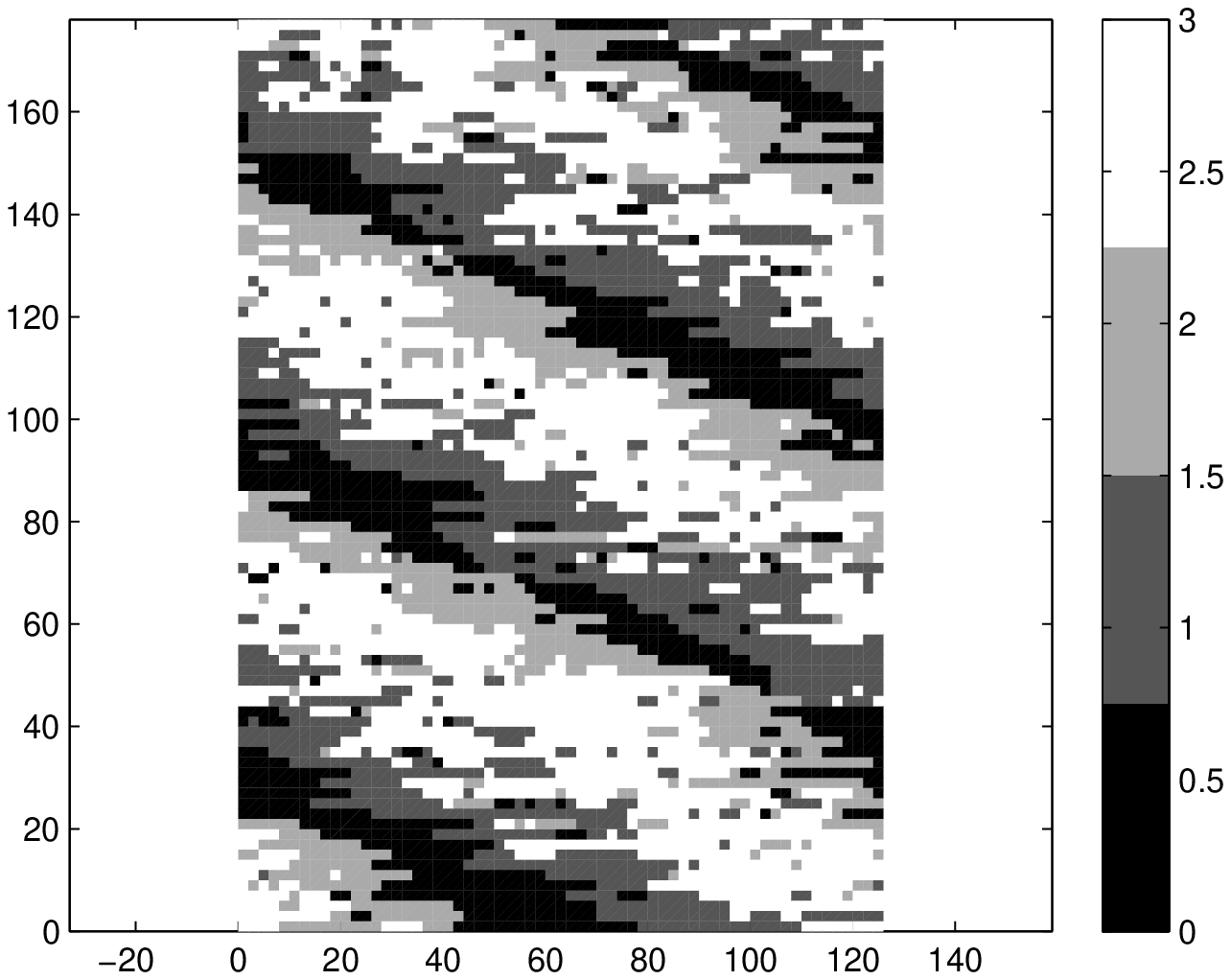}
\EC
\caption{Top: $u_x^2$ in an $x,y$ plane, $L_x=128$ and in a $z,y$
plane, $L_z=180$. Bottom: From left to right, ${\bf u}^2$ in the
$y=-y_{\rm m}$ plane, coarse-grained ${\bf u}^2$ in the $y<0$ domain,
resulting B/W discrimination, and W/G/B discrimination (see text).
$L_x\times L_z=128\times 180$, $R=315$.\label{fig2}}
\EF

The geometry of the experiment is described in Fig.~\ref{fig1}
top-left. The Navier--Stokes equations are written in a reference
frame where $x$, $y$, $z$ are the streamwise, wall-normal, and
spanwise directions respectively. Velocities are made dimensionless
with $U$ the absolute value of the speed at the boundaries $y=\pm h$.
Lengths are rescaled by $h$ and time by $h/U$. The main control parameter is the Reynolds
number $R=Uh/\nu$, where $\nu$ is the kinematic viscosity but
the flow regime also depends on the aspect ratios defined as
$\Gamma_{x,z}=L_{x,z}/2h$, where $L_{x,z}$ are the lateral
streamwise and spanwise dimensions. In the numerics, $h=1$ and the aspect
ratios are $\Gamma_{x,z}=L_{x,z}/2$. The base flow is independent of
$R\,$:
${\bf v}_{\rm b}=y\,{\bf e}_x$. Written for the perturbation to the base
flow ${\bf u}={\bf v}-{\bf v}_{\rm b}$, the Navier--Stokes equations read:
\begin{eqnarray*}
\partial_t u_i+\partial_j (u_i u_ j) &+& y\partial_x u_i+u_y\delta_{i,x}\nonumber\\
&=&\mbox{}-\partial_i p + R^{-1}\partial^2_{jj} u_i\,,\\
\partial_j u_j&=&0\,,
\label{eq1}
\end{eqnarray*}
with no slip boundary conditions at the plates, $u_i(y=\pm1)= 0$, and
periodic boundary conditions at distances $L_x$ and $L_z$ in the
streamwise and spanwise directions, respectively.

{\sc ChannelFlow} \cite{Gibson} implements the Navier--Stokes equations
using a standard pseudo-spectral
scheme with Fourier transforms involving $(N_x,N_z)$ de-aliased modes in
the streamwise and spanwise directions and  $N_y$ Chebyshev polynomials
in the wall-normal direction. As discussed in \cite{MRxx}, our numerical
simulations are deliberately under-resolved: we use $N_y=15$, and
$N_{x,z}/L_{x,z}=8/3$, which preserves all the qualitative features of
the flow at a semi-quantitative level, just shifting the bifurcation
thresholds down to $R_{\rm g}=275\pm5$ and
$R_{\rm t}=345\pm5$, to be compared with experimental or fully resolved
numerical values, $R_{\rm g}\simeq325$ and $R_{\rm t}\simeq415$
\cite{Petal,Detal10}.

In PCF, Prigent {\it et al.} experimentally found
oblique turbulent bands with streamwise period $\lambda_x\simeq 110$ and
variable spanwise period $\lambda_z$ from $85$ around $R_{\rm g}$ to $45$
close to $R_{\rm t}$. The sizes of our numerical domains range from
$L_z=24$ to $192$ and from $L_x=80$ to $170$. Our domains hence remain
rather small since they can contain one to three such spanwise wavelengths
but they are much larger than the
minimal flow unit \cite{mfu} of size $\ell_x\approx6$ and $\ell_z\approx4$, 
below which turbulence cannot self-sustain. They are also much longer in
the streamwise direction than the tilted domains considered by Barkley
{\it et al.} \cite{BT05-07,BTD} but remain smaller than the largest domains considered by
Duguet {\it et al.} \cite{Detal10} or in our preliminary studies
\cite{MRxx} which went
up to $L_x=800$ and $L_z =356$ but at a much lower resolution,
or the latest experiments by Prigent
{\it et al.} with $L_x=770$ and $L_z=340$ \cite{Petal}.

\subsection{Local averaging and related quantities\label{Sb-averaging}}

The square of the perturbation velocity ${\bf u}^2$ is a good
indicator of the local state of the flow. Figure~\ref{fig2} (top)
displays colour level representations of that quantity in typical
wall-normal planes, streamwise $(x,y)$ with height 2 and length
$L_x=128$, and spanwise $(z,y)$ with height 2 and width $L_z=180$,
for $R=315$. The pattern seen from above in the $(x,z)$ plane at a
given wall-normal coordinate $y=-0.57=-y_{\rm m}$ is displayed in
Fig.~\ref{fig2} (bottom, left), the other panels represent the same
image after additional post-treatment to be discussed below.
The value $y_{\rm m}=0.57$ roughly corresponds to the place
where ${\bf u}^2$ is statistically the largest in the range of Reynolds
numbers of interest (see Fig.~5 in \cite{MRxx}).

\begin{figure}[!h]
\BC
\includegraphics[width=0.35\textwidth]{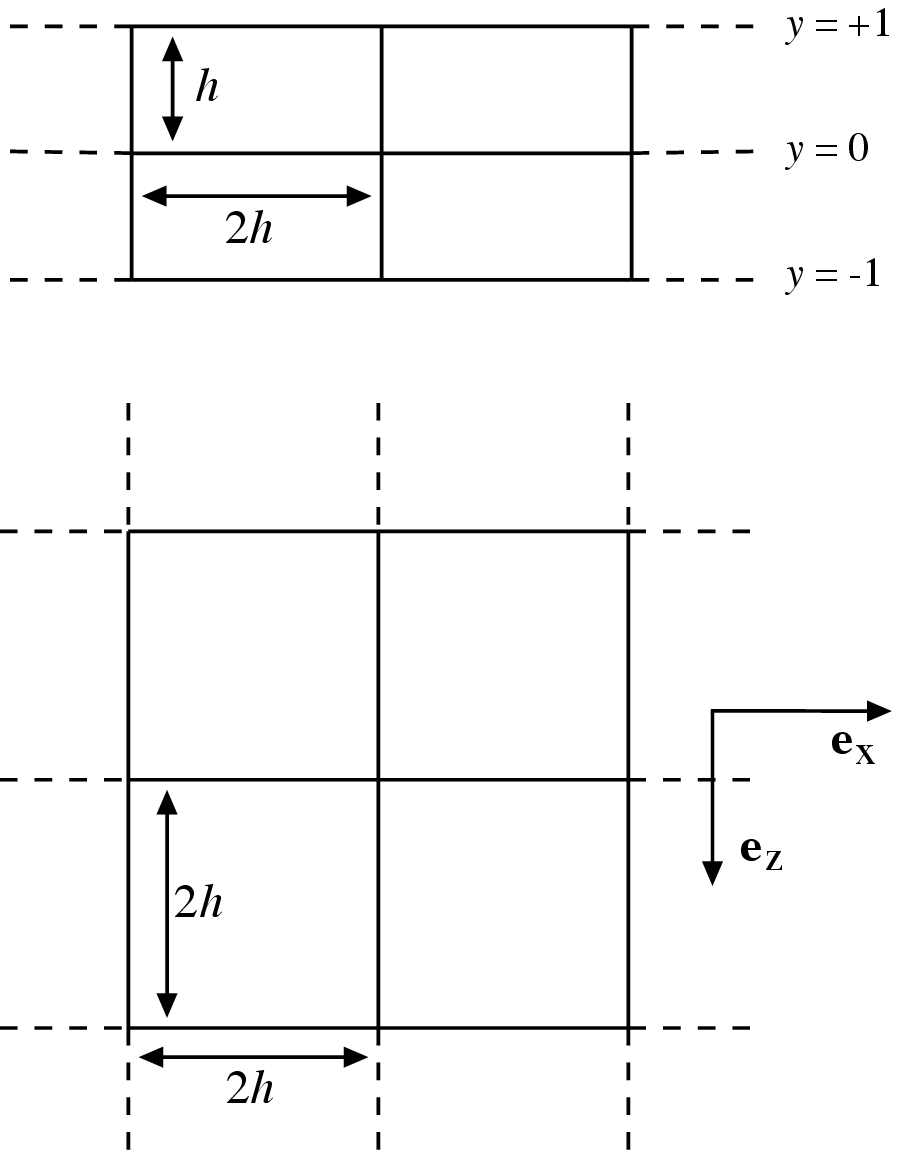}
\EC
\caption{Sketch of the averaging boxes from the side and from above.
\label{fig3}}
\end{figure}

\begin{figure}[!h]
\BC
\includegraphics[height=6cm,clip]{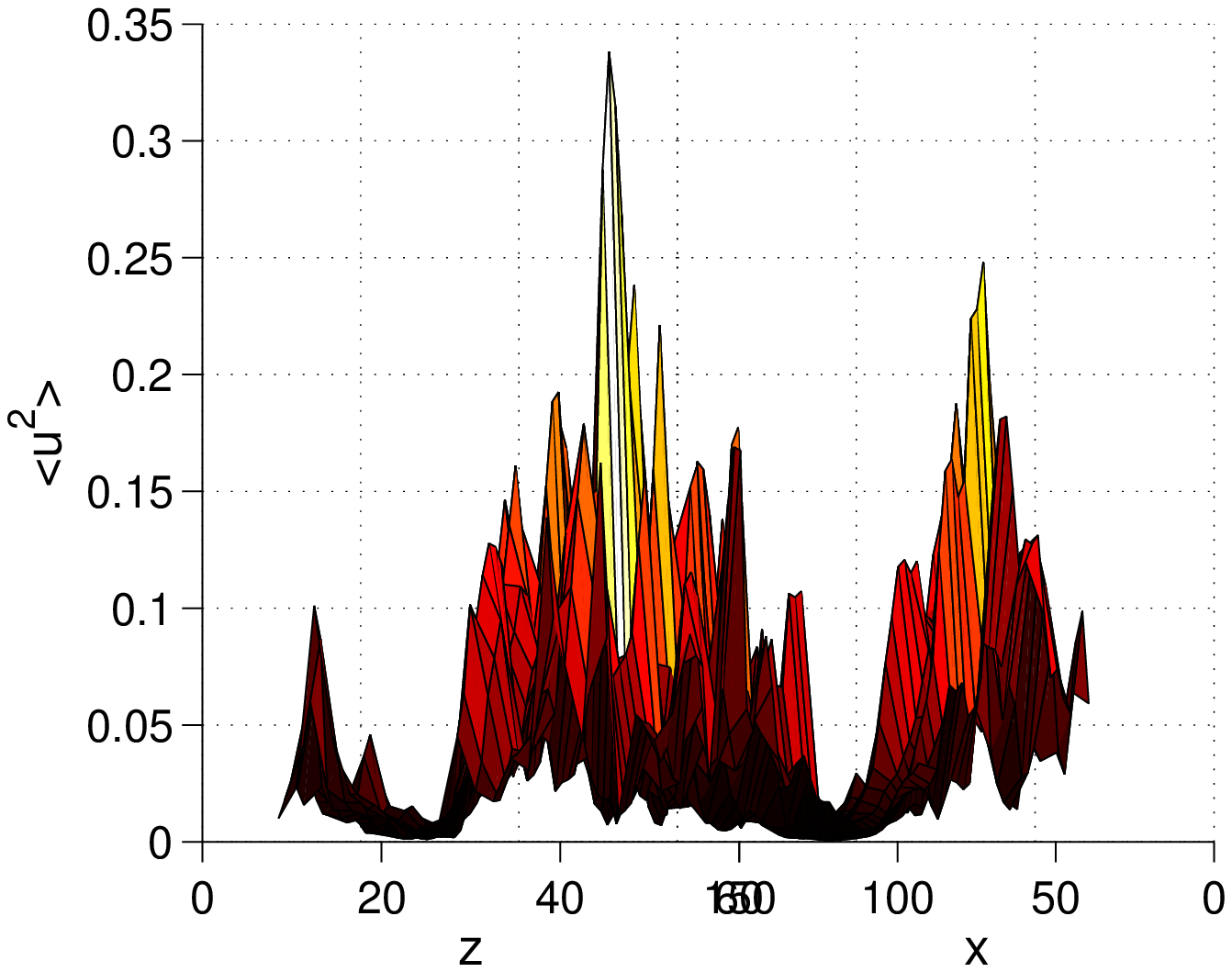}
\EC
\caption{Coarse-grained $u^2$-profile projected along
the direction of the turbulent band ($L_x\times L_z=128\times 64$, $R=315$, $y<0$).\label{fig4}}
\end{figure}

The simplified representations shown in the centre
and right panels of Fig.~\ref{fig2} rest on the coarse-graining of
the ${\bf u}^2$ field introduced in~\cite{RM09}.
This procedure directly stems from the general
organisation of the flow in the band regime already identified in
previous studies \cite{CVA66,BT05-07} and clearly visible in the side
and front views of the flow in Fig.~\ref{fig2} (top).
These pictures suggest to average over the upper layer of the flow ($y>0$) and
its lower layer ($y<0$) separately. Typical experimental observations \cite{Petal,Aetal86,Betal98}, film and pictures, yield an information integrated over the whole gap, which motivates us to compute comparable quantities. As shown in Fig.~\ref{fig3}, the
computational domain is divided in small stacked boxes of size
$l_x\times l_y\times l_z=2\times 1 \times 2$. This size is slightly
smaller than (but related to) that of the minimal flow unit. The width
$l_z=2$ approximately corresponds to the spanwise size of a turbulent
streak. By contrast, $l_x=2$ is much smaller than the typical length of a
turbulent streak, $\tilde{l}\simeq 40$, so that the turbulent intensity
variations along a streak can be captured. Quantity ${\bf u}^2$, henceforth 
called `energy' by a small abuse of language, is then
averaged in each of these cells and a threshold $c$ is chosen according
to which it is laminar or turbulent. The turbulent fraction $f$ is then
the proportion of turbulent cells, and the turbulent energy $e_{\rm t}$
is the energy conditionally averaged in space over the turbulent zone. Conditional averaging of any field can easily be performed in the same
way. The reduction procedure is expected to depend on the value of $c$.
As seen in Fig.~\ref{fig4} which displays the profile of the
coarse-grained energy through the band pattern, the
locally turbulent flow has typical energy higher than 0.1 and locally
laminar flow less than 0.05. The computation of the time-averaged%
\footnote{On general grounds, lower case letters will denote
instantaneous values and upper case letters the corresponding
time averages.}
turbulent fraction $F$ and the time-averaged turbulent energy
$E_{\rm t}$ for values of $c$ ranging from $0.005$ to $0.13$
did not pointed to an optimal value for $c$, as expected from a flow displaying a smooth modulation of turbulence, and $c=0.025$
was eventually chosen with little consequence on the quantitative
information drawn from the procedure.

A typical example of this thresholding is given in Fig.~\ref{fig2},
bottom line:
from a realisation of the flow at $y=-y_{\rm m}$ (left) we compute the
coarse-grained energy for $y<0$ (centre-left) and apply the criterion
to obtain a black-and-white (B=laminar, W=turbulent) representation of
the flow, still for $y<0$ (centre-right).
Distinguishing the $y>0$ layer from the $y<0$ layers allows a refined
representation of the flow as shown in the bottom-right panel of
Fig.~\ref{fig2} which displays the turbulent and laminar areas
using a black/gray/white code: `black' represents laminar cells
of top of each other, `white' turbulent cells on top of each other,
`light grey' $y>0$ turbulent cells on top of $y<0$ laminar cells,
and `dark grey' $y<0$ turbulent cells of top of $y>0$ laminar
cells~\cite{RM09}.
As already seen in the top panels, the streamwise direction going from
left to right, turbulence is to the right of the band for $y>0$ and to
its left for $y<0$, in agreement with previous findings
\cite{CVA66,BT05-07}. This fact could be used to compute properties at
the edge of the bands, for instance velocity or energy profiles. A
quantitative comparison to results of Barkley and Tuckerman
\cite[b]{BT05-07} has not been attempted since the differences in
geometry and resolution shift the Reynolds number correspondence.
\begin{figure}[!h]
\centerline{\includegraphics[height=0.2\textwidth,clip]{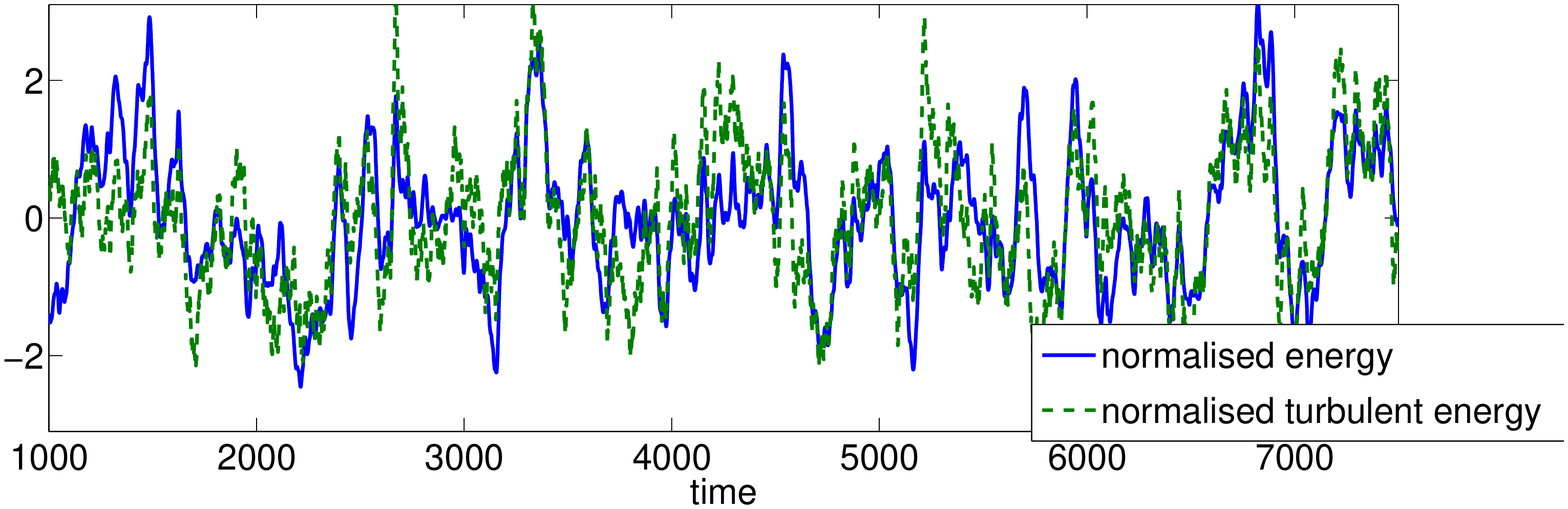}}
\centerline{\includegraphics[height=0.2\textwidth,clip]{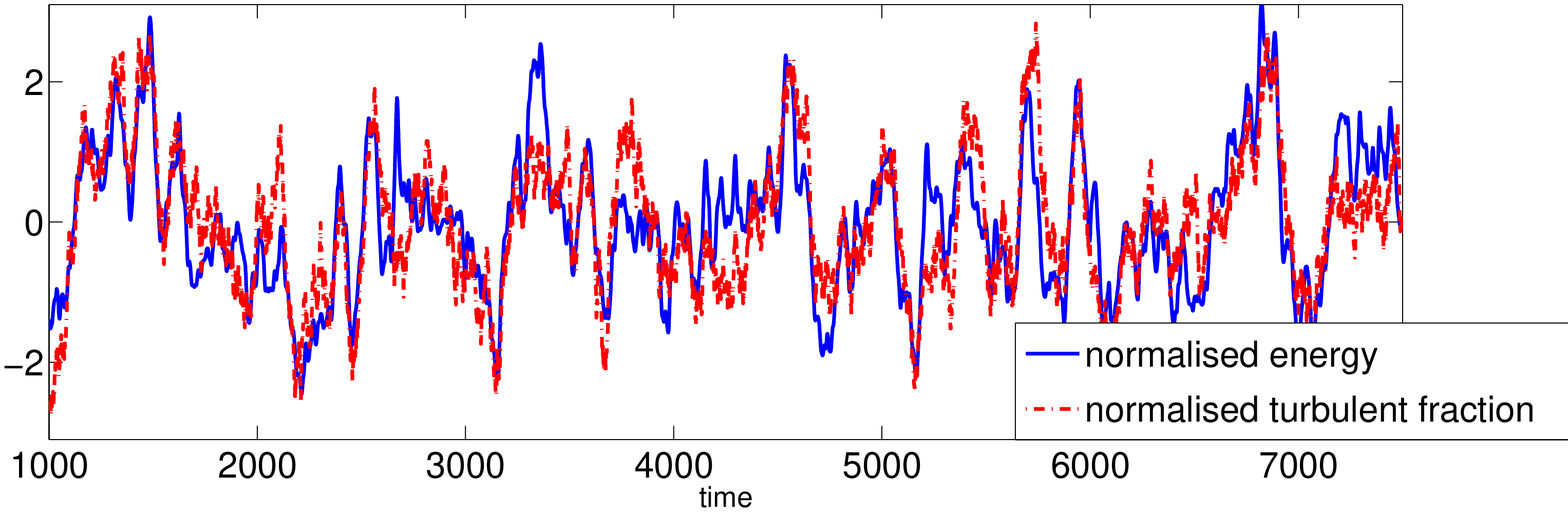}}
\centerline{\includegraphics[height=0.2\textwidth,clip]{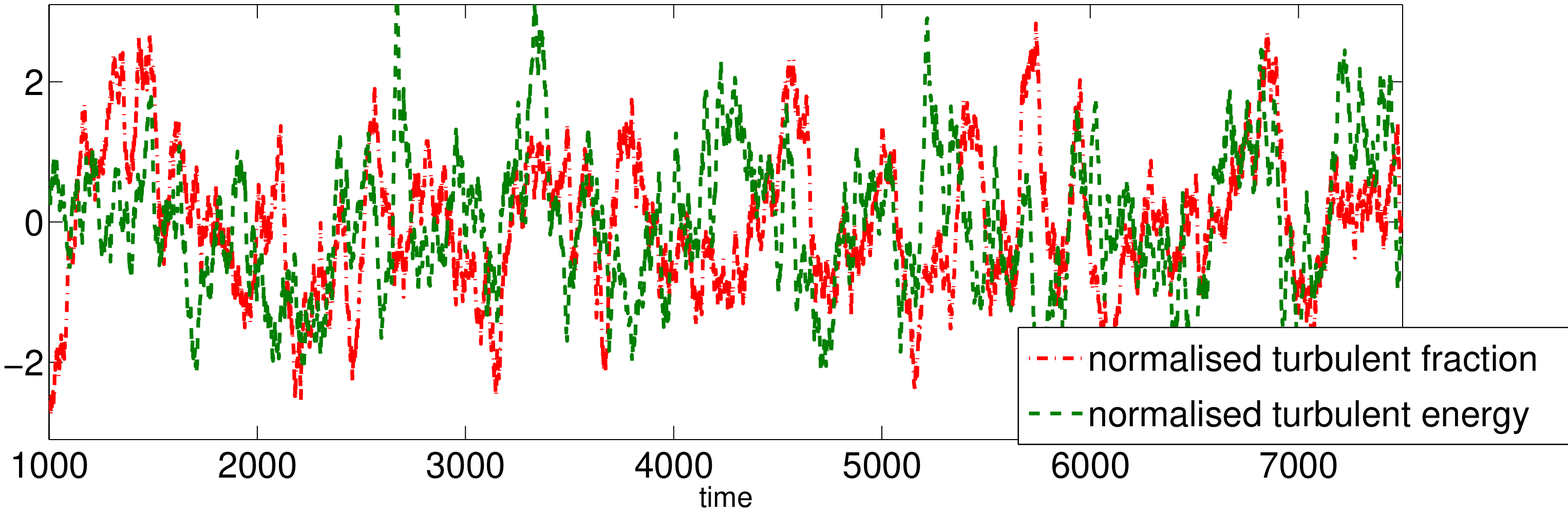}}
\caption{Time series of the normalised average energy $\bar e$,
turbulent energy $\bar e_{\rm t}$ and the turbulent fraction $\bar f$
in a typical numerical experiment for $L_x\times L_z=110\times48$ and $R=315$.\label{fig5}}
\end{figure}

The procedure has been implemented on-line to allow the computation of
time series of the turbulent quantities. Since these quantities fluctuate,
we compute their time-averages $E$, $E_{\rm t}$, and $F$ as
\begin{displaymath}
E=\frac{1}{T}\int_{T_0}^{T_0+T}e(t)\, {\rm d}t\,,\end{displaymath}
where $T_0$ is introduced to take into account the transient
necessary for the flow to reach its permanent regime, and $T$
is taken sufficiently large (typically, over $5000$) to keep the relative fluctuations of $E(T)$ within $0.5\,$\%. The cut-off $c$ being
appropriately chosen,
the energy content of the laminar part is negligible so that
we have $e\simeq f \times e_{\rm t}$, which means that the average energy of the flow is positively correlated to the changes of turbulence intensity
in the bands measured by $e_t$, as well as to the fractional area $f$
occupied by the bands. On the other hand, quantities $e_t$ and $f$ do not
show much correlation. This can be seen in figure~\ref{fig5} in which
normalised quantities,
$\bar{e}=(e-E)/(\langle e^2 \rangle-E^2)^{1/2}$, etc., are displayed.
Computation of the correlation of $\bar{e}$ and $\bar{f}$, as well as
$\bar{e}$ and $\bar{e}_t$ yields $0.5\pm0.1$, on average over all
experiments, whereas $\bar{e}_t$ and $\bar{f}$ are not correlated, yielding
$0\pm 0.1$. Owing to small relative fluctuations, the relation
$e\simeq f \times e_{\rm t}$ implies a similar relation,
$E\simeq F \times E_{\rm t}$, for the averaged quantities.

It turns out that $e(t)$, $e_{\rm t}(t)$, and $f(t)$ are little affected by 
the orientation fluctuations: even when the pattern presents defects,
the surface occupied by turbulence and the turbulence intensity in the
bands remains essentially unchanged. This allows us to perform averaging
regardless of the orientation, but $E$, $E_t$, and $F$ remain sensitive
to the value of the bands' wavelength imposed by the periodic
boundary conditions fixing the in-plane dimensions $L_{x,z}$,
as discussed below.

\subsection{Order parameter\label{Sb-orderparameter}}

\subsubsection{Conceptual framework and operational definitions}

In the theory of phase transitions, an order parameter is an
observable which, at the thermodynamic limit (permanent state at infinite
size), is zero in the non-bifurcated state, here the featureless
turbulent regime, and non-zero in the bifurcated state, here measuring the
amount of coexisting laminar and turbulent domains. The turbulent fraction $F$ (introduced in \cite{Betal98} at a time when
the spatially organised character of the banded regime was not yet
recognised) or rather the laminar fraction $1-F$, partially fulfils this
condition but remains of limited value since it does not account for the
space periodicity of the pattern explicitly, which is what we want to
overcome, inspired by previous work~\cite{Petal,BTD}. In pattern-forming
systems, the bifurcation is generally characterised by the amplitude of
the relevant bifurcating mode and, especially in extended systems, by the
amplitudes of the modes entering the Fourier decomposition of the
structure that develops from the instability mechanism. When fitting the
pattern-forming problem into the phase transition formalism, these
amplitudes are the natural order parameters.

Figure~\ref{fig6}
\BF
\BC
\includegraphics[height=0.3\textwidth,clip]{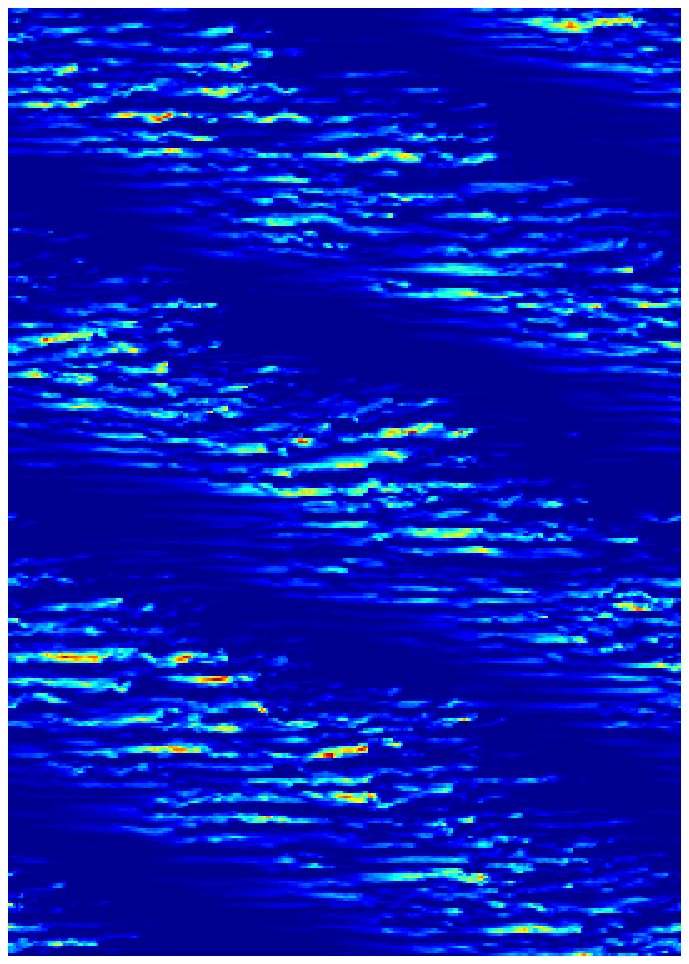}
\hspace{2em}
\includegraphics[height=0.3\textwidth,clip]{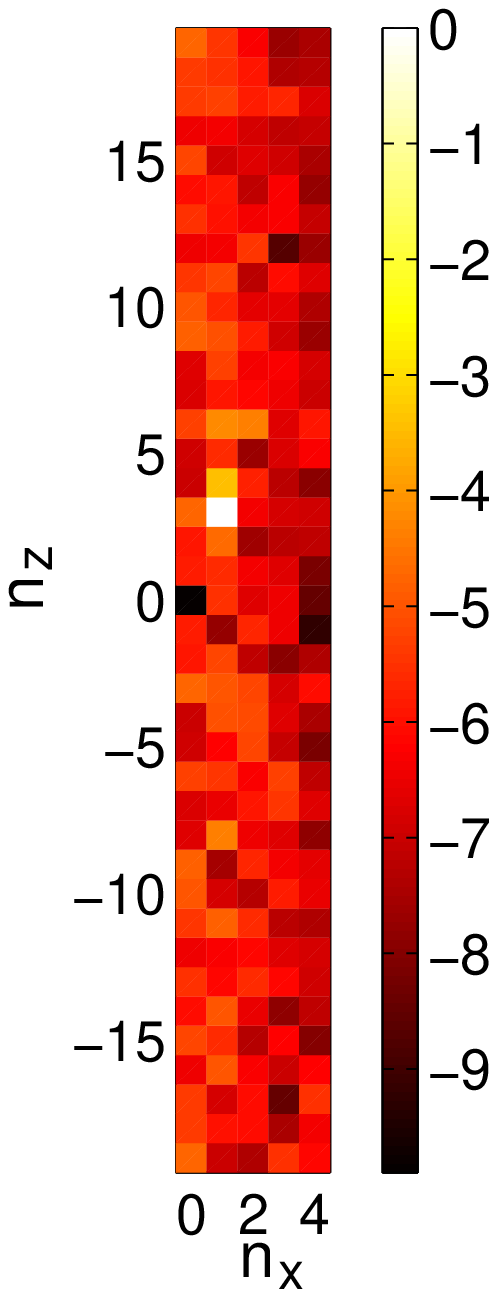}
\hspace{4em}
\includegraphics[height=0.3\textwidth,clip]{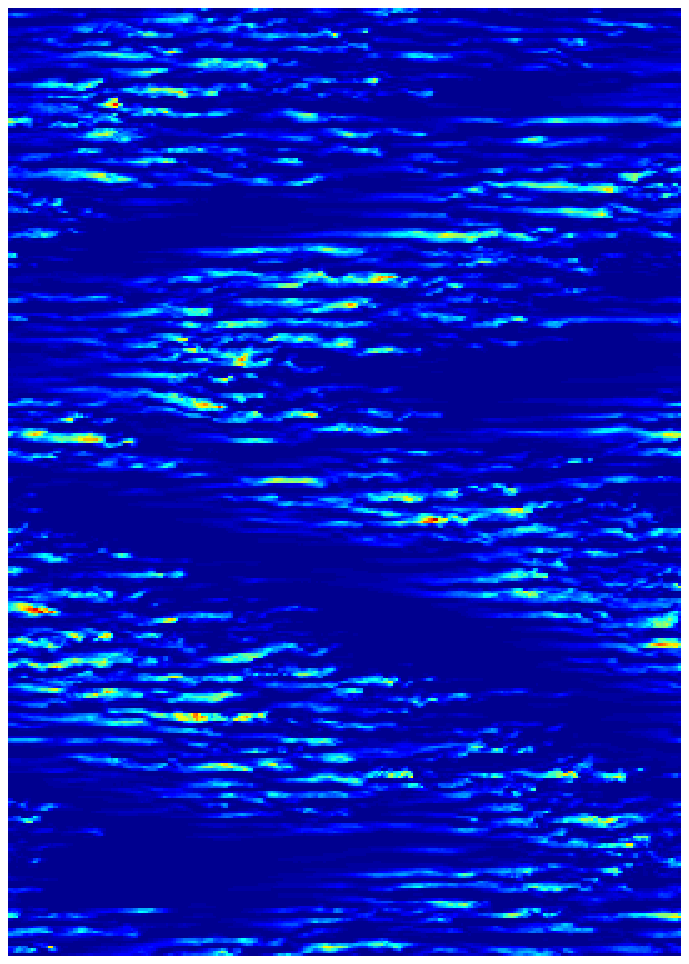}
\hspace{2em}
\includegraphics[height=0.3\textwidth,clip]{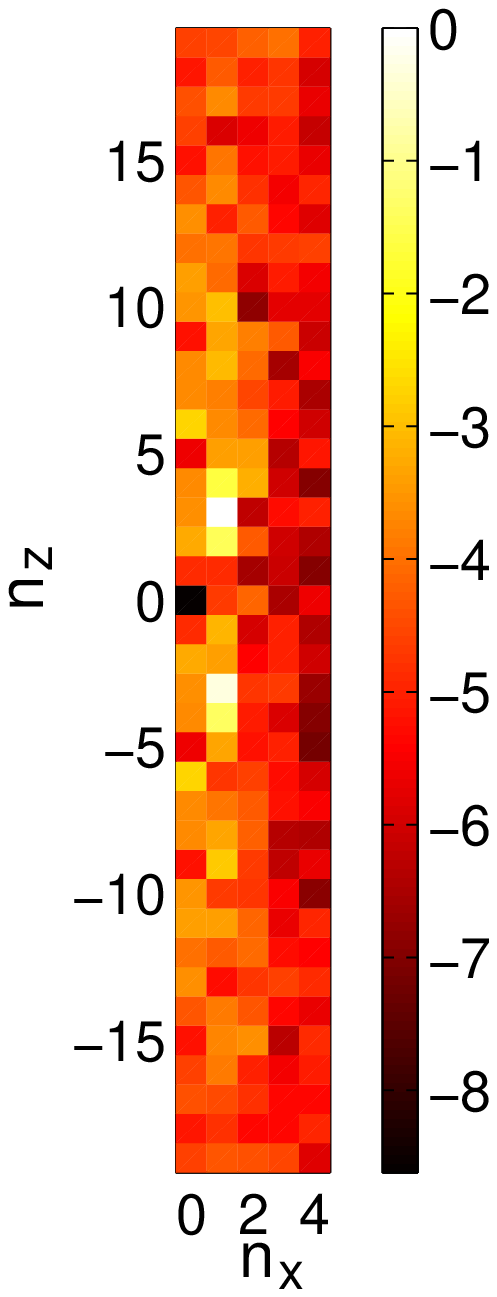}
\EC
\caption{Color plots of patterns and natural logarithm of the
corresponding spectra averaged over $y$ as explained in the text
for a well-formed pattern (left) and for a
pattern with defects (right). In the spectra,
$\hat{u}_x(0,y,0)$ is set to zero, which yields a black dot in the spectra; normalisation by the maximum value
makes its position appear the peak as a white dot.
$L_x\times L_z=128\times 180$, $R=315$.\label{fig6}}
\EF
illustrates the result of a Fourier analysis of patterns with three
bands fitting a domain of size $L_x\times L_z=128\times 180$.
Symmetries in the spectrum allow us to consider wave numbers such that
$0\le n_x\le N_x/2$, $-N_z/2 +1\le n_z \le N_z/2$.%
\footnote{Strictly speaking only $0\le n_x\le N_x/3$ and
$-N_z/3 +1\le n_z \le N_z/3$ since $N_{x,z}$ are the numbers of
de-aliased modes so that, the 3/2-rule being used, the number of
modes truly involved in the dynamics is $2N_{x,z}/3$ and the
corresponding bounds $(2N_{x,z}/3)/2=N_{x,z}/3$. This proviso is
however not essential since we are only interested in centre of the
spectrum with $n_{x,z}$ small.}
The figure displays $(x,z)$-plots of ${\bf u}^2$ at $y=-y_{\rm m}$
(left) and corresponding spectra averaged over the wall-normal direction
(right). The top panels correspond to an ideally formed pattern and the
bottom panels to a defective one. For both flows, the wave numbers corresponding to the peak are $n_x=1$ and $|n_z|=3$.
The spectra are zoomed on the smallest wave numbers so that modulations
at the scale of the streaks are outside the reframed graphs. 
When the pattern is well formed, a single mode corresponding to the
fundamental of the modulation clearly emerges, about two orders of magnitude larger than the other modes. These  background modes account
for small irregularities at a given time and not to steady anharmonic
corrections to a basically sinusoidal profile: the average ratios
$m_{n_z\ne3}/m_{n_z=3}$ are at most $0.1$ and the harmonics have no
definite phase relation  with the fundamental,
corroborating the observation by Barkley and Tuckerman that the
modulation is quasi-sinusoidal \cite[b]{BT05-07}. 
In the defective case (figure~\ref{fig6}, bottom), two peaks emerge,
corresponding to the two orientations. Their amplitude is smaller,
and other harmonics have non negligible amplitudes, accounting for
the spatial modulations of the pattern. Envelopes can be defined,
one for each orientation, obtained by standard demodulation.

The picture shown corresponds to a case with three bands, showing that there is enough room for a grain boundary. For smaller systems with one or two bands, defects correspond to the coexistence of
laminar and turbulent regions without conspicuous organisation.
In fact, the pattern can be observed only when the domain is
above some minimal size $L_{x,z}^{\rm{min}}$. Our simulations suggest
$L_z^{\rm{min}}\sim24$ and $L_x^{\rm{min}}\sim70$ (a precise
determination of the minimal size is still under study). This is
much smaller than in experiments because periodic boundary conditions 
tend to stabilise the pattern: only a tendency to form oblique turbulent patches was observed in laboratory experiments with
$L_x\times L_z = 280 \times 72$ \cite{Betal98}, where the ideal
simple shear flow was achieved with sufficient accuracy only in the centre
of the set-up due to lateral boundary effects.

Following Prigent {\it et al\/}~\cite[c]{Petal}, both orientations being
equivalent, we expect that the pattern can be characterised by two complex
quantities $A_\pm$:
\BE
u_{x}=\sum_{\pm} A_\pm\left(\tilde x,\tilde z,\tilde t\,\right)
\exp i(k_x^{\rm c}x\pm k_z^{\rm c} z) +{\rm c.c.}\,,
\EE
where $A_\pm\in \mathbb{C}$ describe {\it slow modulations\/} at
scales much larger than $\lambda_{x,z}^{\rm c}=2\pi/k_{x,z}^{\rm c}$,
the `optimal' streamwise and spanwise wavelengths. 
Variables ${\tilde x}$ and ${\tilde z}$ denote the corresponding space
coordinates. Despite the highly fluctuating nature of the turbulent flow,
the pattern being time-independent, there is just a possible {\it slow evolution\/} at an effective time $\tilde t$ linked to wavelength
selection and defect dynamics. The modulus of
$A_\pm$ gives the amplitude of the turbulent intensity modulation, and
the phase fixes the absolute position of the pattern in the domain.

Near the threshold $R_{\rm t}$, introducing
$\epsilon=(R_{\rm t}-R)/R_{\rm t}$, $A_\pm $ are guessed to fulfil
Ginzburg--Landau--Langevin equations in the form~\cite{Petal}:
\BA
\tau_0\partial_{\tilde t} A_\pm&=&
(\epsilon+\xi_x^2\partial_{\tilde x\tilde x}^2
+\xi_z^2\partial_{\tilde z\tilde z}^2)
A_\pm\nonumber\\
&&\quad\mbox{}- g_1|A_\pm|^2A_\pm-g_2|A_\mp|^2A_\pm + \alpha\zeta_\pm
\label{E-rnd}
\EA
where the $\alpha\zeta_\pm$ are additive noise terms expressing the
local fluctuations caused by intense small scale turbulence, $\alpha$
being the strength of the physical noise.
Though this noise is both more intense than thermal fluctuations
(see \cite{conv} and references therein) and much more correlated
since the featureless turbulent state is not without structure
\cite{vls}, terms $\zeta_\pm$ are tacitly taken as
independent normalised delta-correlated Gaussian white noise processes
($\langle\zeta_\pm(t)\zeta_\pm(t')\rangle=\delta(t-t')$).

Periodic boundary conditions determine accessible wavelengths in a given
domain: $\lambda_{x,z}=L_{x,z}/n_{x,z}$, where the  integers $n_{x,z}$
are the wave numbers. In the computational
domains considered here, with $L_{x,z}$ not so large,
it turns out that states with $n_x=1$ or $2$ and $n_z=\pm1$ up to
$n_z=\pm3$ can be observed, depending on the precise value of $L_x$ and
$L_z$. When the wavenumbers are small enough, the partial differential
model (\ref{E-rnd}) can be reduced to a set of ordinary differential equations for
scalar complex amplitudes, and when there is no wavelength competition
but only an orientation competition playing with the $\pm$, just by two
amplitudes $A_{\pm,n_x,n_z}$ corresponding to a specific pair of
wavenumbers ($n_x,\pm n_z$). These amplitudes are then governed by:
\BA
\tau_0\partial_{\tilde t} A_{\pm,n_x,n_z}&=& \tilde\epsilon_{n_x,n_z}
A_{\pm,n_x,n_z}\nonumber\\
&&\mbox{}\hspace*{-8em}
-\left(g_1|A_{\pm,n_x,n_z}|^2+g_2|A_{\mp,n_x,n_z}|^2\right)
A_{\pm,n_x,n_z} + \alpha\zeta_\pm
\label{E-rnd1}
\EA
with $\tilde\epsilon_{n_x,n_z}=\epsilon-\xi_x^2\delta k_x^2-
\xi_z^2\delta k_z^2$, $\delta k_{x,z}=k_{x,z}-k_{x,z}^{\rm c}$, and
$k_{x,z}=2\pi n_{x,z}/L_{x,z}$, so that the dependence of
the pattern on the value of the wavevectors can be studied by changing
the size of the domain.

When a single wavelength and a single orientation are selected, a
single complex amplitude can
serve to characterise the corresponding pattern. This was precisely the
case considered by Barkley \& Tuckerman \cite{BT05-07} who defined the
order parameter from a single Fourier amplitude by sampling
its probability distribution function (PDF) and averaging over its
phase \cite{BTD}. So doing, they were able to detect the bifurcation
to the band regime from the change in the PDF as $R_{\rm t}$ was
crossed. In our simulations a single orientation is selected
only deep enough in the band regime, i.e. sufficiently below $R_{\rm t}$ but above $R_{\rm g}$. The pattern is then well installed and its
orientation remains fixed but its lateral position in the domain
can fluctuate, which strictly corresponds to the phase fluctuations
alluded to above. When this is the case, symmetry considerations
underlying (\ref{E-rnd1}) imply that the phase is dynamically neutral,
hence constant in a deterministic context, while it is expected to
evolve as a random walk in a noisy context \cite{BTD}.
Figure~\ref{fig7} shows that this is indeed the case.
\BF
\BC
\includegraphics[height=6cm]{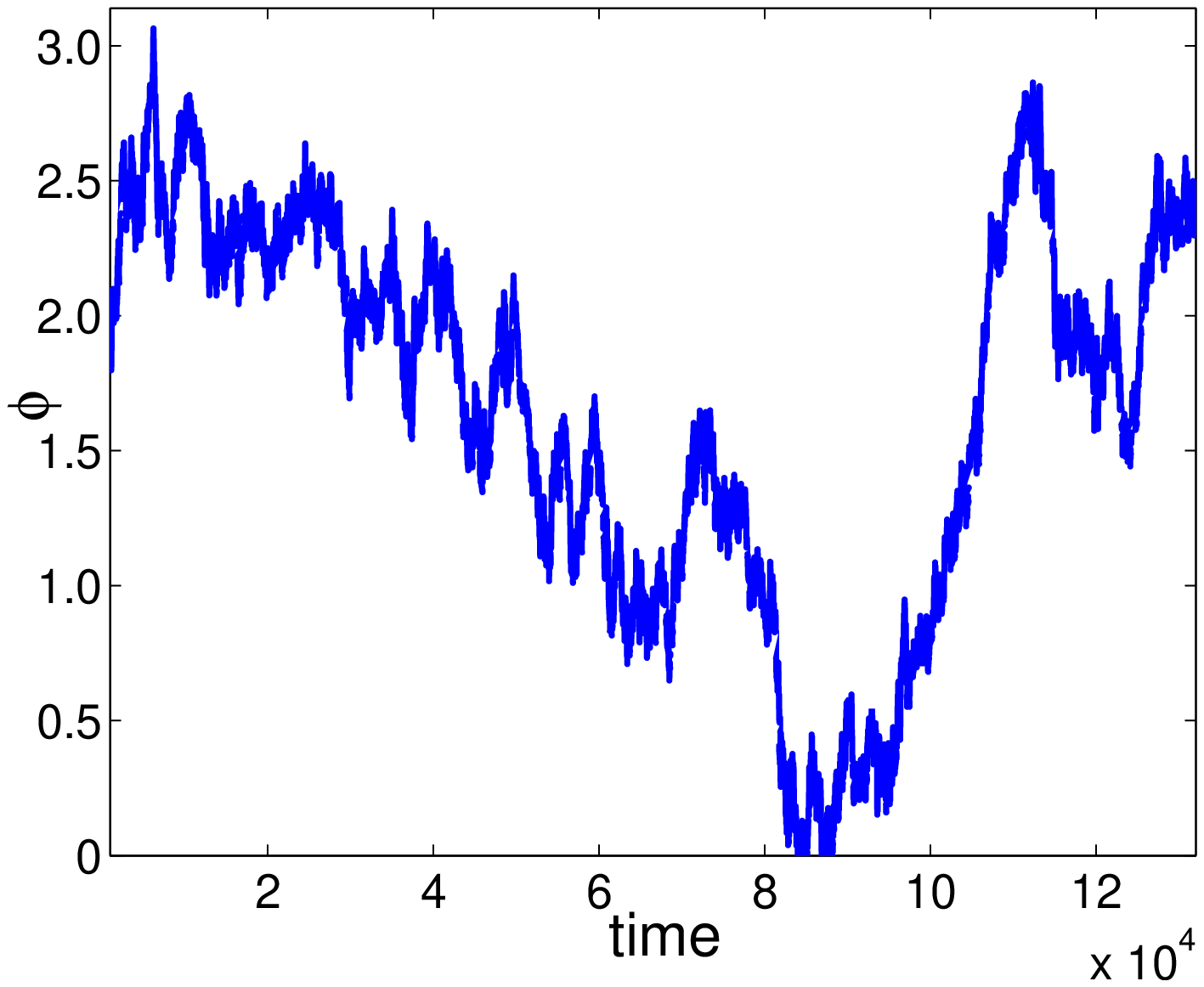}\hspace{0.1cm}
\includegraphics[height=6cm]{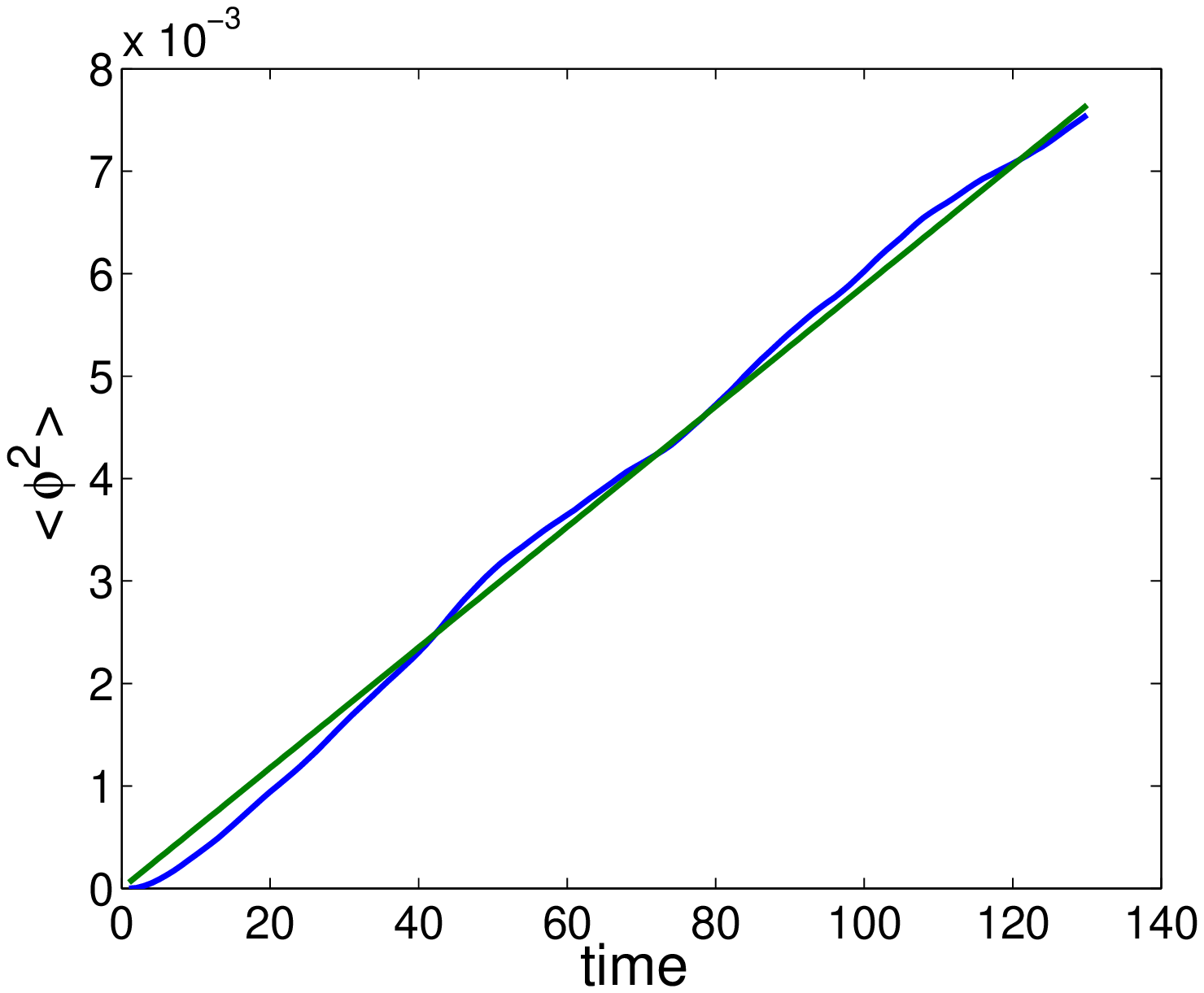}
\EC
\caption{Top: Variation with time of the phase of the main Fourier
component of $\hat u_x$ for $L_x=110$, $L_z=48$, $R=290$. Bottom:
growth of the variance of the ensemble-averaged
fluctuations as a function of time ($T_{\rm e}=130$, $N_{\rm e}=1000$), with linear fit.\label{fig7}}
\end{figure}
The top panel illustrates the variations of the phase of
the Fourier amplitude $\hat u_x$ of the streamwise velocity component
at $y=-y_{\rm m}$ for $R=290$ in a domain of size
$L_x\times L_z=110\times48$. The expected property is illustrated in
the bottom panel which displays the linear growth of the variance of
the phase fluctuations as a function of time after appropriate
ensemble averaging: From the initial time series we define an ensemble
of $N_{\rm e}$ successive sub-series of duration $T_{\rm e}$ as:
\BM
\phi_i(t)=\phi(t+(i-1)T_{\rm e})-
\phi\left((i-1)T_{\rm e}\right)\,,
\quad t\in\left[0,T_{\rm e}\right]
\EM
for $i=1\dots N_{\rm e}$. We next define the ensemble average:
\BM
\langle\phi\rangle(t)=\frac{1}{N_{\rm e}}
\sum_{i=1}^{N_{\rm e}}\phi_i(t)\,,\quad t\in\left[0,T_{\rm e}\right]\,.
\EM
which always remains of order $10^{-3}$, while the
variance:
\BM
\sigma_\phi^2(t)=
\frac{1}{N_{\rm e}}\sum_{i=1}^{N_{\rm e}}
\left(\phi_i(t)-\langle\phi\rangle(t)\right)^2\,,\quad
t\in\left[0,T_{\rm e}\right]
\EM
is indeed seen to grow linearly with time (Fig.~\ref{fig7}).

When the wavelength and/or the orientation can fluctuate, as is now the
case of interest, the practical definition of an order parameter is
less straightforward since a single complex amplitude is not enough.
Here, we forget about the information contained in the phase of the
relevant complex amplitudes and focus on their modulus. We then define
the {\it instantaneous\/} order parameter $m_{n_x,n_z}(t)$ as the
modulus at time $t$ of the fundamental Fourier mode $(n_x,n_z)$
accounting for the pattern as featured by the streamwise velocity field
$u_x$ averaged along the wall-normal direction:
\BE
m_{n_x,n_z}(t)=\left(\frac{1}{2}\int_{-1}^1
|\hat u_x(n_x,y,n_z,t)|^2\,{\rm d}y\right)^{1/2}
\label{E-op}
\EE
but equivalent results are obtained from the other velocity
components, with or without wall-normal averaging.

\subsubsection{Typical experiments and the order-parameter time-averaging issue}

Owing to the linear stability of the laminar flow, turbulence has to be triggered by finite amplitude disturbances. 
A typical experiment consists of creating a random initial condition
and evolving it at a Reynolds number for which uniform turbulence is expected, here $R=450$ ($\gg R_{\rm t}\simeq 345$ at the resolution chosen in the present work). That state is next used as an initial condition for a
simulation at the targeted value of $R$ for which the pattern of interest is expected, hence $R_{\rm g}<R<R_{\rm t}$. Such experiments were named {\it quench\/} in \cite{Betal98,BC98}. Variations of turbulent quantities
$e(t)$, $e_{\rm t}(t)$, $f(t)$ and of the order parameters $m(t)$
at the beginning of a typical experiment are shown
in Fig.~\ref{fig8}:
\BF
\BC
\includegraphics[height=0.4\textwidth,clip]{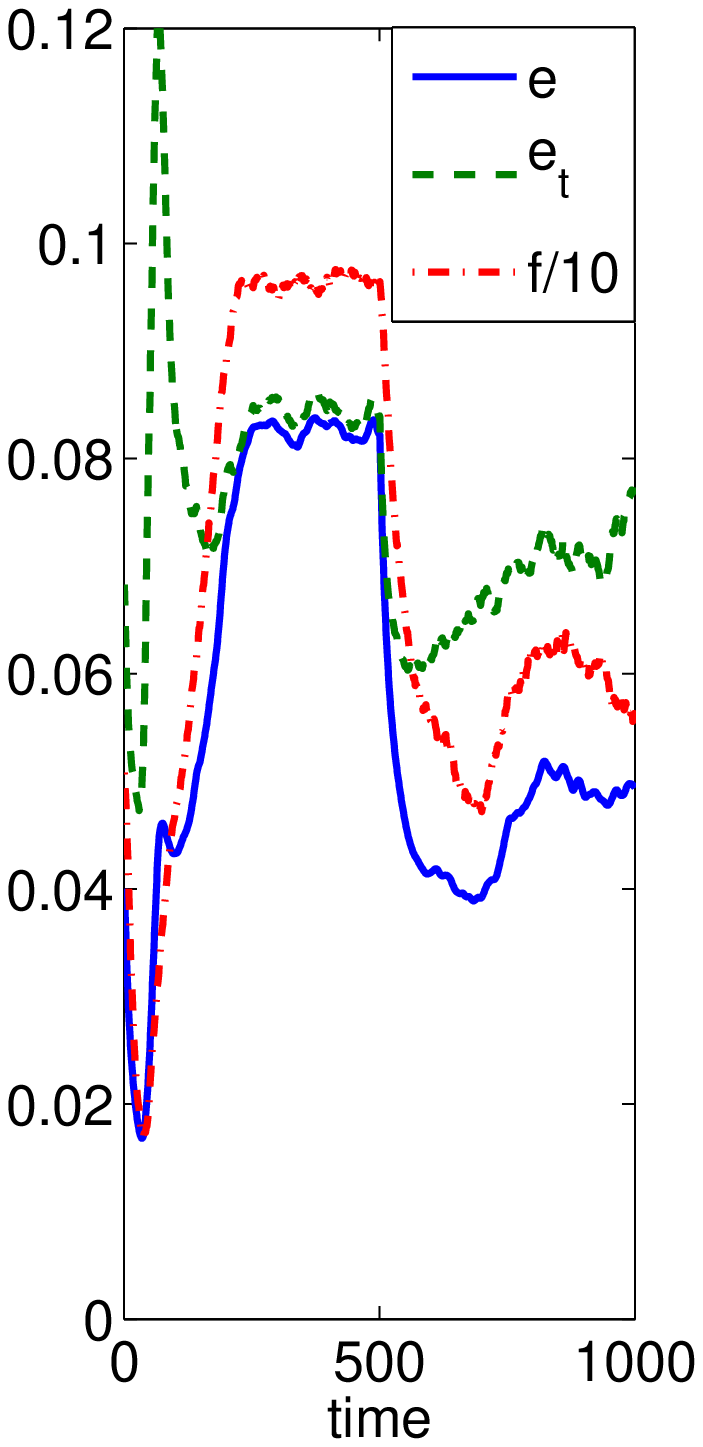}
\includegraphics[height=0.4\textwidth,clip]{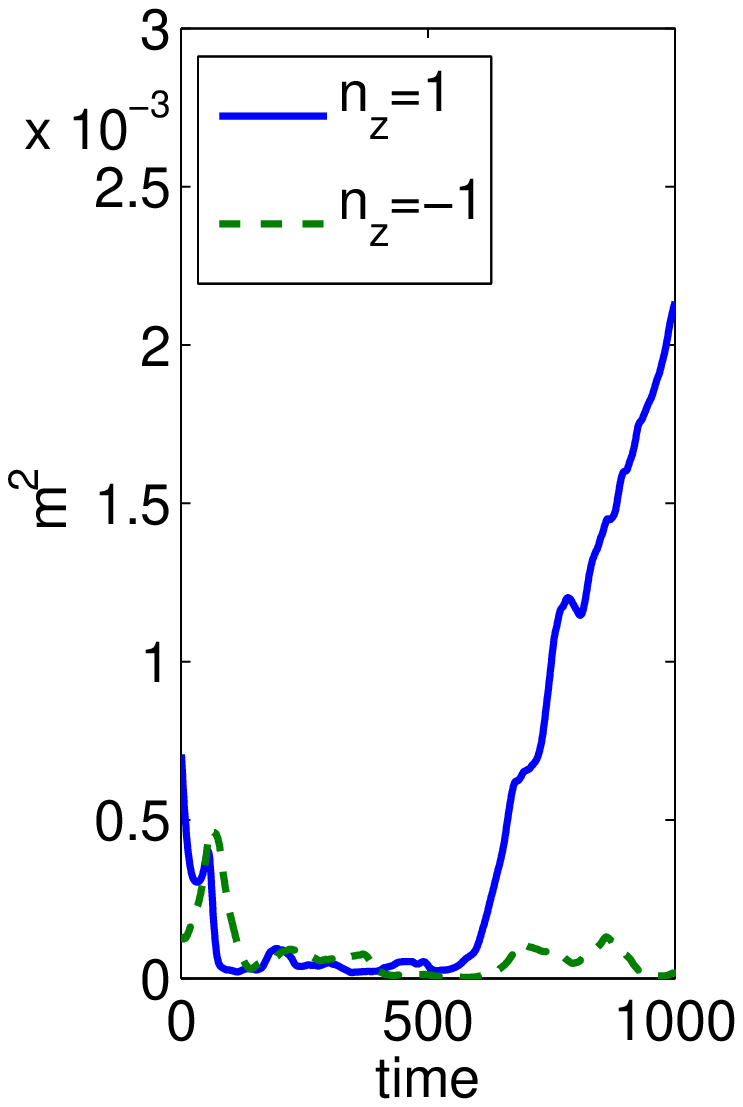}
\EC
\caption{Time series of turbulent quantities (left) and $m$ (right)
during the initial stage of a typical experiment. Here, for
$L_x\times L_z=110\times 48$, $R$ is initially set at 450 and
switched to 315 at $t=500$.\label{fig8}}
\EF
The stabilisation of the featureless regime
at $R=450$ is clearly visible with $e\simeq e_{\rm t}$, $f/10\simeq0.1$
(left), and $m\sim0$ (right). The subsequent quench at $t=500$, $R=315$
is seen to produce some undershoot of $e$, $e_{\rm t}$ and $f$, while $m$
grows slowly, which corresponds to the formation of bands.
After a short period of exponential growth, the order parameters saturate as shown in Fig.~\ref{fig9} for a series of 6 independent runs in the same conditions where a band is expected, pointing out the selection of the
orientation, with one of the order parameters larger than the other by
typically one to two orders of magnitude.
\BF
\BC
\includegraphics[width=0.7\textwidth,clip]{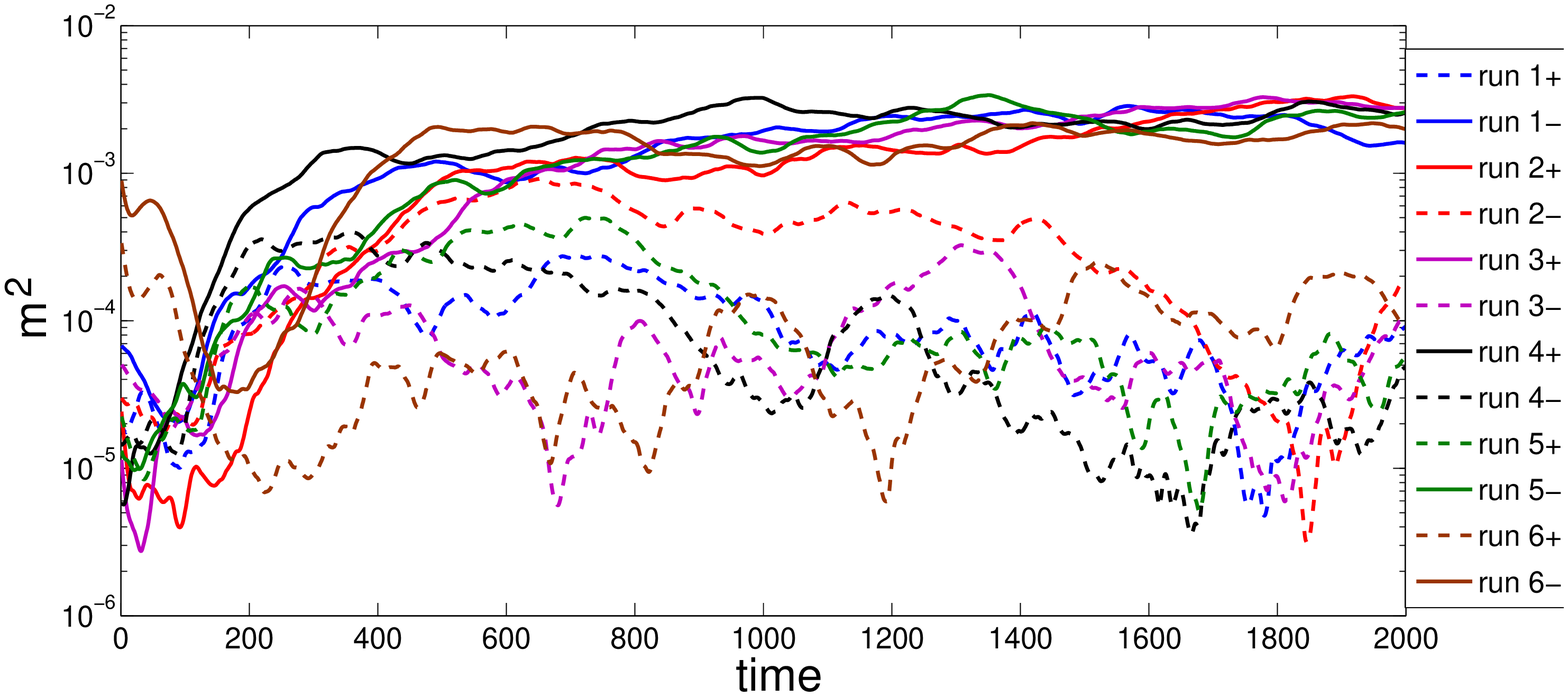}
\EC
\caption{Time series of $m^2$ for six different
runs at $R=315$, with $L_x\times L_z=128\times48$, starting from uniform
turbulence (time is reset upon quenching).  $+$ and $-$ refer to $n_z=+1$ and $n_z=-1$. Mode $+1$ is selected in runs 2, 3, 4, and mode $-1$ in
runs 1, 5, 6. (colour online) \label{fig9}}
\EF
The simulation is continued during at least $5000$ time units in order
to ensure good convergence of the time averages $E$, $E_t$, $F$ and $M$.
The same procedure is repeated for all the values of $L_x$, $L_z$,
and $R$ considered, except in \S\ref{Sb-R} where an adiabatic procedure
is adopted to vary $R$.

Like the turbulent quantities $e$, $e_{\rm t}$ and $f$,
order parameters $m_{n_x,n_z}$ fluctuate in time but, since
orientation changes are now of interest, care is required when
computing their averages.
Figure~\ref{fig10} displays a typical example of
\BF
\BC
\includegraphics[width=0.7\textwidth,clip]{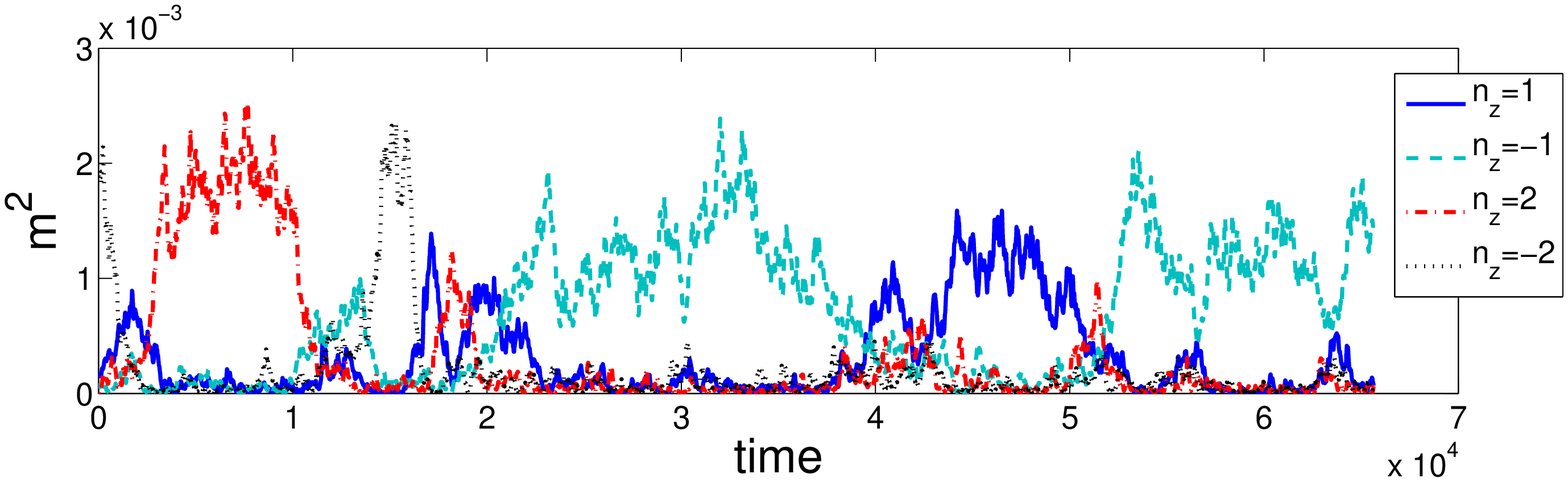}
\EC
\caption{Time series of $m^2$ for
$L_x\times L_z=128\times 84$ at $R=315$. Well-formed patterns with
$n_z=+1$ appear for $t\in[43,51]$, $n_z=-1$ for $t\in[22,37]$ and $t>52$,
$n_z=+2$ for $t\in[3,11]$, and $n_z=-2$ for $t\in[14,16]$. Defective
patterns are obtained for $t\in[18,22]$ or $[38,43]$. (All times to be multiplied by $10^3$.)
\label{fig10}}
\EF
long-lasting time series of $m^2$ for $L_x=128$, $L_z=84$, and $R=315$, which produces patterns with $n_x=1$ and $|n_z|=1$
and $2$, so that modes $n_z=\pm1$ and $\pm2$ dominate in turn.
As long as the instantaneous state of the system is close to ideal,
$m$ fluctuates around a specific mean value which depend only on
$|n_z|$ as expected from symmetry considerations.
Defects may appear and disappear, involving several
modes with similar amplitudes. For the data in Fig.~\ref{fig10},
$L_z=84$ lies in a range $L_z\in]80,96[$ where the competition
between different values of $|n_z|$ is particularly intense (see below).
When it is the case, a proper definition of order parameters implies conditional averaging over periods during which the pattern is
well formed with the chosen value of $|n_z|$. For example, in
Fig.~\ref{fig10}, $|n_z|=1$ is present during about 3/4 of the time
window and $|n_z|=2$
less than 1/4 of it. Since $m_{\pm2}>m_{\pm1}$ when the corresponding
modes dominate the pattern, one gets $M_2>M_1$, but it would be
meaningless to make a blend of the two and define a single
order parameter for the system. A detailed study of this special case
is deferred to \cite{crossref}.

However, outside cases of strong wavelength competition, a single
value of $|n_z|$ is selected, which makes things somewhat easier and
allows us to simplify the notation: $m_{n_x,\pm n_z}\mapsto m_\pm$.
An example is displayed in Fig.~\ref{fig11} for $L_z=32$ where
only $|n_z|=1$ shows up.
\BF
\BC
\includegraphics[width=0.7\textwidth,clip]{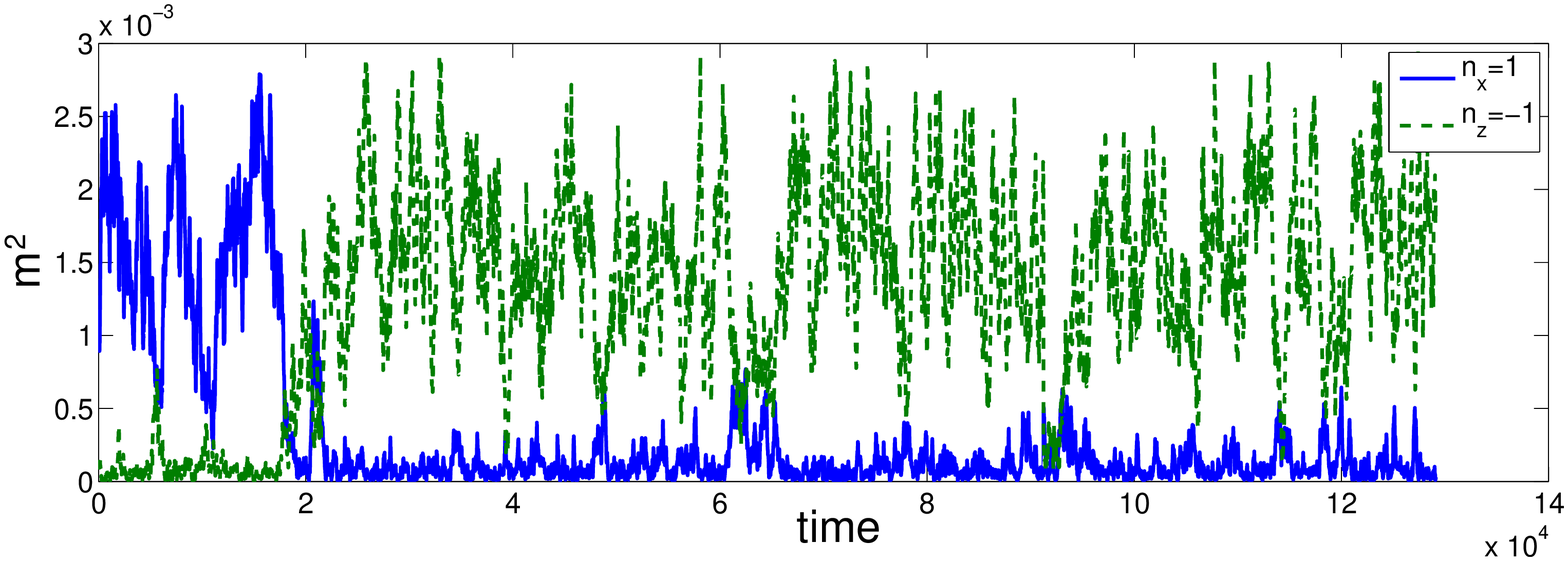}
\EC
\caption{Time serie of $m^2(t)$ for
$L_x\times L_z=110\times 32$, $R=330$}
\label{fig11}
\EF
Averaging can then be performed from two-dimensional probability
distribution functions (PDF) $\Pi_{\rm e}(m_+,m_-)$,
where subscript `e' means `empirical'.%
\footnote{In contrast with what was defined by Barkley {\it et al.}
\cite{BTD} who chose to scale out the pre-exponential factor, having ${\rm d}P(a)=a{\rm d}a\, \rho(a)$, where $a$ is the modulus of the dominant Fourier mode, corresponding to one of our $m_\pm$, we have
here ${\rm d}\Pi(m_+,m_-)={\rm d}m_+{\rm d}m_-\, \Pi(m_+,m_-)$, as a consequence: $\Pi(0,m_-)=\Pi(m_+,0)=0$.}
\BF
\centerline{
\includegraphics[width=0.28\textwidth,clip]{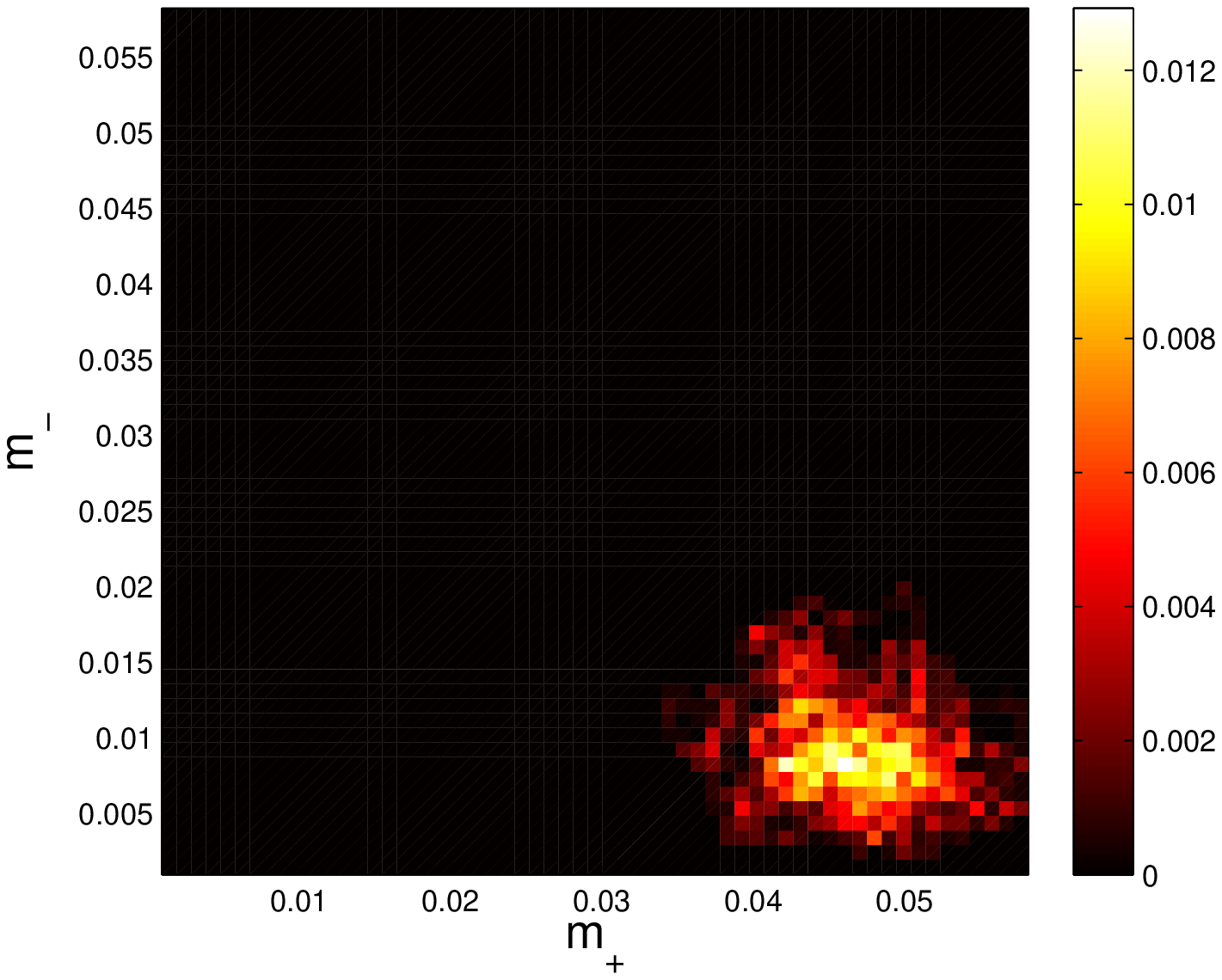}
\includegraphics[width=0.28\textwidth,clip]{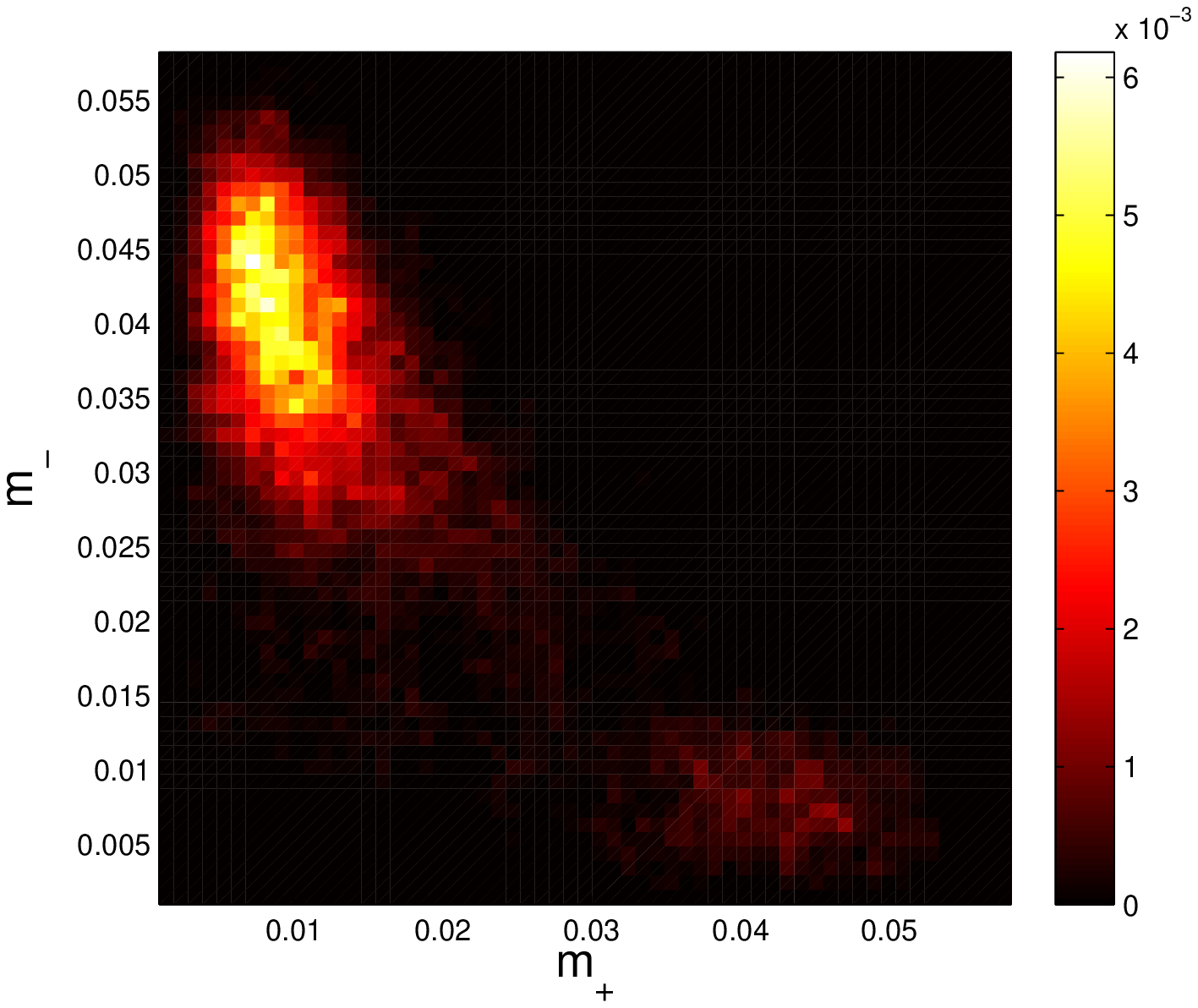}
\includegraphics[width=0.28\textwidth,clip]{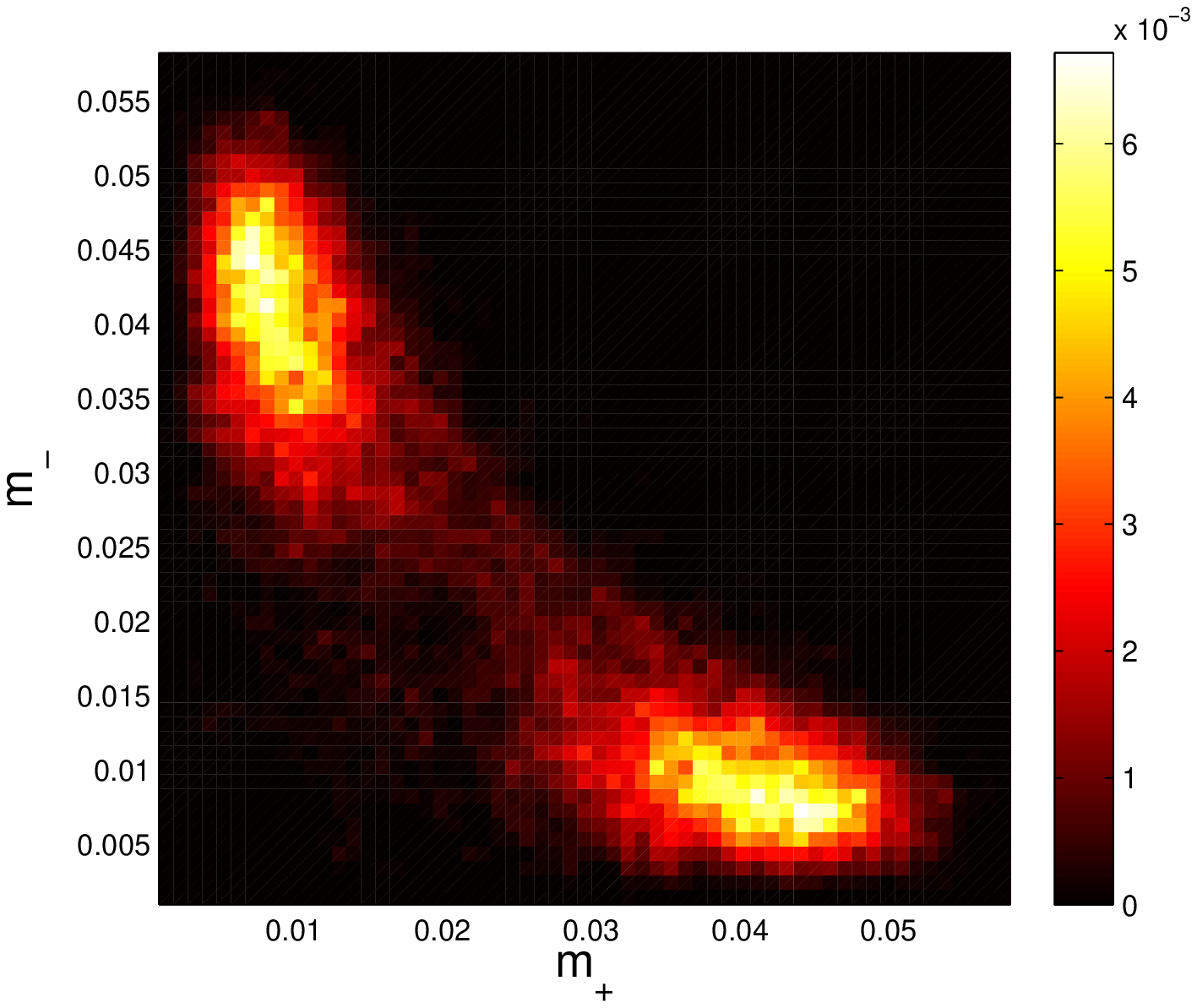}
\includegraphics[width=0.28\textwidth,clip]{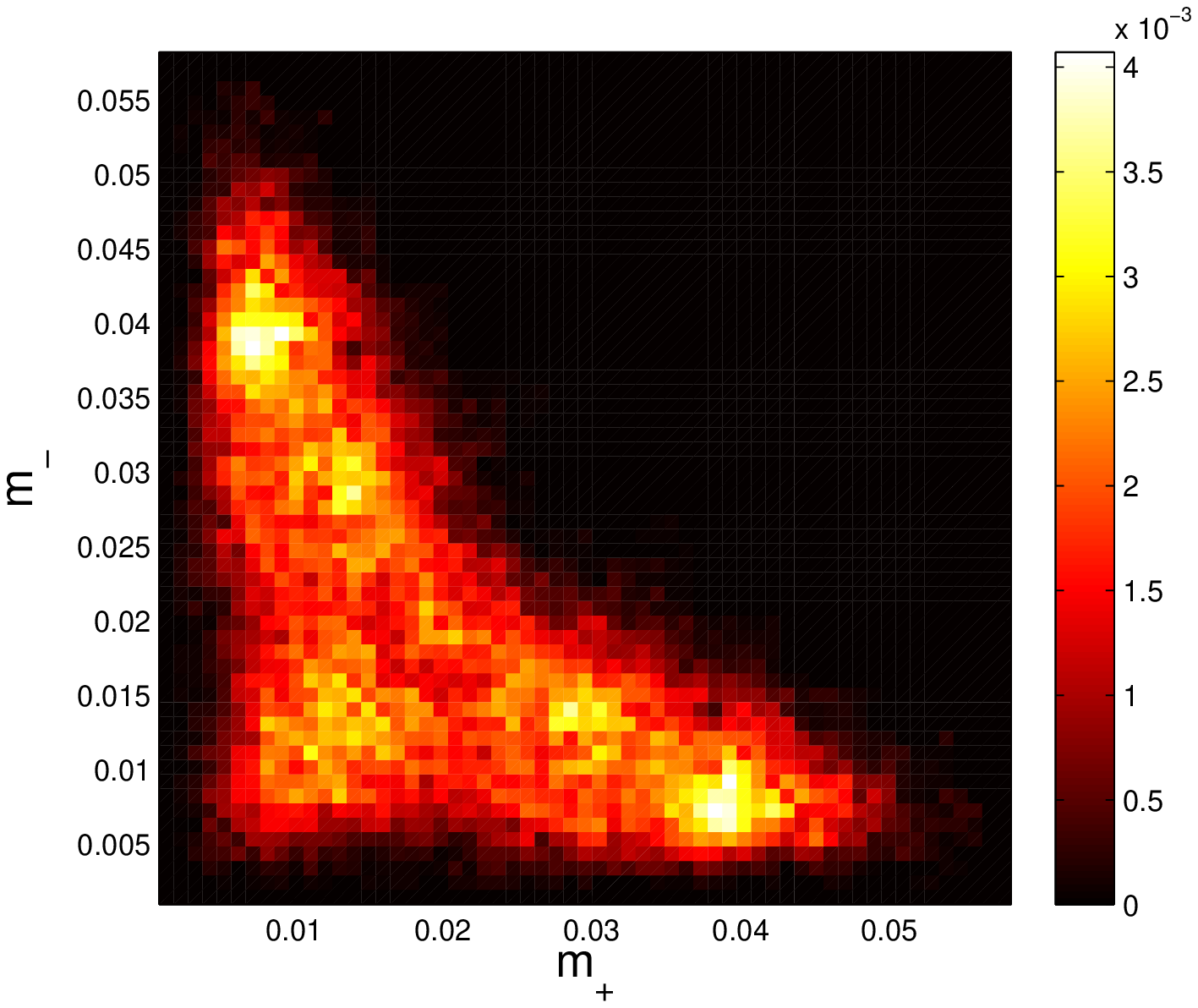}}
\caption{PDF of $m$ for $L_x\times L_z=110\times 32$,
$R=290$ (left), $R=330$ (centre-left),  $R=330$,
symmetrised (centre-right), $R=337$, symmetrised (right).}
\label{fig12}
\EF
Far away from $R_{\rm t}$ the orientation does not fluctuates and the
pattern remains without defects, which yields a one-hump PDF such as the
one in  Fig.~\ref{fig12} (left) for $R=290$, but closer to $R_{\rm t}$
the orientation fluctuates and defects are present. Two humps are then
obtained as in Fig.~\ref{fig12} (centre-left) which derives from
the time series for $R=330$ shown in Fig.~\ref{fig11}. Due to the finite length of
the time series, the PDF is not symmetrical with respect to the diagonal
but since, for symmetry reasons, the two orientations should be present
with the same weight, one may improve the statistics by constructing
$\Pi_{\rm e,s}(m_+,m_-)=\frac12\left(\Pi_{\rm e}(m_+,m_-)+
\Pi_{\rm e}(m_-,m_+)\right)$, where the additional
subscript `s' means `symmetrised', 
which is done in  Fig.~\ref{fig12} (centre-right).
Averages can then be extracted from the `symmetrised' PDF, which works
fine as long as the orientation fluctuates but neither the wave numbers
$n_x$ nor $|n_z|$. We thus define:
\BM
M=2\int_{m''<m'} m' \Pi_{\rm e,s}(m',m'')\, {\rm d}m'{\rm d}m''\,.
\EM
The right panel in Fig.~12 displays the (symmetrised) PDF for $R=337$, when
re-entrant featureless turbulence intermittently bursts in, which
manifests itself as a secondary hump close to the origin, see below
\S\ref{Sb-R}, and especially the discussion related to Fig.~\ref{fig17}. 

\section{Results\label{S2}}

\subsection{Theoretical expectations\label{Sb-conceptual}}

The statistically steady states (permanent regimes)
obtained in the DNS and characterised by the time-averaged empirical
order parameters $m_{n_x,n_z}$ defined through (\ref{E-op})
can then be compared to the equilibrium states predicted by model
(\ref{E-rnd1}), the deterministic part of which can be written as
deriving from the potential:
\BA
{\cal V} &=& -\mbox{$\frac12$}\tilde\epsilon\left(|A_+|^2 + |A_-|^2\right)
+\mbox{$\frac14$} g_1 \left(|A_+|^4 +|A_-|^4\right)\nonumber\\
&&\mbox{}
+\mbox{$\frac12$} g_2  |A_+|^2  |A_-|^2\,,
\label{E-pot}
\EA
where $\tilde\epsilon$ is a short hand notation for
$\tilde\epsilon_{n_x,n_z}$, computed from the values of $n_x$ and
$\pm n_z$ relevant to the pattern of interest,
again with a single pair of modes present in the system.

Assuming Gaussian noises of strength $\alpha$, the theoretical
expression of the PDF reads \cite{vk83}:
$$
\Pi_{\rm t}(m_+,m_-)= Z^{-1}\, m_+ m_- \exp(-2{\cal V}/\alpha^2)\,,
$$
where subscript `t' means `theoretic' and
$$
Z=\int_0^\infty\int_0^\infty
 m_+ m_- \exp(-2{\cal V}/\alpha^2)\,{\rm d}m_+ {\rm d} m_-
$$
is a normalisation factor called the {\it partition function\/} in
statistical physics. For values of $\tilde\epsilon$ that are not too small, the most probable
values  $m_\pm^0$ corresponding to the maxima of $\Pi_{\rm t}$ give a good
estimate of expected mean values $\langle m_\pm\rangle$ (mean-field
approximation). They are given by the solutions to:
\BM
0=-\tilde\epsilon m_\pm^2+g_1 m_\pm^4
+g_2m_\pm^2 m_\mp^2
-\alpha^2/2\,.
\EM
At lowest non-trivial order in $\alpha^2$, we have:
\BE
m^0_\pm\propto\frac{\alpha}{|\tilde\epsilon|}\,,
\label{E-A00}
\EE
which corresponds to the trivial solution of the deterministic problem,
just shifted by the effects of noise. The non-trivial
solutions read:
\BE
m^0_+=m^0_-=\sqrt{\tilde\epsilon/(g_1+g_2)}
\label{E-sq}
\EE
and
\BE
m^0_\pm=\sqrt{\tilde\epsilon/g_1}\,,\quad
m^0_\mp =\frac{\alpha}{\sqrt{2\tilde\epsilon(g_2-g_1)/g_1} }
\label{E-pm}
\EE

For $\tilde\epsilon<0$, solution (\ref{E-A00}) is stable and the other
solutions do not exist. For $\tilde\epsilon>0$, solution  (\ref{E-A00}) is
unstable and the symmetric solution (\ref{E-sq}) is a saddle point since,
in order to get a stripe pattern we assume $g_2>g_1$; otherwise a stable
rhombic pattern would be obtained but is observed neither in the experiments
nor the numerical simulations.
This solution lies on the boundary of the attraction
basins of solutions (\ref{E-pm}), which exist for $\tilde\epsilon>0$ and
are stable. They represent the amplitude of the turbulence modulation for
$R<R_{\rm t}$. The mean amplitude of the installed mode varies as 
$\sqrt{\tilde\epsilon}$ as is typical of a supercritical bifurcation.
The other mode, expected to be zero in the deterministic case, is present
with small amplitude due to noise. Noise is also responsible for a switch from the `$\pm$' situation to the `$\mp$' one when fluctuations make the
system leave the well corresponding to an installed `$+$' mode to reach
the other one where the `$-$' mode is installed and vice versa, going
through the potential barrier at the saddle solution (\ref{E-sq}). The asymptotic expressions above agree with the values computed from the PDFs
obtained by direct simulations of model (\ref{E-rnd1}) and
will be plotted together with our results in Figures~\ref{fig18}
and~\ref{fig19}. When $\tilde\epsilon$ is very small, fluctuations around
the most probable values have to be taken into account. The mean field
approximation is no longer valid and a behaviour in the form
$\langle m_\pm\rangle\propto|\tilde\epsilon|^\beta$ is expected, where
$\beta$ is the {\it critical exponent\/} describing the variations of
the order parameter with the control parameter in the theory of phase
transitions.
We shall restrict to the mean-field approximation as a first
guess since the nature and extent of this specific regime, called
{\it critical\/} in statistical physics, are not yet clearly
characterised in the present case (see \cite{conv} and references therein for examples where fluctuations have thermal origin).
\BF
\centerline{
\includegraphics[height=0.28\textwidth,clip]{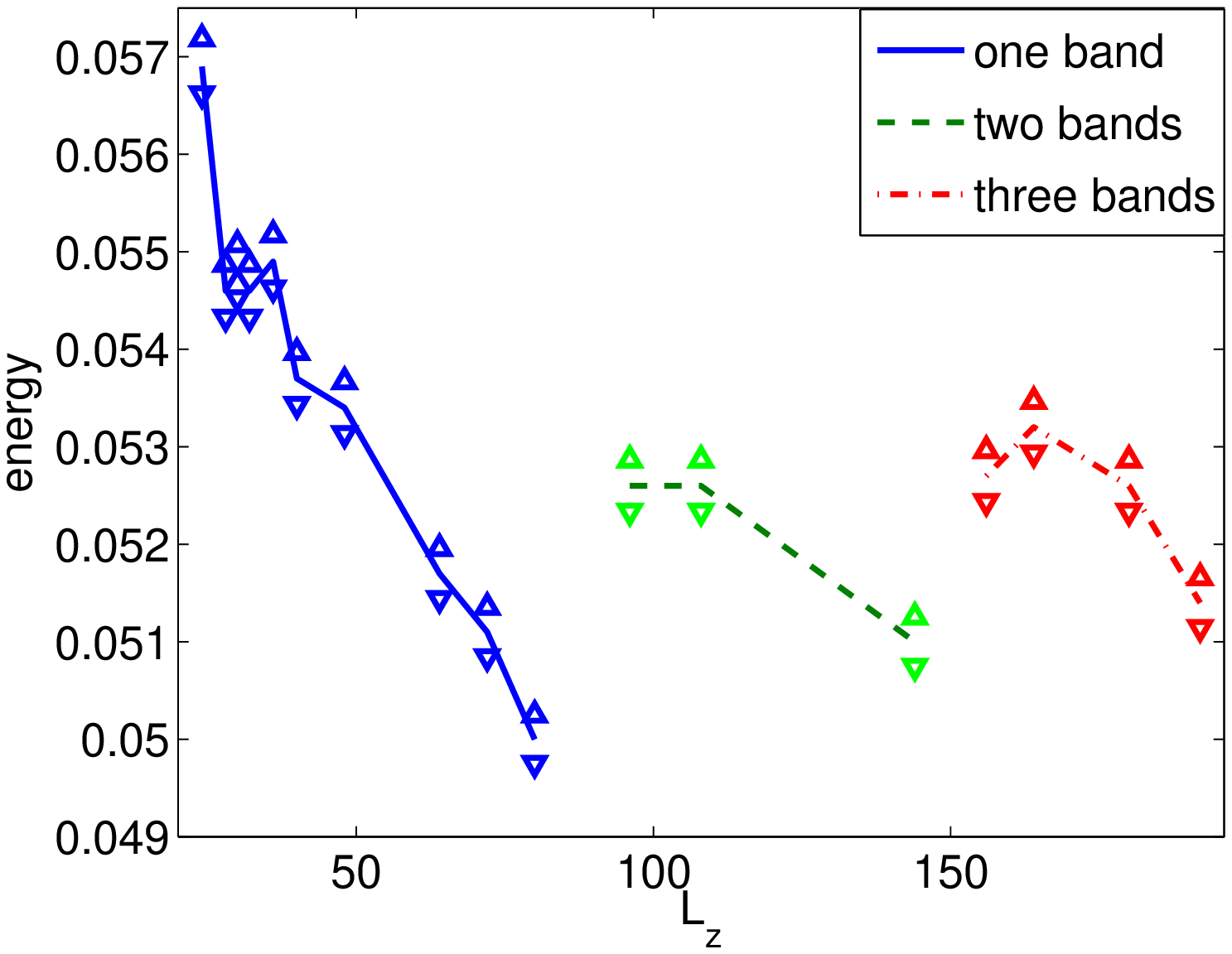}
\includegraphics[height=0.28\textwidth,clip]{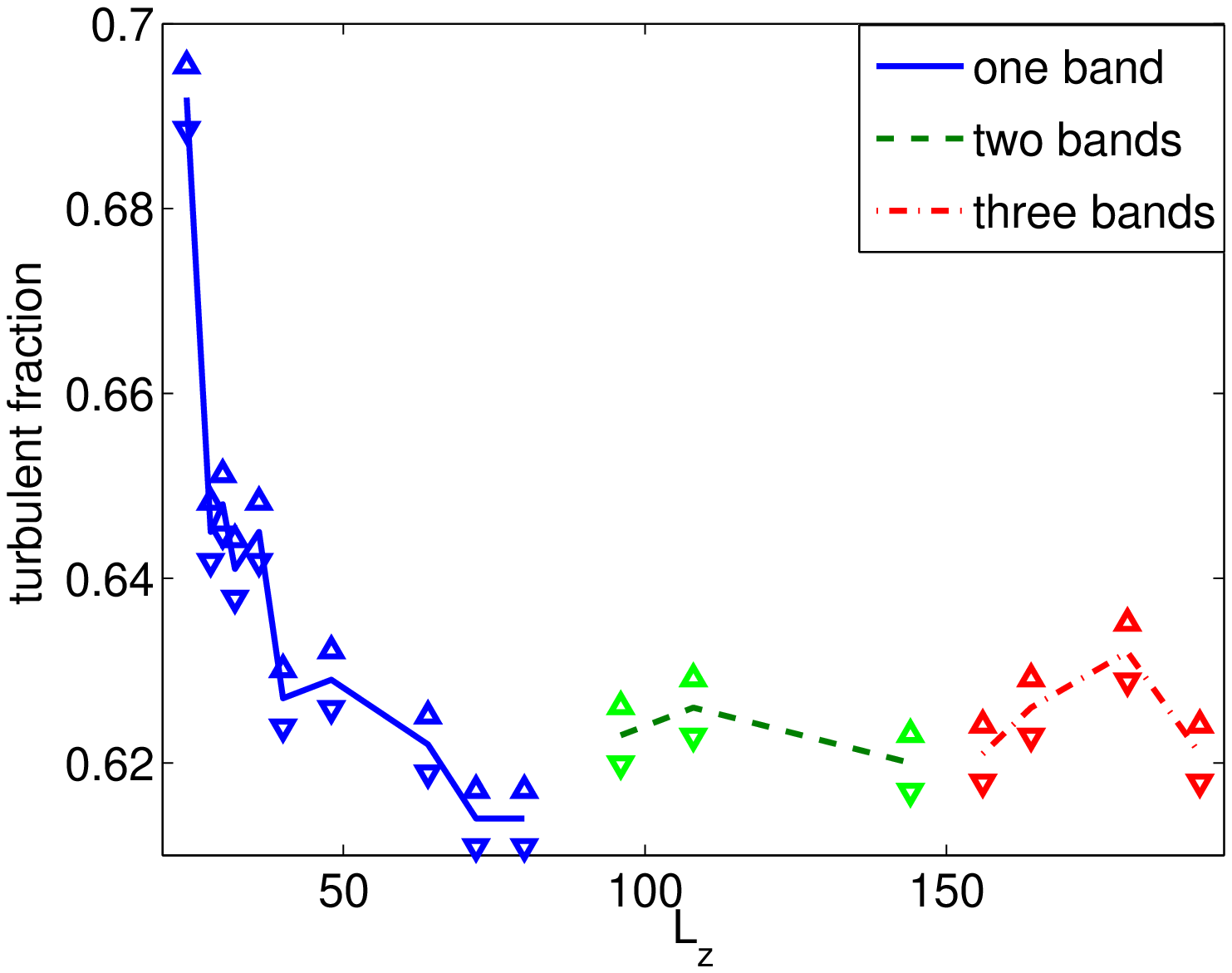}
\includegraphics[height=0.28\textwidth,clip]{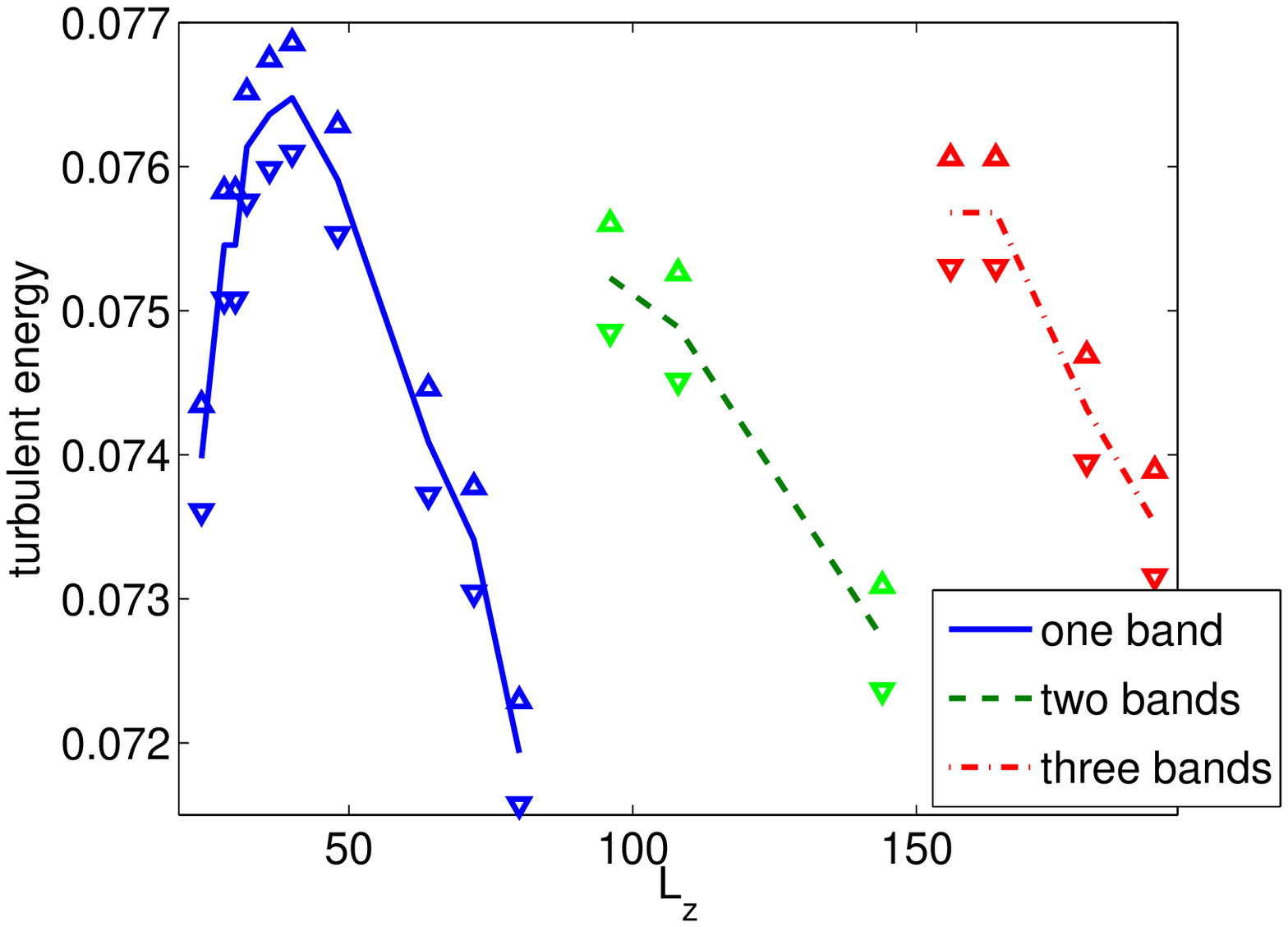}
}
\caption{Perturbation energy $E$ (left), turbulent fraction $F$ (center)
and turbulent energy $E_t$ (right) as functions of $L_z$ for $L_x=128$
and $R=315$.\label{fig13}}
\EF

The deterministic part of model (\ref{E-rnd1}) is invariant against phase
changes of the complex amplitudes $A_\pm=m_\pm\exp(i\phi_\pm)$, implying
that the $\phi_\pm$ are dynamically neutral. They are indeed
governed by:
\BE
\tau_0\partial_{\tilde{t}} \phi_\pm={\cal I}m \left[\exp(-i\phi_\pm) \alpha\zeta_\pm\right]/m_\pm\,,
\label{E-phi}\EE
i.e. a stochastic process, the strength of which depends on
the instantaneous value of $m_\pm$. In fact, the right hand side of
(\ref{E-phi}) is another random Gaussian process
$\tilde\zeta(t)\alpha/m(t)$ with zero mean and variance
$\alpha^2\delta(t-t')/(\langle m\rangle^2+\sigma_m^2)$ where $\sigma_m^2$
is the variance of $m(t)$, which can be checked numerically using model (\ref{E-rnd1}). Results in Fig.~\ref{fig7} above can be quantitatively
rendered by taking $\alpha/\tau_0=4\times 10^{-4}$.

Coherence lengths $\xi_x$ and $\xi_z$ in (\ref{E-rnd1}) control how
strictly the wavevectors $k_x$ and $k_z$ are bound to their optimal
values $k_x^{\rm c}$ and $k_z^{\rm c}$. The anisotropy of the base
flow leads to expect different values for $\xi_x$ and $\xi_z$.
For PCF, experimental data \cite{Petal} suggests that $\lambda_x$ and
therefore $k_x$ do not depend on the Reynolds number, whereas $\lambda_z$
decreases with $R$. Prigent {\it et al.} also report a decrease of the
effective value of $\xi_z$ as $R$ is increased but the experiment did
not give access to $\xi_x$. In the following, we determine most of
coefficients in model (\ref{E-rnd1}) from the dependence of
$E$, $E_{\rm t}$, $F$ and $M$ on $k_z$, $k_x$, by varying $L_x$,
$L_z$ and $R$ using the quench protocol explained above. The
dependence upon the Reynolds number analysed next is obtained from simulations in which $R$ is varied adiabatically.

\subsection{Dependence on $k_z$\label{Sb-kz}}
\BF
\BC
\includegraphics[width=0.42\textwidth,clip]{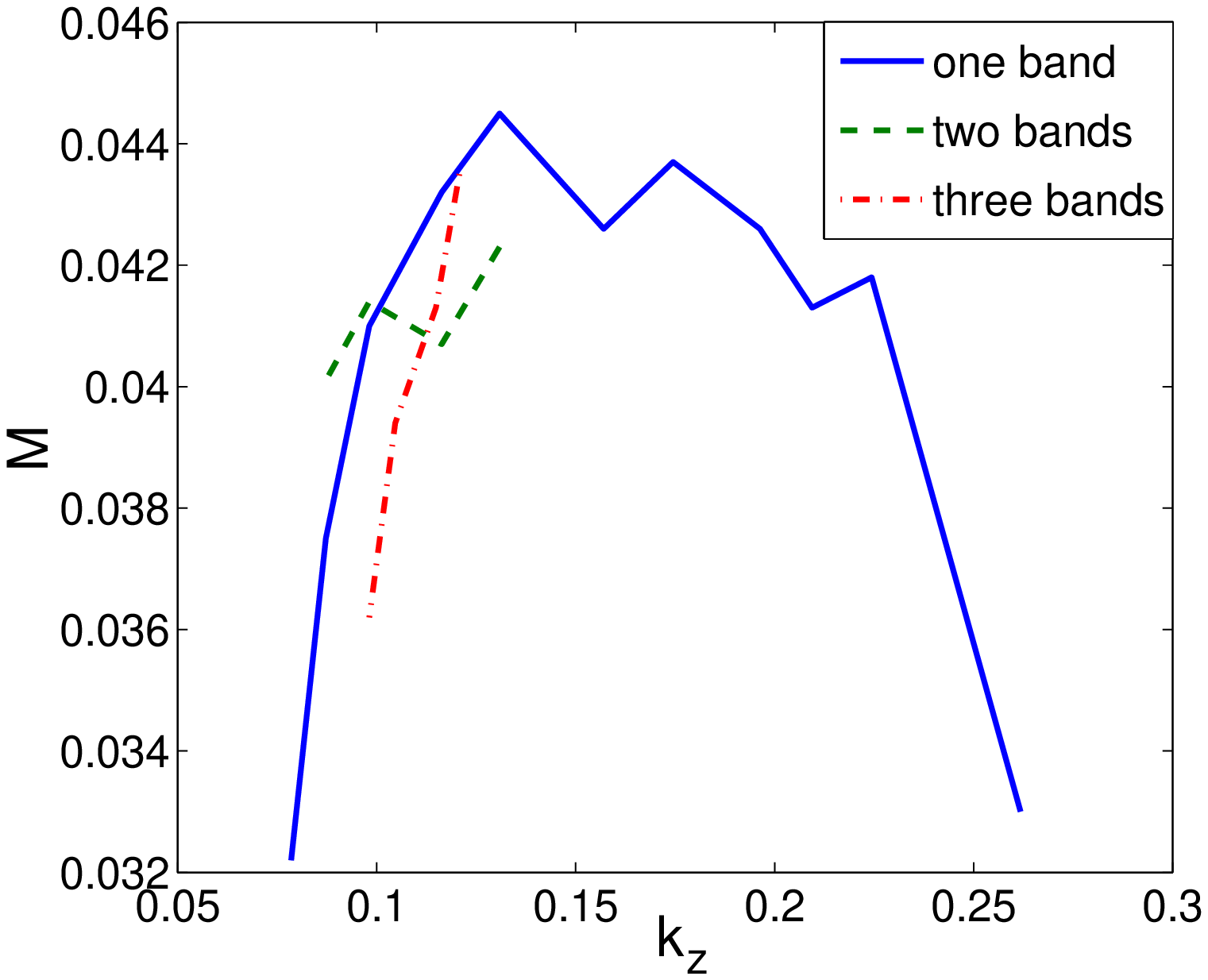}\hspace{0.1cm}
\includegraphics[width=0.42\textwidth,clip]{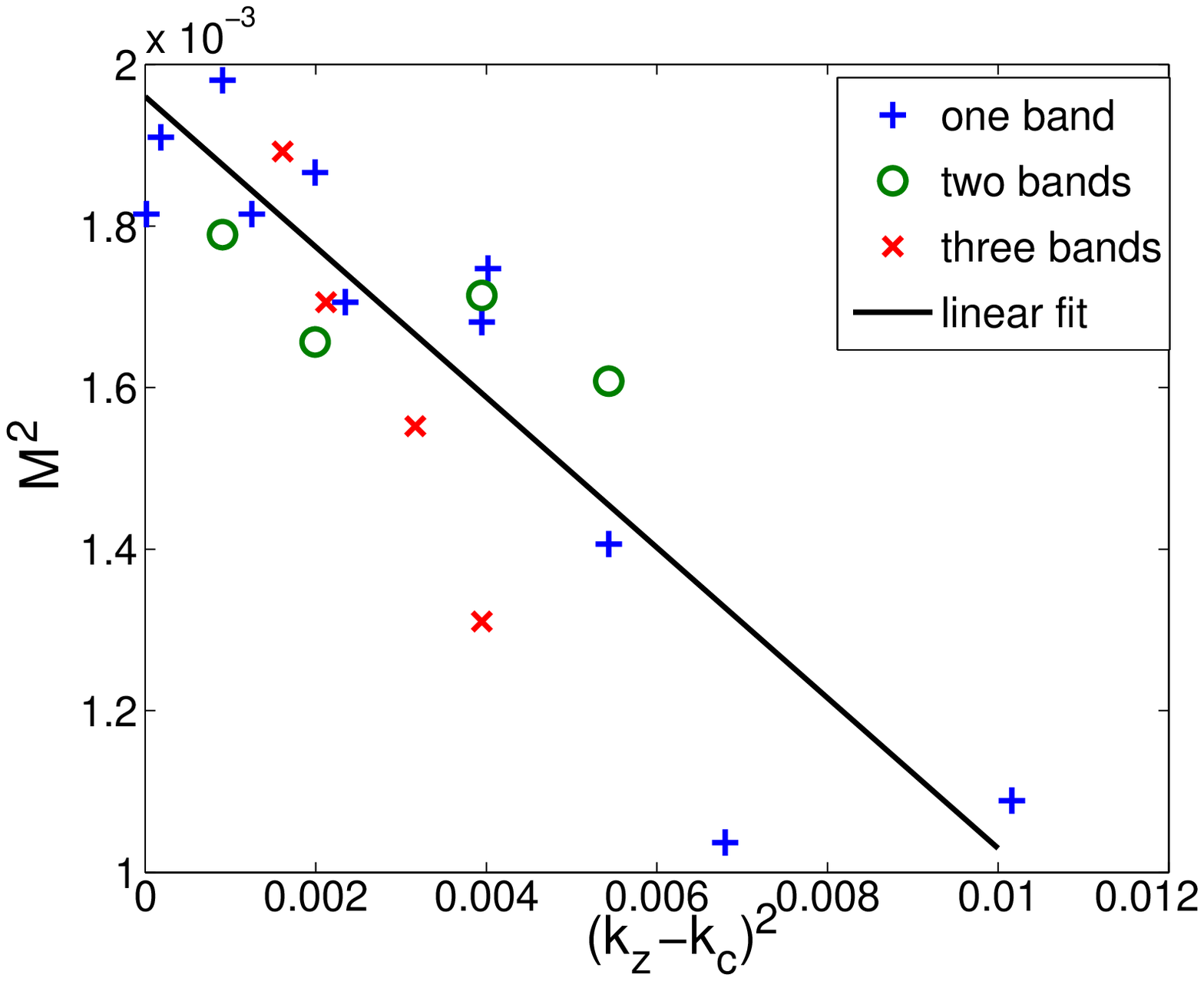}
\EC
\caption{Order parameter squared $M$ as a function of $k_z$ (left) and
of $(k_z-k_z^{\rm c})^2$ (right) for $L_x=128$ and $R=315$.\label{fig14}}
\EF

We fix $R=315$, in the middle of the range where bands are expected at the resolution that we consider~\cite{MRxx}, and $L_x=128$ so that a single
streamwise period is obtained ($n_x=1$, $k_x=2\pi/L_x$). We take
values of $L_z$ ranging from 24 to 192. Taking the number $|n_z|$ of spanwise periods into account,
we have $k_z=2\pi|n_z|/L_z$. Figure~\ref{fig13} displays $E$, $E_t$,
and $F$
as functions of $L_z$, showing that $|n_z|$ increases with $L_z$: one band
for $24\le L_z \le 80$, two bands for $96\le L_z \le 144$ and three bands
for $156 \le L_z \le 192$. In these ranges, $|n_z|$ stays fixed during
the simulation. In contrast, patterns with $|n_z|=1$ and $|n_z|=2$
alternate in time for $80 < L_z < 96$, here for $L_z=84$ (see figure~\ref{fig10}) and $L_z=90$.
This special case is studied more thoroughly in \cite{crossref}. A similar
competition between $|n_z|=2$ and $|n_z|=3$ is expected to occur for
$144\lesssim L_z\lesssim150$.
Taken together, the results in Fig.~\ref{fig13} illustrate confinement
effects when $L_z$ is small. Turbulence is featureless for $L_z<24$
and the turbulent fraction $F$ (central panel) rapidly decreases
from~1 down to $\simeq0.63$ which therefore represents some kind of
optimum at $R=315$.

The results also suggest to check cases with $n_z>1$ against case $n_z=1$.
Figure~\ref{fig14} (top) displays $M^2$ as a function of $k_z=2\pi n_z/L_z$ 
and $n_z=1$ as a full line. Data obtained with two and three bands are also
shown as dashed and dash-dotted lines, respectively. For them no points at
large wavevectors are obtained because the corresponding patterns are not
stable enough to be observed. The parabolic shape expected from the theory
(\S\ref{Sb-conceptual}) is reasonably well reproduced by the data. The
maximum is reached at $k_z^{\rm c}\simeq0.16$, that is
$\lambda_z^{\rm c}=2\pi/k_z^{\rm c}\simeq39$, as determined from a fit
against a parabola. The so-obtained value of $k_z^{\rm c}$ can next be
used to determine $\xi_z^2$ from the slope of a linear fit of $M^2$
against  $(k_z-k_z^{\rm c})^2$. The result is displayed in
Figure~\ref{fig14} (bottom) where data corresponding to one band
are shown with
`$+$' signs. From it one derives $\xi_z^2/g_1=0.1$. In turn, the constant
term in the fit is a compound accounting for the dependence on $R$ and
$k_x$, namely $(\epsilon-\xi_x^2\delta k_x^2)/g_1=0.002$. Data for two
and three bands, respectively shown with `$\circ$' and `$\times$' symbols,
are seen to be consistent with these estimates. Here a single value of
$R$ has been considered. In the CCF case, Prigent {\it et al.}
found for $\xi_z^2/g_1$ values of the same order of magnitude,
decreasing with $R$ from $0.5$ to $0.1$ \cite[c]{Petal}. 

\subsection{Dependence on $k_x$\label{Sb-kx}}

The dependence of the pattern's characteristics on $k_x$ is studied
for $R=315$, $L_z=48$, and $L_x\in [80,170]$. In this range, only
$n_x=1$ is obtained, except for $L_x=170$ where $n_x=2$ can also be
observed. Figure~\ref{fig15} shows that, as a function
of $k_x$ (top), $M^2$ displays a maximum at $k_x^{\rm c}=0.058$, hence
$\lambda_x^{\rm c}=110$, whereas
fitting $M^2$ against $(k_x-k_x^{\rm c})^2$ (bottom)
yields $\xi_x^2/g_1=2.7$. The same study at $R=330$ (closer
to $R_{\rm t}=345$) gives $\lambda_x^{\rm c}=110$ and $\xi_x^2/g_1=3.9$,
while at  $R=290$ (closer to $R_{\rm g}=275$) we get
$\lambda_x^{\rm c}=125$ and $\xi_x^2/g_1=2.2$, which is a rough estimate
since the lack of symmetry in the exchange
$\delta k_x\leftrightarrow-\delta k_x$ visible
in the top panel of Fig.~\ref{fig15} has not been taken into account.

The variation
of $\lambda_x^{\rm c}$ with $R$ that we obtain here is not observed
in the the plane Couette flow experiments but remains compatible
with the trend seen in CCF case \cite[a]{Petal}. Rather
than to the role of rotation or curvature, this observation points to the
role of streamwise periodic boundary conditions enforced by the cylindrical
geometry or the numerical implementation.
\BF
\BC
\includegraphics[width=0.42\textwidth,clip]{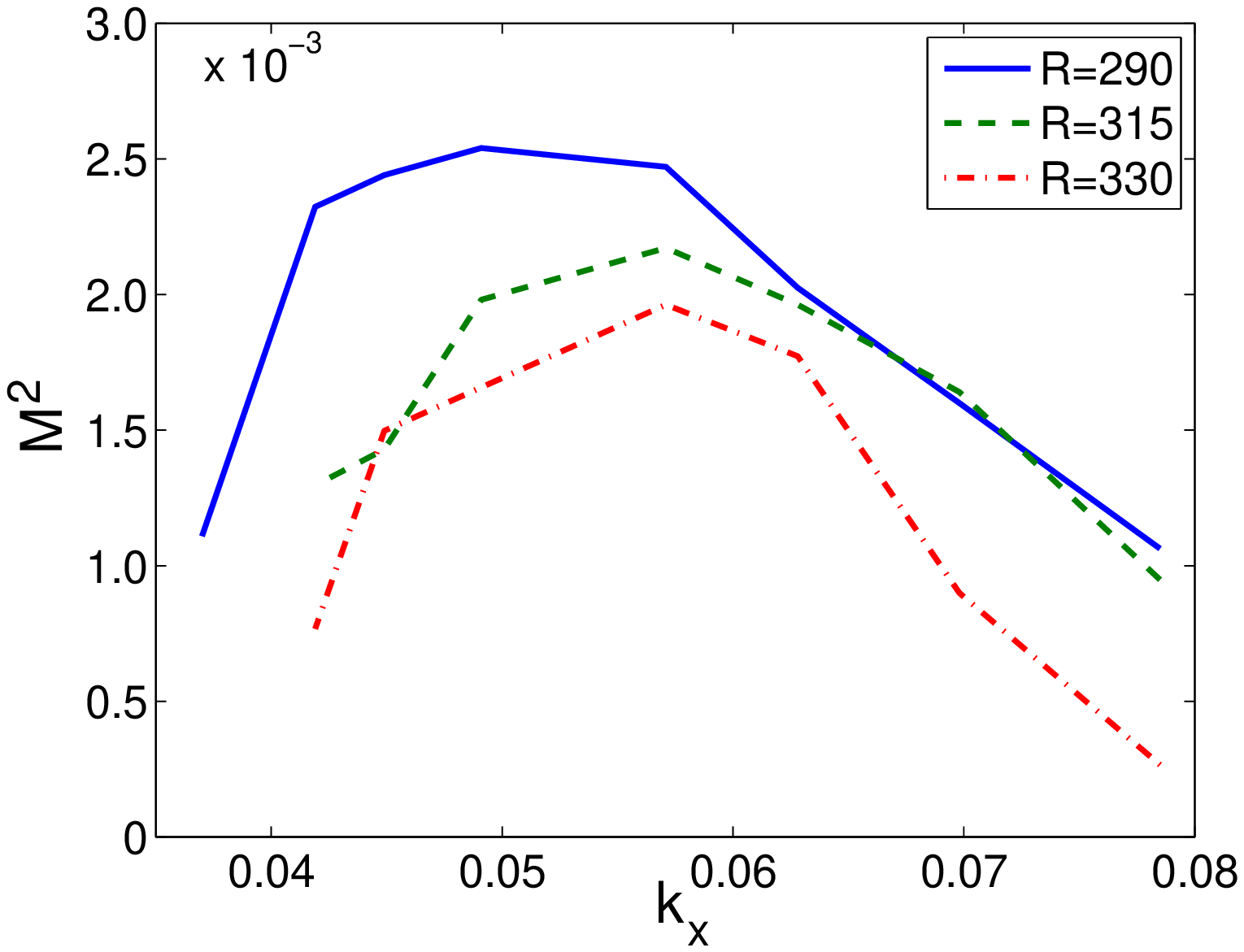}\hspace{0.1cm}
\includegraphics[width=0.42\textwidth,clip]{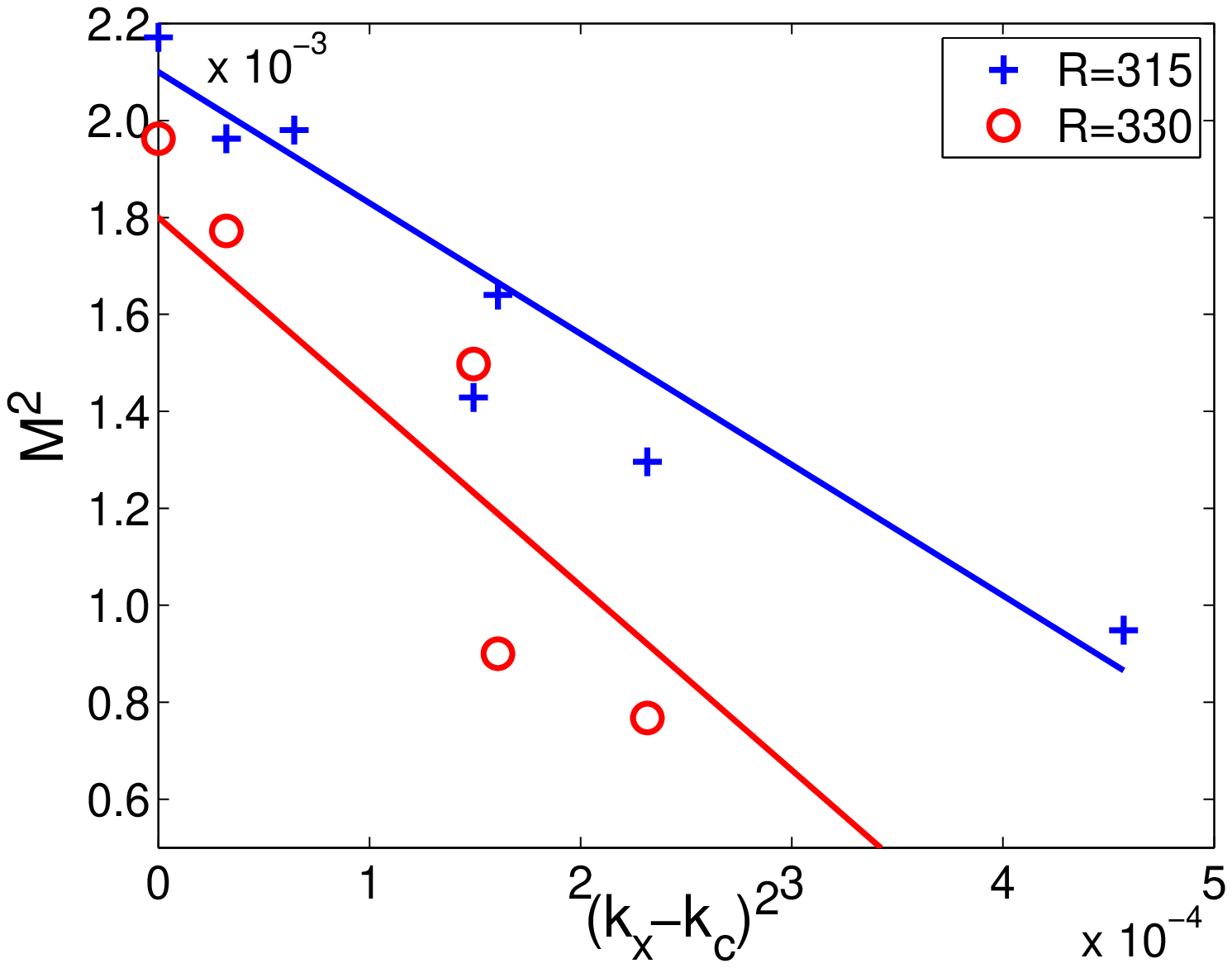}
\EC
\caption{Order parameter squared $M^2$ as a function of $k_x$ (left)
and of $(k_x-k_x^{\rm c})^2$ (right) for $L_z=48$ and
$R=315$.\label{fig15}}
\EF

\subsection{Dependence on $R$\label{Sb-R}}

Variations of $E$, $F$, $E_{\rm t}$, and $M$ against $R$ are
studied using a different protocol. Two sizes are considered:
$L_x\times L_z=128 \times 64$ and $110\times 32$. From the study in
previous sections, both domains are expected to fit one elementary
band $\lambda_x\times\lambda_z$. The pattern should feel ``at ease'' in
the first domain and more ``spanwise-confined'' in the second one.
A first simulation at $R_0=315$ serves to prepare initial conditions for
simulations at higher and lower Reynolds numbers by increasing or
decreasing $R$ by steps $\Delta R=5$. The flow is integrated over $5000$
time units at each value
of $R$ and the so-obtained state is used as an initial condition for
the next value of $R$ in the range $[260,350]$.
Additional values $R=333$, $336$, $337$, and
$R=370$ and $390$ outside the interval are also considered. At given
$R$ statistical results involve time integration over at least
$1.5\times 10^4$ time units.

The main effect of increasing $R$ seems to be an expansion of the
turbulent part of the band pattern as illustrated in Fig.~\ref{fig16}.
When $R$ is close enough to $R_{\rm t}$ the orientation of the pattern
fluctuates: destroying a well-established ideal pattern, turbulence
invades the laminar band, stays featureless for a while, before another
pattern grows, which may or may not have the same orientation.
\BF
\centerline{
\includegraphics[width=0.33\textwidth,clip]{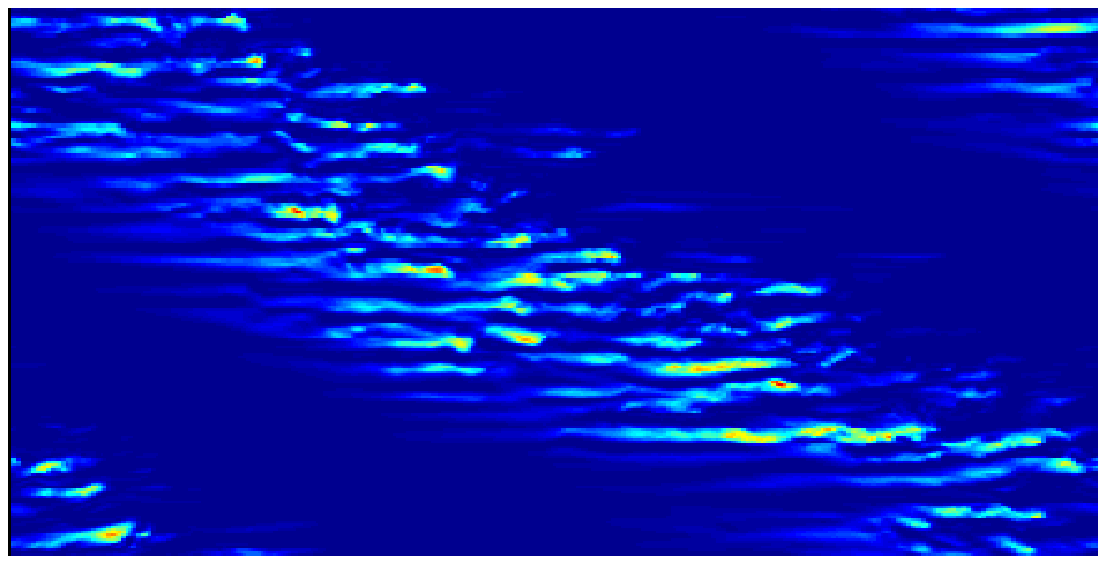}\hspace{0.15cm}
\includegraphics[width=0.33\textwidth,clip]{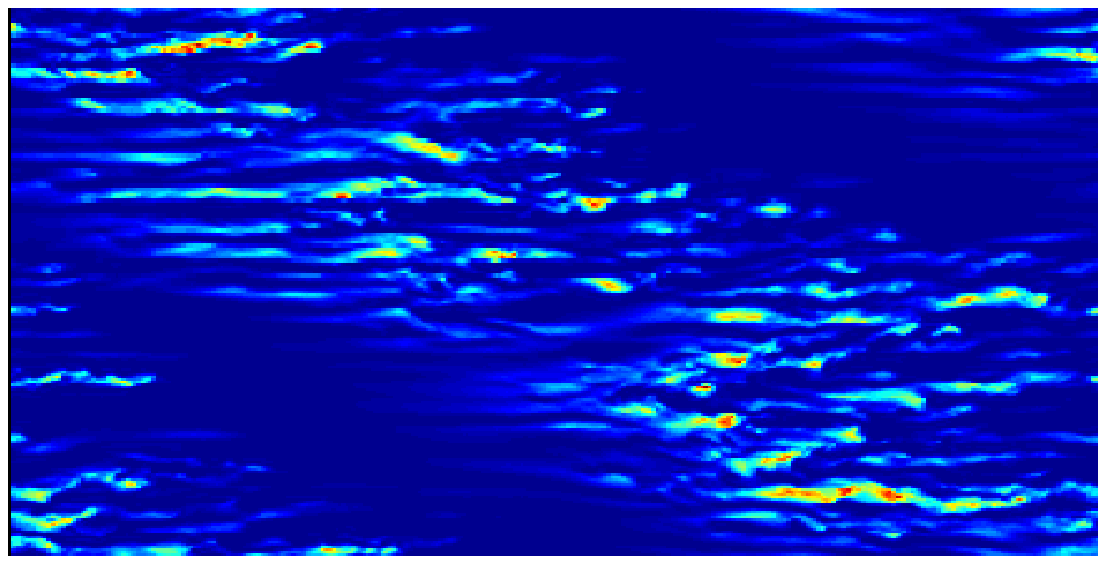}\hspace{0.15cm}
\includegraphics[width=0.33\textwidth,clip]{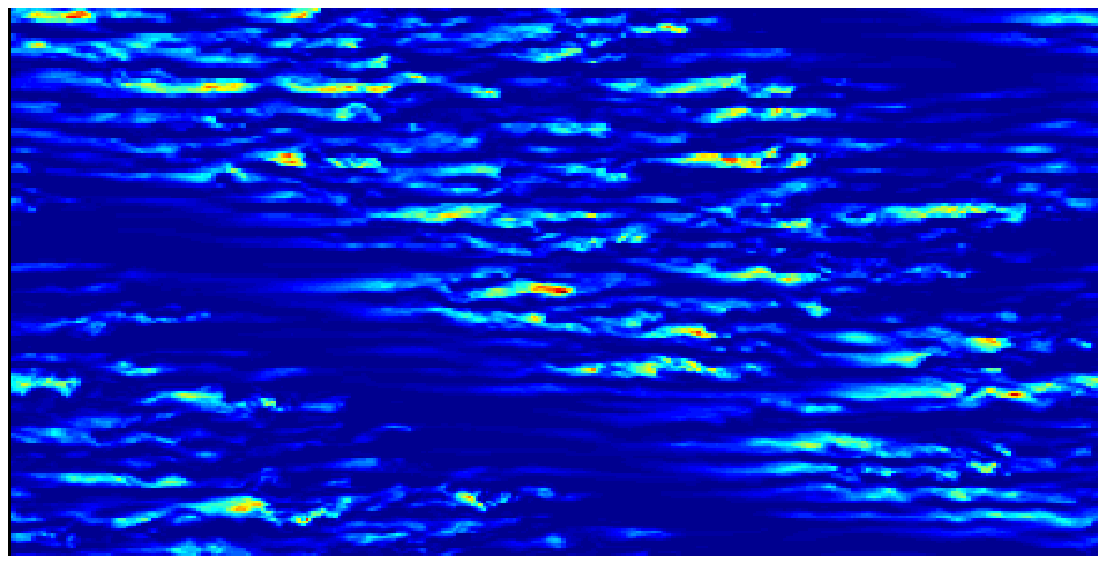}
}
\caption{Colour plot of ${\bf u}^2$, from left to right: $R=280$,
$R=300$ and $R=330$. $L_x\times L_z=128\times 64$. \label{fig16}}
\EF
Figure~\ref{fig17} displays a featureless turbulent episode for
$L_x\times L_z=110\times 32$ and $R=335$, during which $m(t)$ stays
close to $0$ (left panel),
the turbulent fraction approaches one, indicating the decrease of the
size of the laminar domain and commanding the variation of the total
energy (central panel), while the intensity of turbulence inside the turbulent domain does not changes (right panel). Such events cannot
be mistaken with the transient occurrence of a defect in the pattern
since both $m_{+1}$ and $m_{-1}$ remain simultaneously close to zero
for a relatively long period of time, which is characteristic of the featureless state.
\BF
\BC
\includegraphics[width=0.32\textwidth,clip]{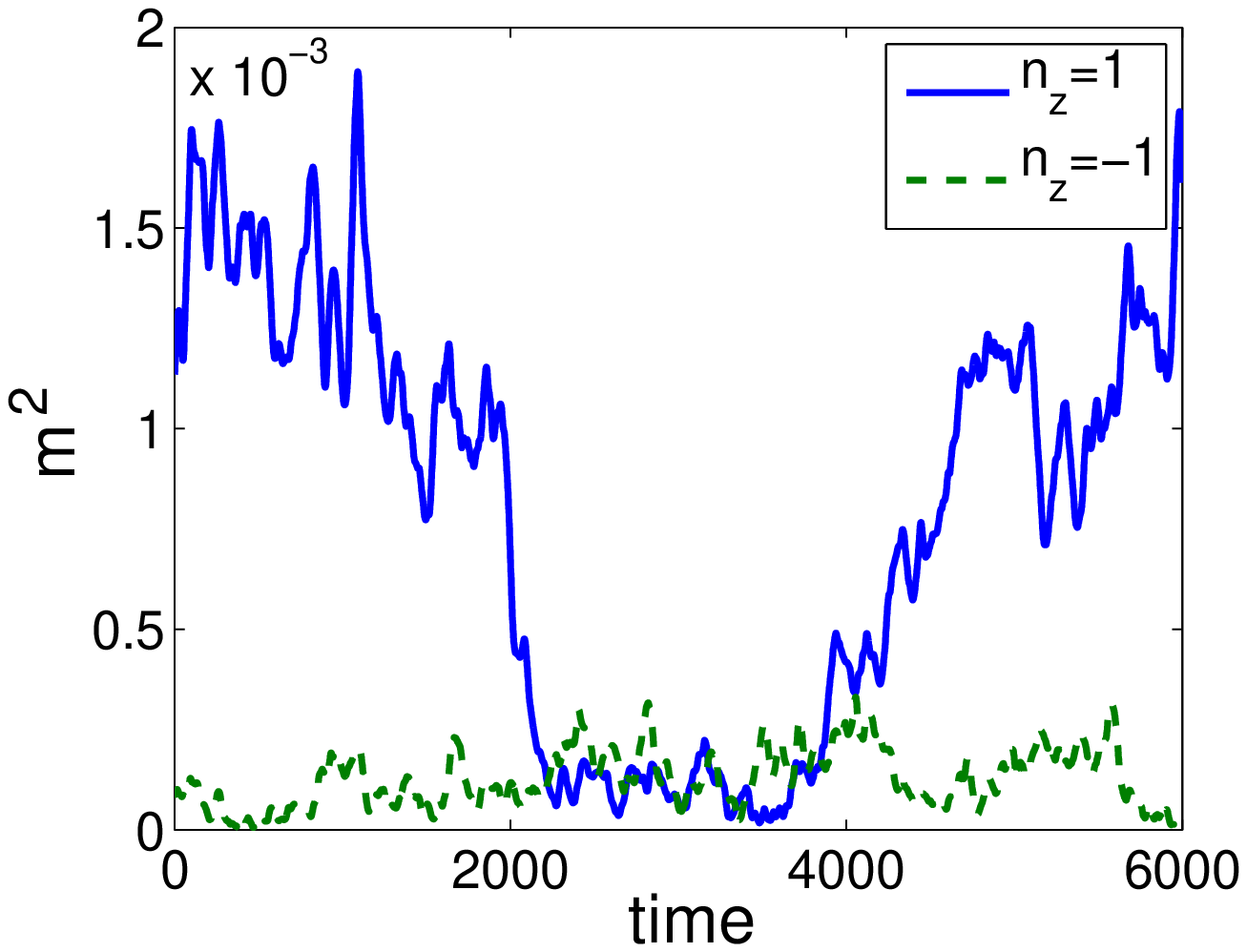}\hfill
\includegraphics[width=0.32\textwidth,clip]{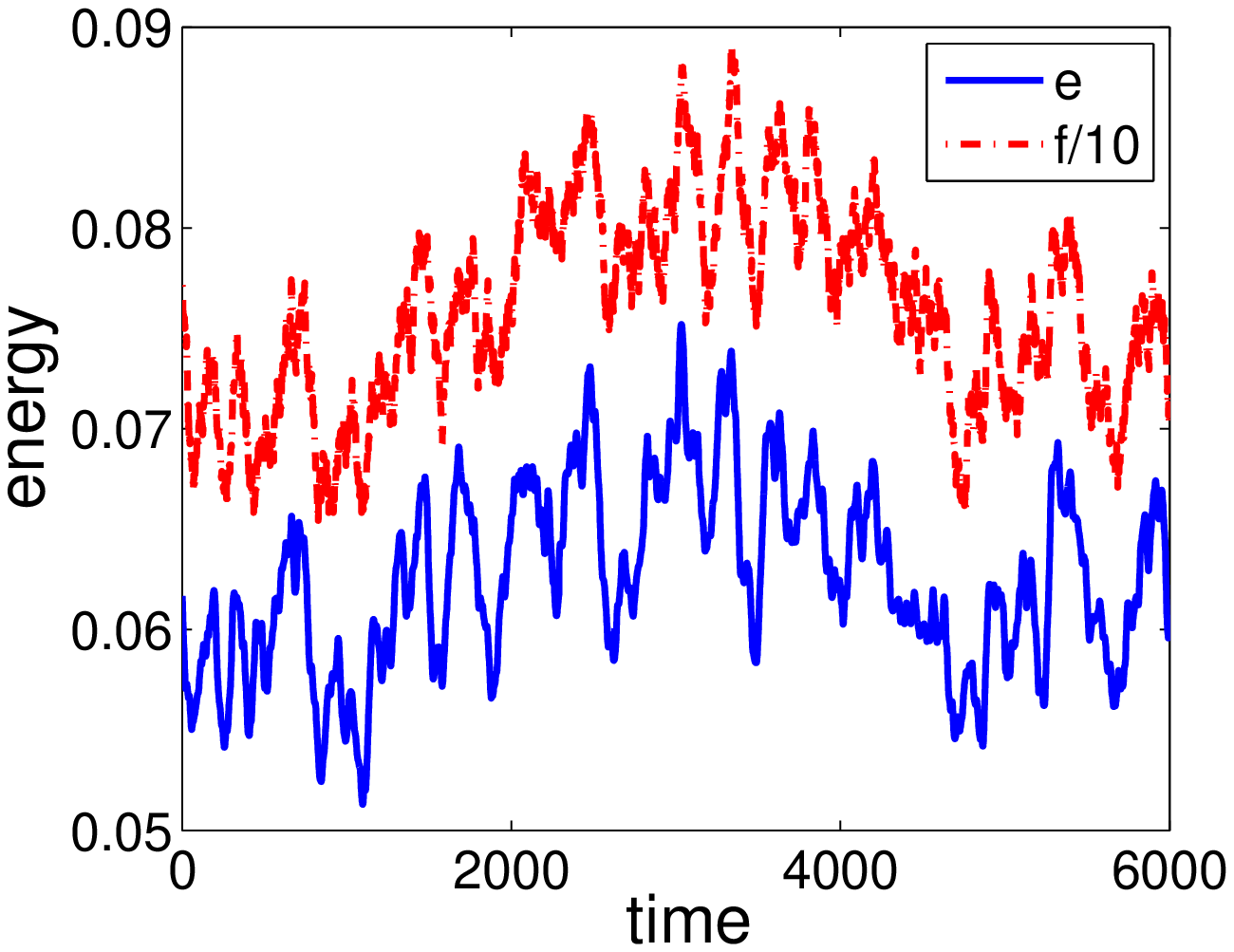}\hfill
\includegraphics[width=0.32\textwidth,clip]{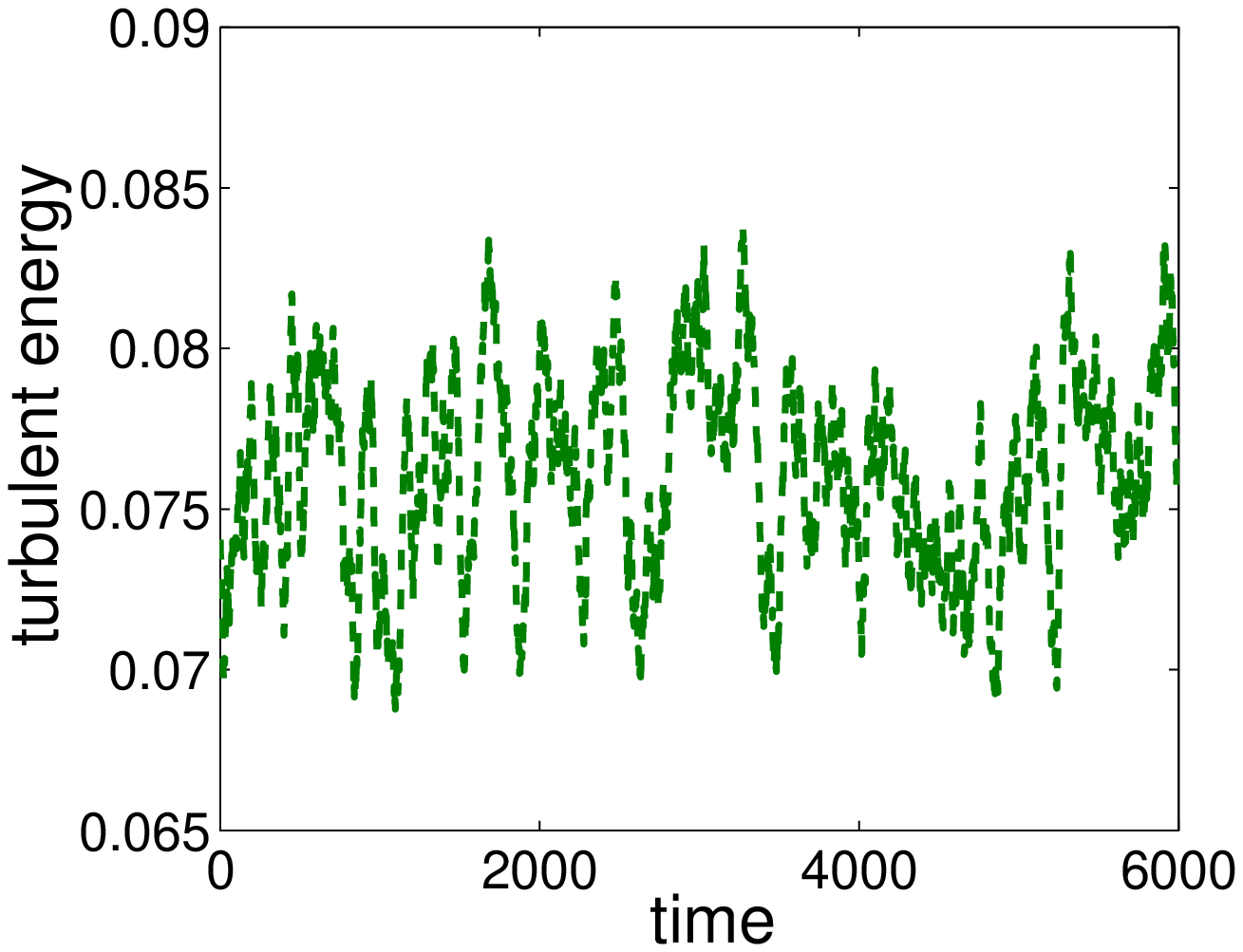}
\EC
\caption{Time series of $m$ (left), $e$ and $f/10$ (center) and $e_{\rm t}$ (right) zoomed on the appearance and disappearance on
an uniformly turbulent flow for $L_x\times L_z=110\times 32$ and $R=335$. 
\label{fig17}}
\EF
They are not observed for $R\le333$, and go from extremely rare at
$R=335$ and $337$ to common at $R=340$, to the most common state at
$R=345$ (though a trace of modulation persists).

\begin{figure}[!h]
\BC
\includegraphics[width=0.38\textwidth,clip]{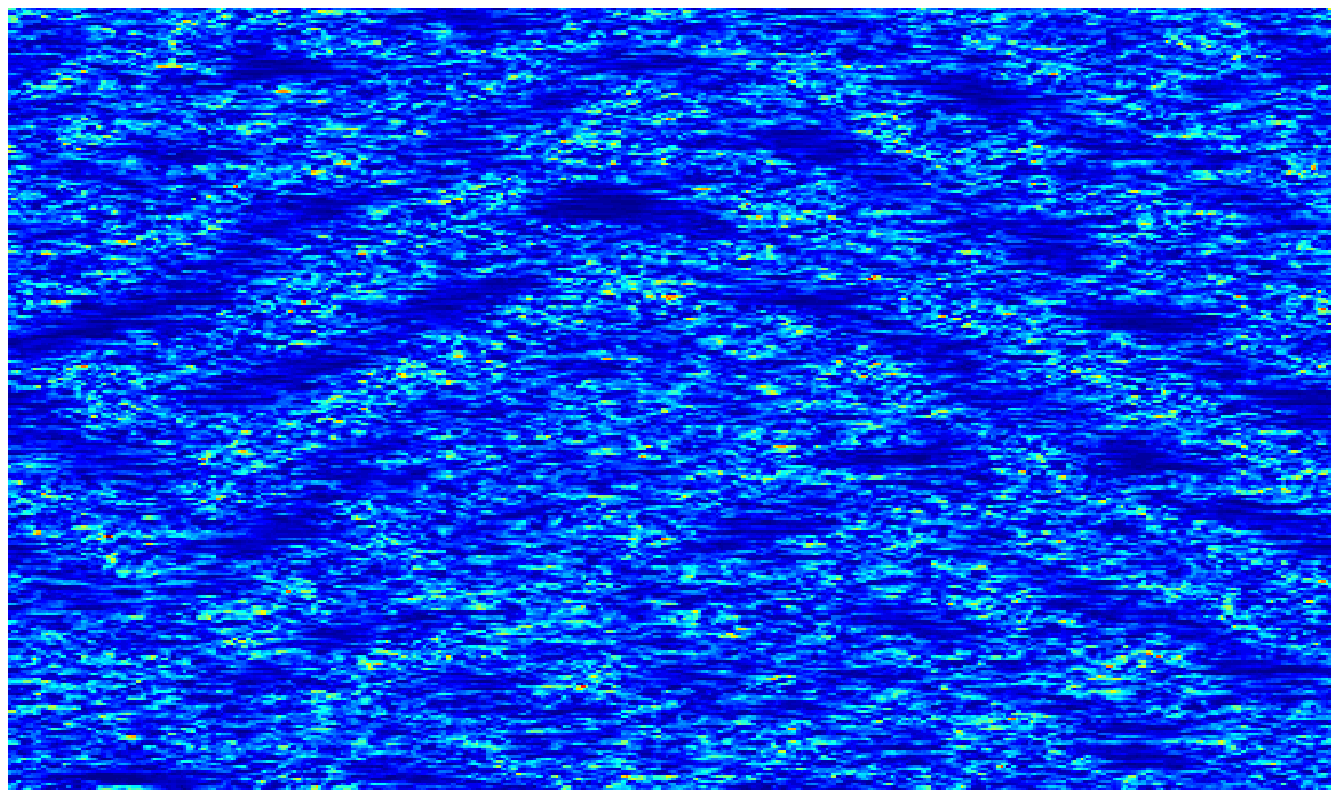}
\EC
\caption{Snapshot of the numerical solution for $R=340$ at $t=20000$;
same simulation conditions as in Fig.~\ref{fig1} (bottom-right), in particular $L_x=432$, $L_z=256$.\label{fig17p}}
\end{figure}

Since it is a three-state jump process instead of a two-state one, this feature should be treated appropriately following the same procedure as for orientation fluctuations. However, it cannot be accounted
for by the plain model (\ref{E-rnd1}) since empirical PDFs for $R=337$ (Fig.~\ref{fig12}, right) and higher clearly present three maxima, one of
which is close to the origin ($m_+\simeq0\simeq m_-$). The phenomenon can
however be treated within the same conceptual framework by assuming a
slightly modified potential with an additional relative minimum at the
origin separated by saddles from the main minima corresponding to the
pattern installed in one or the other orientation, justified by the appearance of a third maximum in the PDFs. The splitting
probability between the featureless regime and the pattern would then
be controlled by the relative depths of the three wells \cite{vk83},
which could be studied by following the procedure for orientation
fluctuations \cite{crossref}. This complication has however not been
explored further because the phenomenon is likely a size effect: In the
upper transitional regime at large aspect-ratio, bands form out of
scattered elongated
regions where turbulence is depleted, see Fig.~\ref{fig17p}.
The computational domains
considered here are just sufficient to contain a pattern cell of size
$\lambda_x\times\lambda_z$. It is therefore not surprising that the
spatiotemporal intermittence of laminar troughs comparable in size to
that cell be turned into temporal intermittence
of well-formed laminar bands recurrently destroyed by featureless
turbulence. The improved modelling suggested above would transform
the supercritical bifurcation into a slightly subcritical one, with
associated coexistence of featureless and patterned states, as expected from system where a spatial and temporal cohabitation of different states is possible. This
would explain the shape of the PDFs once noise is introduced as for
the original model. The same explanation,
if correct, would explain the presence of the `intermittent regime'
described, although not fully investigated, by Barkley \& Tuckerman~\cite{BT05-07,BTD} since turbulence modulations around $R_{\rm t}$ are also much longer than the width
of the oblique computational domain they considered.

\begin{figure}[!h]
\BC
\includegraphics[width=0.42\textwidth,clip]{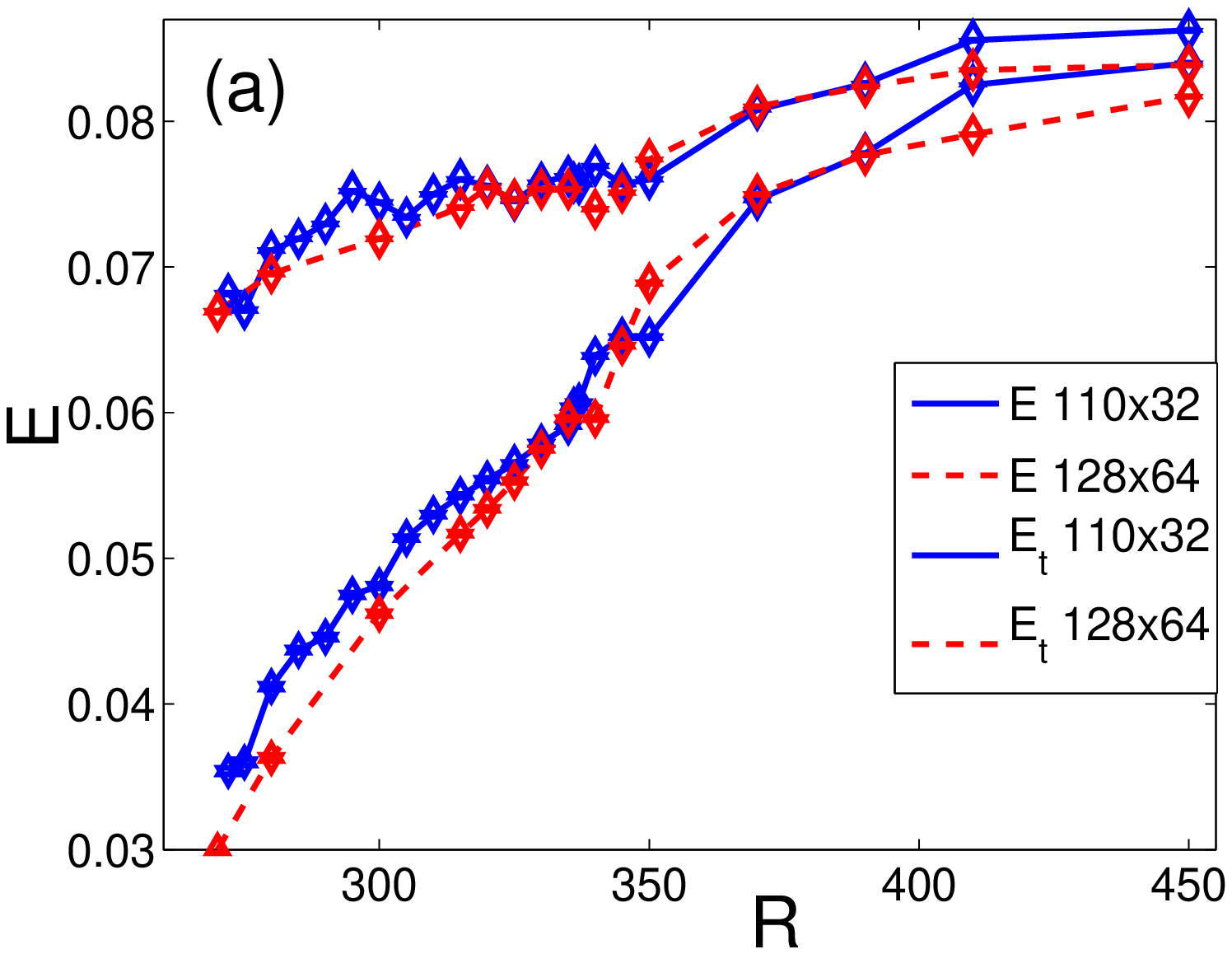}
\includegraphics[width=0.42\textwidth,clip]{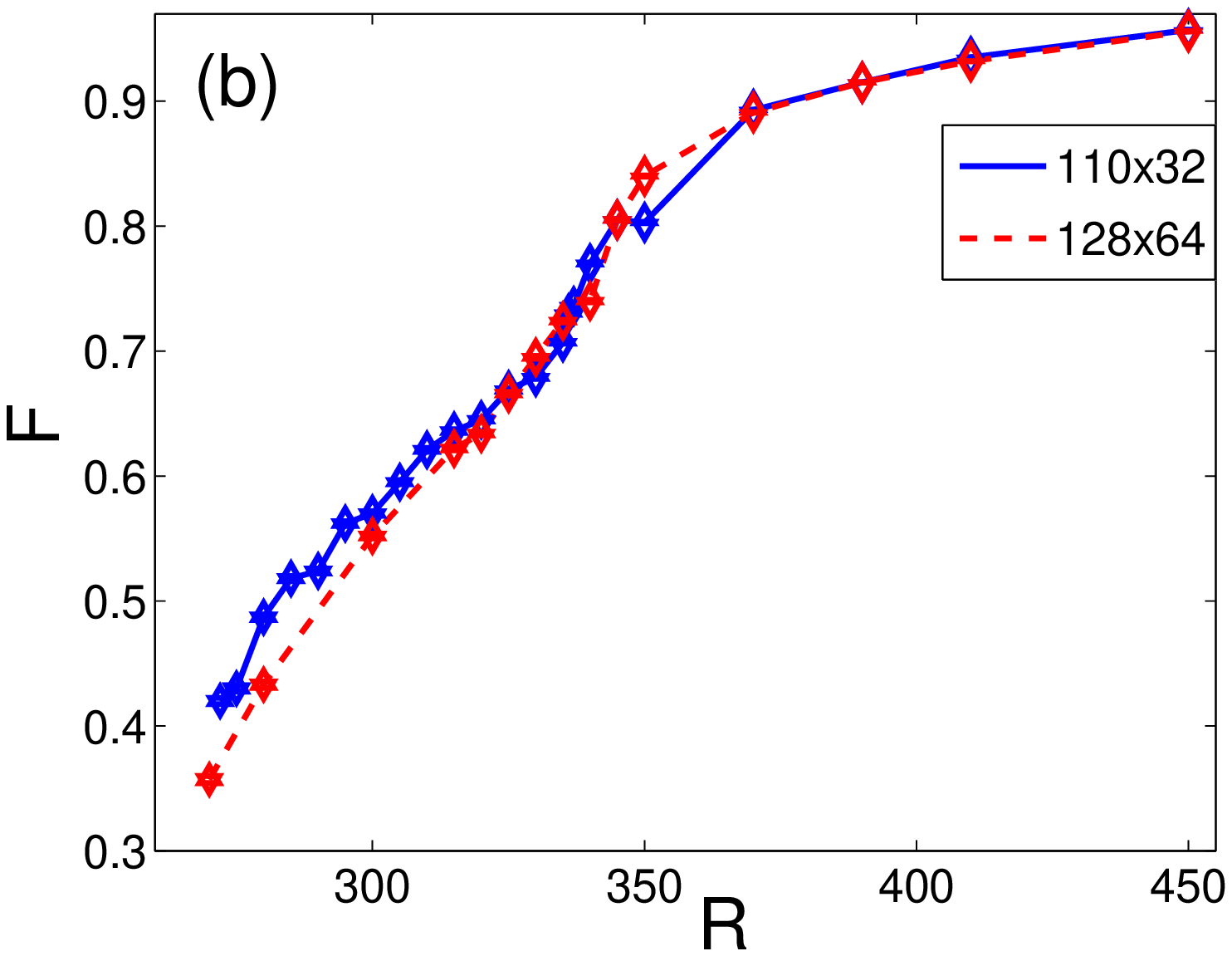}\\
\includegraphics[width=0.42\textwidth,clip]{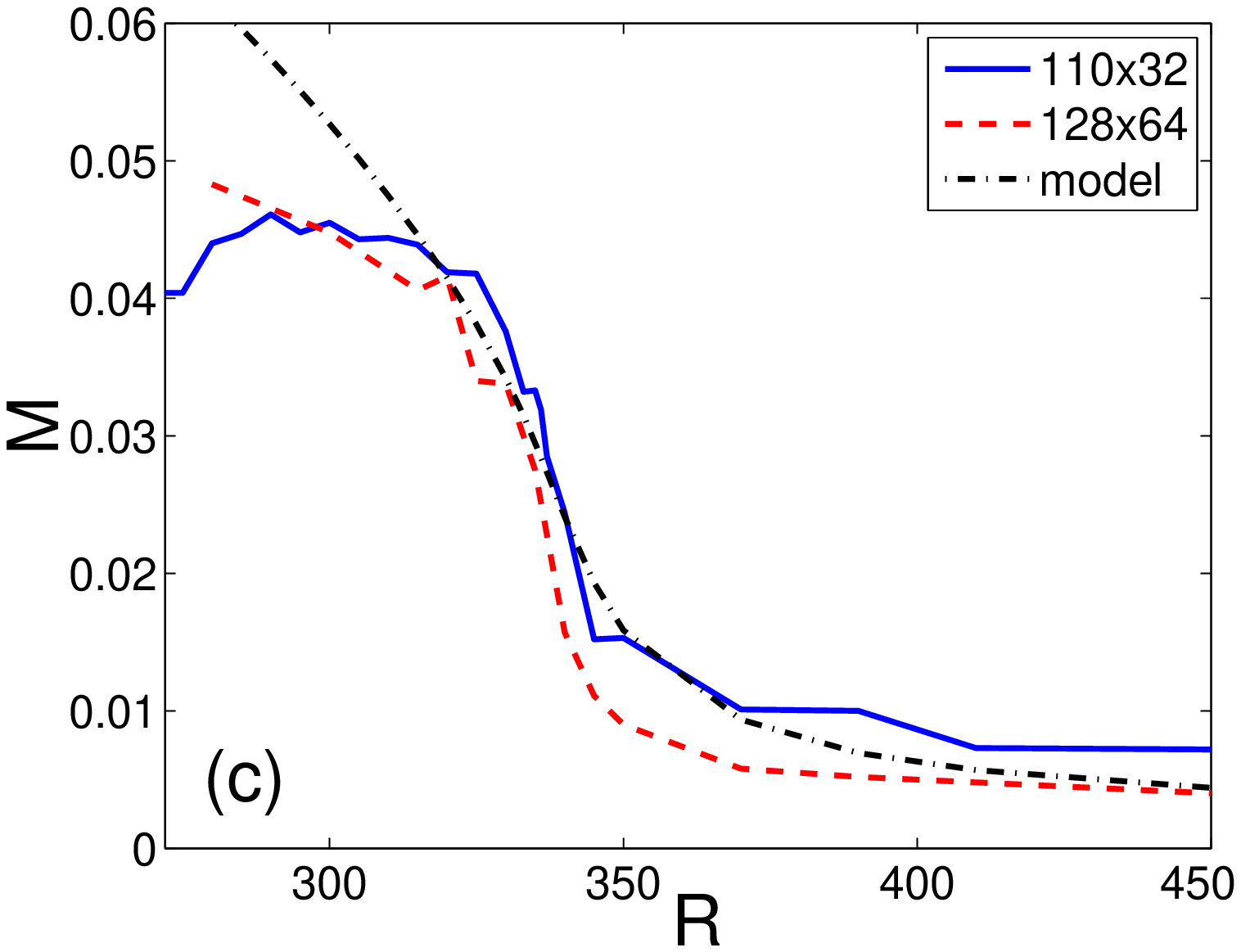}
\includegraphics[width=0.42\textwidth,clip]{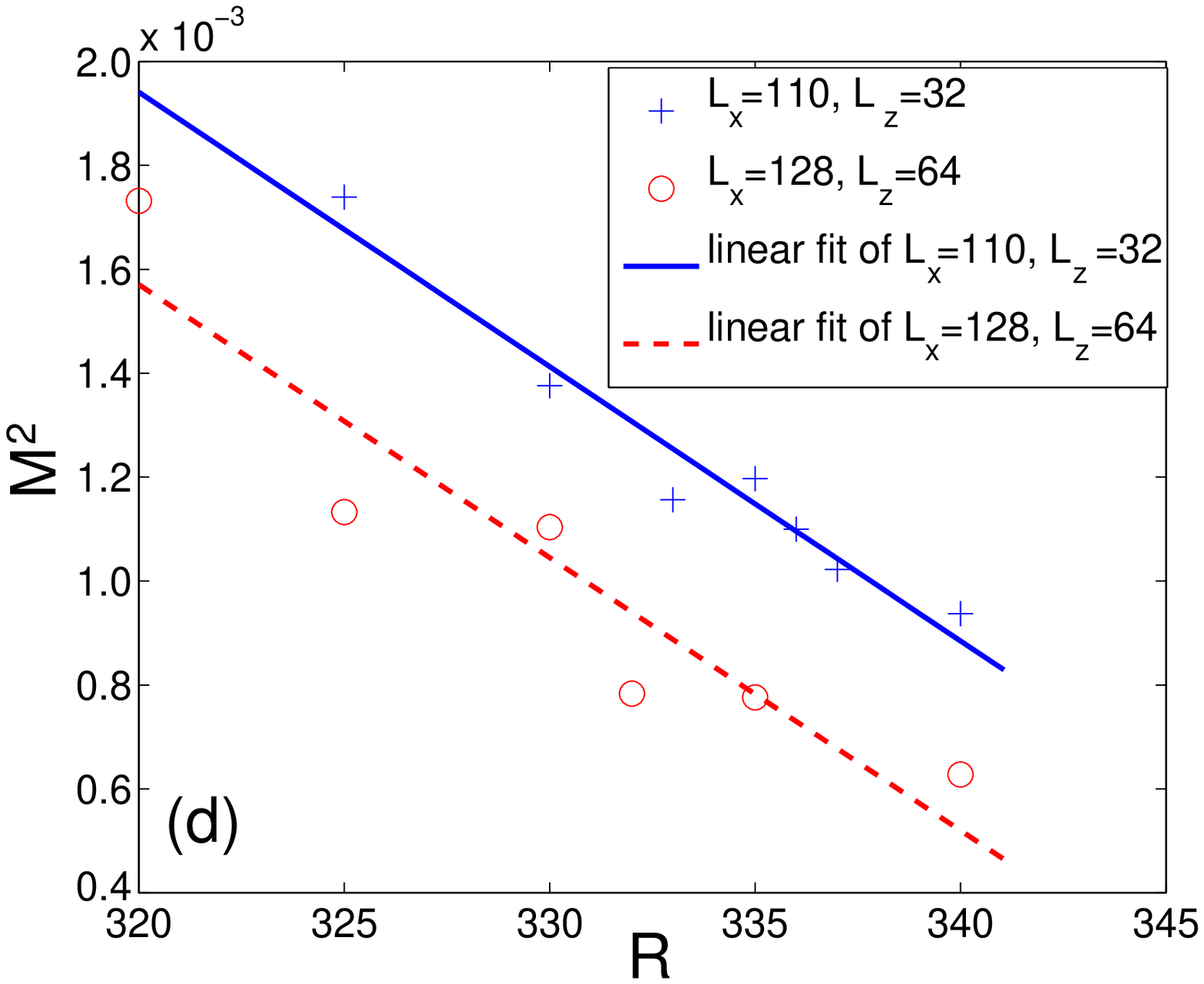}
\EC
\caption{Variation with $R$ of the different averaged quantities for
domains $L_x\times L_z=110\times 32$ (blue on line) and $128\times 64$
(red on line): (a)
Perturbation energy $E$ (full line) and turbulent energy $E_{\rm t}$
(dashed line). (b) Turbulent fraction $F$.
(c) Order parameter $M$ and corresponding values from the model
(dash-dotted line, $g_1=55$, $g_2=250$, $\alpha=0.002$, $R_{\rm t}=355$).
(d) Square of the order parameter $M^2$, observations ($\circ: 128\times 64$ and $+ : 110\times 32$)
and linear fits (lines).\label{fig18}}
\end{figure}

Figure~\ref{fig18} displays the variations of the different observables of interest with $R$.
The growth of the width of the turbulent domain illustrated in
Fig.~\ref{fig17} is clearly reflected by the increase of $F$ with $R$ in
Fig.~\ref{fig18}~(b). Quantity $F$ varies roughly linearly with
$R$ for $R<R_{\rm t}$, and with a much smaller slope above. This increase mostly
explains the growth of the perturbation energy since the turbulent
energy depends more weakly on $R$, with no singular behaviour visible
at $R_{\rm t}$ (Fig.~\ref{fig18}, a): the intensity of
turbulence inside the turbulent domains does not seem sensitive
to the global organisation in an oblique pattern. The slope
discontinuity at $R_{\rm t}\simeq 350$ marking the bifurcation was used
as a criterion in our previous study \cite{MRxx}.
Values obtained for $E$, $E_t$, and $F$ are slightly different
for the two sizes considered, which is related to lateral
confinement effects already illustrated in Fig.~\ref{fig13} (left)
for $R=315$.

In Figure~\ref{fig18} (c), it can be seen the order parameter $M$
departs from the expected classical $\tilde\epsilon^{1/2}$ behaviour and
tends to saturate in the lowest part of the transitional range. For
$L_x\times L_z=110\times32$, it even decreases as $R$ is lowered further,
which is again a confinement effect since, from the experiments
\cite{Petal} as well as from our earlier (less well resolved) numerical
results \cite{MRxx}, the spanwise wavelength $\lambda_z$ is expected to
increase up to about $80$ as $R$ decreases: this implies a less optimal
pattern and a weaker modulation for $L_z=32$, while for
$L_x\times L_z=128\times64$, with a more favourable $\lambda_z$, $M$
continues to increase as $R$ is lowered in agreement with
the Ginzburg--Landau picture. In the upper part of the transitional
range, $M$ decreases quickly as $R$ increases. The decay of the
modulation corresponds to the increase of the width of the turbulent
domain. Again in line with the Ginzburg--Landau interpretation, the
variation of $M^2$ with $R$ (Fig.~\ref{fig18}, d) appears to be linear
with a slope $1/(g_1R_{\rm t})\simeq-5.3\times 10^{-5}$. Meanwhile,
the extrapolation of $M^2$ to zero gives a value of
$R_{\rm t}\approx 355$ or $348$ depending on whether one takes
the data from case $L_x\times L_z=110\times32$ or $128\times64$, respectively. In contrast with what happens for $R\sim R_{\rm g}$,
here the estimate with $L_z=32$ is likely the best one
since experiments suggest $\lambda_z\simeq37.5$
either by extrapolation for plane Couette flow or from measurements
in the CCF \cite{Petal}.
Taking $R_{\rm t}\simeq 355$, we get $g_1\simeq55$, which is consistent
with Prigent's value $g_1\sim 100$ in the CCF case.
This value of $g_1$ further yields
$\xi_z\simeq2.3$ (measured values for CCF range
between $0.9$ and $3.2$) and $\xi_x\simeq11$ for $R=315$.

In Fig.~\ref{fig18} (c), it can be noticed that $M$ remains finite
for $R>R_{\rm t}$, as the result of intrinsic fluctuations in the
featureless turbulent regime, in contrast with what would happen
in the deterministic case. Fluctuations indeed gives a finite
background level to modes $m_\pm$, a fact which is well accounted for
by the model in the mean-field approximation represented by
dash-dotted line in Fig.\ref{fig18} (c).

\begin{figure}
\centerline{\includegraphics[width=7cm,clip]{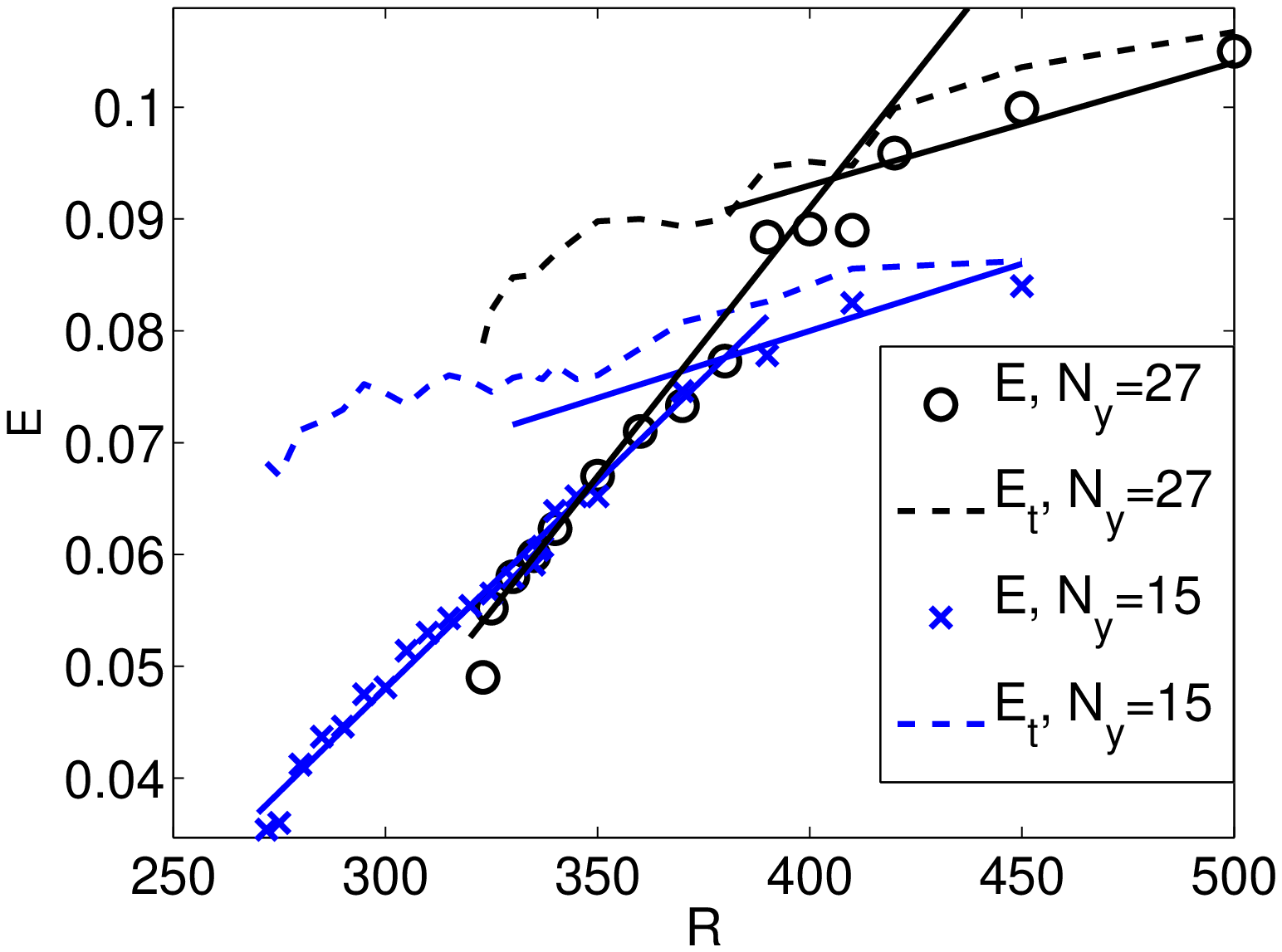}\includegraphics[width=7cm,clip]{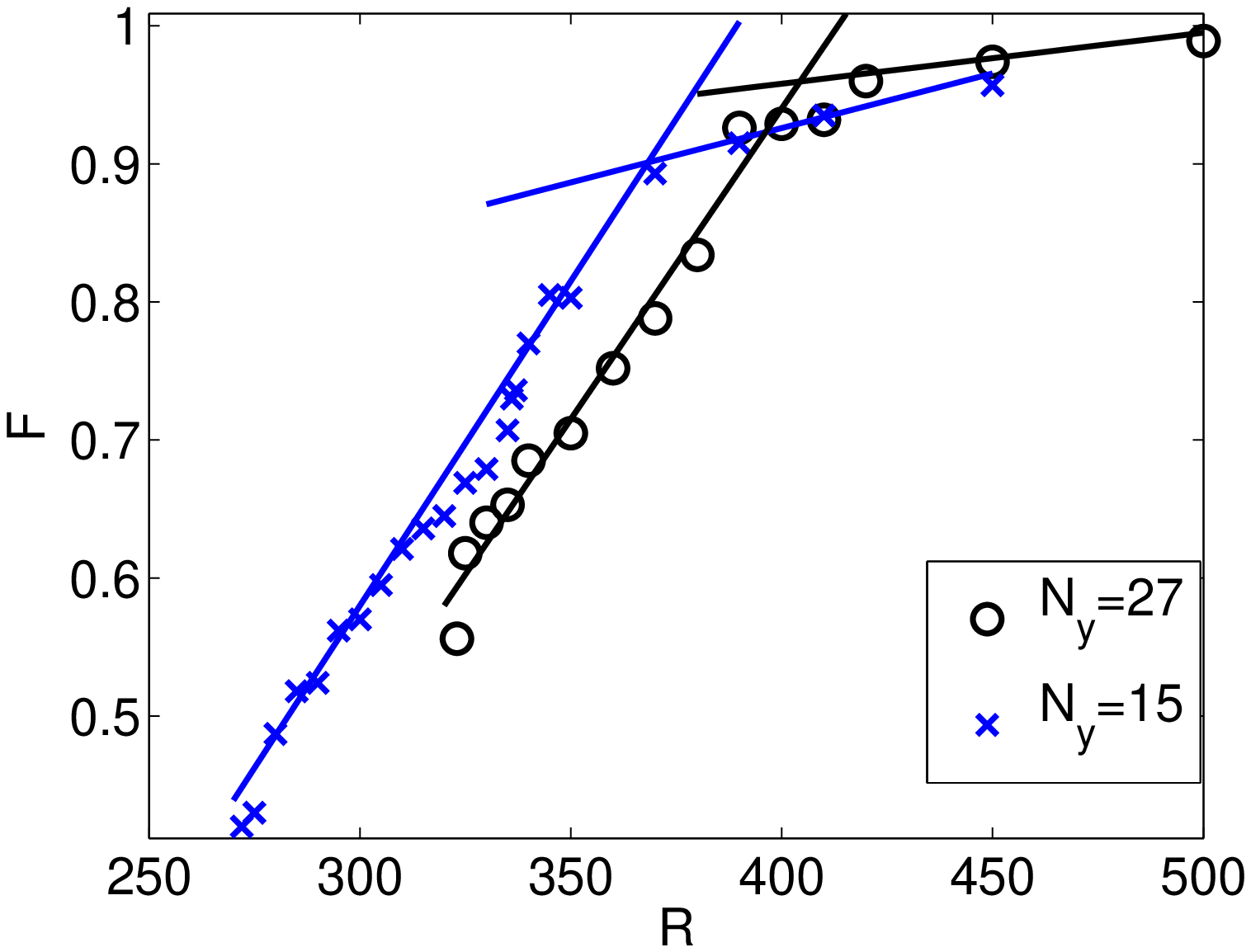}
\includegraphics[width=7cm,clip]{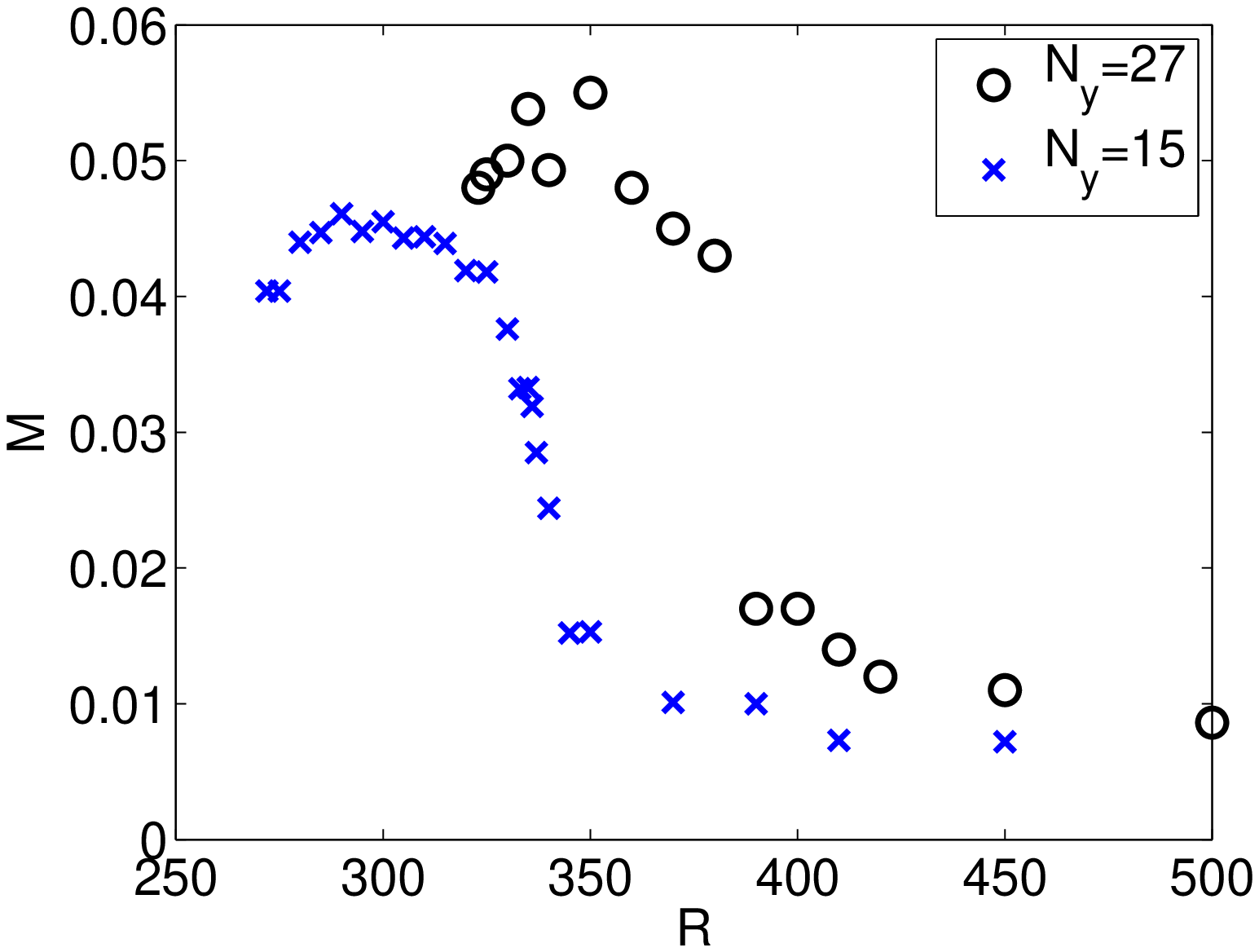}}\caption{Turbulent quantities $E$, $E_t$ (left) and $F$ (center), order parameter $M$ as function of $R$ for a domain of size $L_x\times L_z=110\times 32$ (right), for two resolutions : $N_y=27$, $N_{x,z}/L_{x,z}=6$ (circles) and $N_y=15$, $N_{x,z}/L_{x,z}=4$ (crosses)}\label{fig18b}
\end{figure}

  These result are not qualitatively affected by the increase of resolution from $N_y=15$ and $N_{x,z}/L_{x,z}=8/3$ to $N_y=27$ and $N_{x,z}/L_{x,z}=4$ as can be seen in Figure~\ref{fig18b} which compares the results for both resolutions at size $L_x\times L_z=110\times32$. The quantitative change is minor and in the expected fashion\cite{MRxx}. The thresholds $R_{\rm g}$ and $R_{\rm t}$ move to approximately $321$ and $390$. The square of the perturbation undergoes an increase of about $10\%$. Apprt from the threshold shift, the turbulent fraction $F$ is little affected by the resolution change. Both $E$ and $F$ display the expected slope-break. Given the uncertainty on the values of $M$ near $R_{\rm t}$, the value $g_1\simeq 30$ at $N_y=27$ is acceptable. This reassert the validity of our semi-quantitative approach.

Parameter $g_2$ has little influence and reasonable results are
obtained from $0.001\lesssim \alpha \lesssim 0.003$. This estimate is consistent with the value
obtained from the fit of the phase dynamics fit $\alpha/\tau_0\sim 4\times 10^{-4}$, (Fig.~\ref{fig7}) if we
accept Prigent's finding $\tau_0 \sim 30\, h/U$ \cite[(a,c)]{Petal}. The variance of the fluctuations
of $m$ in the vicinity of $R_{\rm t}$ is also of interest. Let us define:
\BF
\BC
\includegraphics[width=0.42\textwidth,clip]{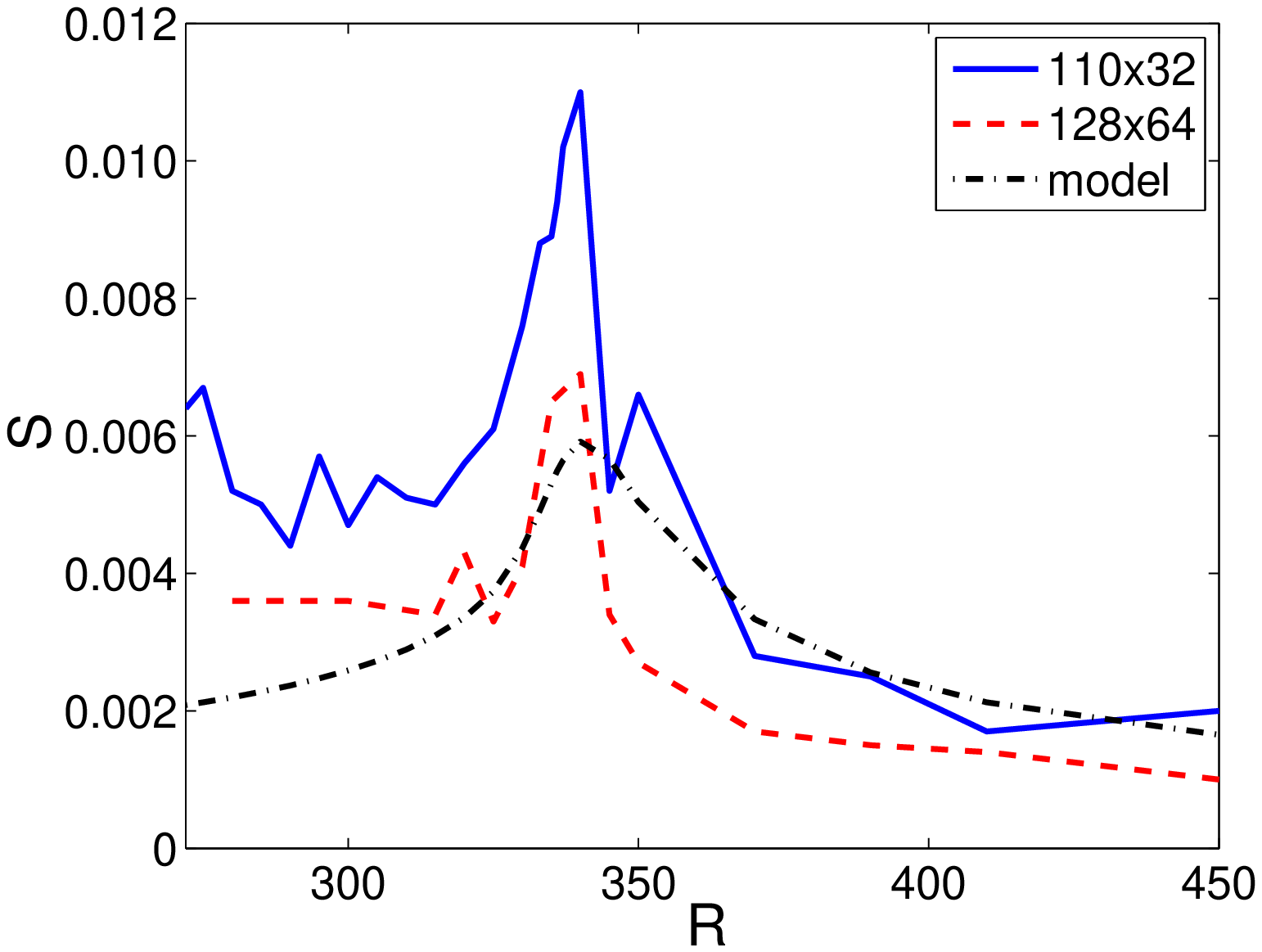}
\EC
\caption{Average root-mean-square fluctuation of $m$ as a function of
$R$ for $L_x\times L_z=110\times 32$ (full line), $128\times 64$ (dashed),
and the model  with $g_1=55$, $g_2=250$, $R_{\rm t}=355$,
$\alpha=0.002$ (dash-dotted).\label{fig19}}
\EF
\BM
S^2=2\int_{m''<m'} (m'-M)^2 \Pi(m',m'')\,
 {\rm d}m'{\rm d}m''\,.
\EM
Figure~\ref{fig19} displays the variation of $S$ as a function of~$R$.
Fluctuations appear to be strongly enhanced in the vicinity of
$R_{\rm t}$, due to orientation changes and re-entrance of featureless
turbulence. Though model (\ref{E-rnd1}) does not account for the latter
phenomenon, it already explains a large part of the enhancement.
Its parameter $g_2$ controls the amplitude of fluctuations that bring
about orientation changes. The position of the maximum of $S(R)$
strongly depends on it. With $g_1=55$, satisfactory agreement is found
for $g_2\gtrsim120$. Results obtained with $g_2=250$ are represented as
a dash-dotted line in Fig.~\ref{fig19}. Including the re-entrance of
featureless turbulence would certainly increase the variability but this
would still not be the whole story since, like for second order phase
transitions, one would expect a divergence of $S$ in the form
$S\propto |\tilde\epsilon|^\gamma$, $\gamma$ being the critical exponent
attached to the susceptibility of the order parameter, just rounded
off by finite-size effects. Even at reduced numerical resolution, improving
the statistics to study the pattern's fluctuations in the simulations
seems presently out of reach.
\section{Summary and Conclusion\label{S3}}

Prigent {\it et al.} \cite{Petal} have put the problem of the emergence
of turbulent bands in wall-bounded flows within the Ginzburg--Landau framework of pattern formation, adding noise to account for background
turbulence. Doing so, they were able to extract most of the coefficients
in the model equation from laboratory experiments in the case of
circular Couette flow, while restricting themselves to threshold localisation and wavelength measurements for PCF. In a similar vein,
Barkley {\it et al.} \cite{BTD} later performed simulations of PCF,
detecting the formation of bands from Fourier analysis of the pattern.
They considered a quasi-one-dimensional configuration excluding
orientation fluctuations expected to play a role close to $R_{\rm t}$
for symmetry reasons. Though having the model in mind, they did not
attempt any quantitative fit. Our work has been mostly devoted to
overcome these two limitations, to check the validity of the noisy Ginzburg--Landau framework, and to compare finding for PCF to those for
CCF. Previous result \cite{MRxx} were reasserted,
showing that controlled under-resolution gives excellent qualitative
agreement with experiments and good quantitative results once corrected
for a general shift of the range $[R_{\rm g},R_{\rm t}]$ where the bands
are present. We performed numerical experiments in domains of sizes able
to contain one to three bands in the spanwise direction and one or two
bands in the streamwise direction, while letting the pattern's orientation
fluctuate. Under-resolution reducing the computational load, we could
carry out long duration simulations in order to accumulate reliable
statistics.
 
The emergence of bands was first quantitatively characterised using
standard statistical quantities such as the total perturbation
energy $E$, the turbulent fraction $F$, and the average energy
contained in turbulent domains $E_{\rm t}$. These quantities quickly
converge to their steady-state values but do not give information on
orientation or wavelength fluctuations. This limitation has been next overcome by
defining order parameters measuring the amplitude of the modes involved
in the Fourier series decomposition of the patterns, appropriately
amending the Barkley {\it et al.} definitions and procedure.
The full nonlinear dispersion relation describing the formation of bands
could be studied by varying the Reynolds number and the size of the
computational domain which controls the allowed wavevectors. The
coefficients of the relevant Ginzburg--Landau equation and the intensity
of the noise were estimated, showing the overall consistency of the
approach. In particular, two coherence lengths, spanwise and streamwise,
were evaluated and the square of the modulation amplitude was shown to
vary linearly with $R$ far enough from $R_{\rm t}$, while its fluctuations
and the intermittent re-entrance of featureless turbulence were strongly
enhanced close to $R_{\rm t}$. It has been argued that the re-entrance of
featureless turbulence was a side effect of the limited size of the
system, probably explaining the `intermittent regime' of Barkley
\& Tuckerman \cite{BT05-07} by the same token, and that this observation
should be better replaced on a spatiotemporal footing in more extended
domain, in relation to patterns with mixed orientations
observed near $R_{\rm t}$ in CCF experiments \cite{Petal}
or PCF simulations in Fig.~\ref{fig17p}.  
Finally, comparing our results with those obtained in CCF
we obtain satisfactory general agreement, but with the supplementary
information that the streamwise coherence length $\xi_x$ is significantly
larger than the spanwise coherence length $\xi_z$ indicating that the
selection of the streamwise wavelength~$\lambda_x$ is more effective
than that of the spanwise wavelength~$\lambda_z$.

As a whole, the emergence of oblique bands from featureless turbulence
upon decreasing $R$ has been seen to fit the conventional framework of
a pattern-forming instability. However, the very fact that the base
state is turbulent calls for the introduction of a large noise in the
picture. These numerical studies are performed with the hope that they will
contribute to the understanding of the cohabitation of turbulent and
laminar flow typical of the transition to/from turbulence in wall-bounded
flows, the detailed mechanism of which is still largely unknown.

\end{document}